\documentclass[twocolumn,aps,prb,showpacs]{revtex4}
\usepackage{amsmath}  
\usepackage{nccmath}
\usepackage{amssymb} 
\usepackage{graphics}
\usepackage{graphicx} 
\usepackage{epsfig}
\usepackage{epstopdf}
\usepackage{color}
 \usepackage{rotating}
 \usepackage{bm}
 \usepackage{xspace}
 \usepackage{bbold}
 \newcommand{\Mo}{Mo$_3$S$_7$(dmit)$_3$\xspace}
 \newenvironment{mbmatrix}{\begin{medsize}\begin{bmatrix}}%
    {\end{bmatrix}\end{medsize}}%
 \begin{document}
 \title{ Effects of anisotropy in spin molecular-orbital coupling on effective spin models of trinuclear organometallic complexes}
 \author{J. Merino}
\affiliation{Departamento de F\'isica Te\'orica de la Materia Condensada, Condensed Matter Physics Center (IFIMAC) and
Instituto Nicol\'as Cabrera, Universidad Aut\'onoma de Madrid, Madrid 28049, Spain}
  \author{A. C. Jacko, A. L. Khosla}
 \affiliation{School of Mathematics and Physics, The University of Queensland,
Brisbane, Queensland 4072, Australia}
 \author{B. J. Powell}
 \affiliation{School of Mathematics and Physics, The University of Queensland,
Brisbane, Queensland 4072, Australia}
 \begin{abstract}
We consider layered  decorated honeycomb lattices at two-thirds filling, as realized in some trinuclear organometallic complexes.  Localized $S=1$ moments with a single-spin
anisotropy emerge from the interplay of Coulomb repulsion and spin molecular-orbit coupling (SMOC).  
Magnetic anisotropies with bond dependent exchange couplings occur in the honeycomb layers
when the direct intracluster exchange and the spin molecular-orbital coupling are both present. We find that the effective spin exchange 
model within the layers is an XXZ + 120$^\circ$ honeycomb quantum compass model.
 The intrinsic non-spherical symmetry of the  multinuclear complexes
leads to very different transverse and longitudinal spin molecular-orbital couplings, which greatly enhances the single-spin 
 and exchange coupling anisotropies. 
The interlayer coupling is described by a XXZ model with anisotropic biquadratic terms.   
As the correlation strength increases the systems becomes increasingly one-dimensional. 
Thus, if the ratio of SMOC  to the interlayer hopping is small this stabilizes the Haldane phase. However, as the ratio increases there is a quantum phase transition to  
the topologically trivial `$D$-phase'. We also predict a quantum phase transition from a Haldane phase 
to a magnetically ordered phase at sufficiently strong external magnetic fields.
 \end{abstract}
 \date{\today}
 \maketitle
 
 \section{Introduction}
 
The interplay of strong Coulomb interaction and spin-orbit coupling (SOC) can lead to emergent quantum phases \cite{balents2014} 
and new phenomena which remain poorly understood. The conventional Mott transition can be strongly affected by SOC leading
to a topological Mott insulator with bulk charge gap but fractionalized surface states carrying spin but no charge 
\cite{pesin2010}.  Such states may be realized in Ir-based transition metal oxides such as Sr$_2$IrO$_4$.
In contrast to conventional Mott insulators, the spin exchange interactions arising in Mott insulators with SOC are typically 
anisotropic with quantum compass \cite{nussinov2015} instead of the 
conventional Heisenberg interactions.  A possible realization of  a quantum compass model on a hexagonal lattice, {\it i.e.}, 
a Heisenberg-Kitaev model,\cite{jackeli2009,perkins2014a,perkins2014b} may be found in Na$_2$IrO$_3$ and Li$_2$IrO$_3$
materials in which SOC removes the orbital degeneracy of the 5d electrons leading to effective $S=1/2$ pseudospins.
Interestingly, the Kitaev model is exactly solvable: it sustains a spin liquid ground state whose low energy excitations are Majorana 
fermions \cite{kitaev2006}. In other iridates with strong SOC such as Sr$_2$IrO$_4$, an antisymmetric  
Dzyaloshinski-Moriya (DM) interaction arises associated with the lack of an inversion symmetry center.

There are several  strongly correlated molecular materials in which spin-orbit coupling
is relevant including, metal-organic frameworks,\cite{oshikawa2016}  layered organic salts,\cite{winter2017,elise2017} and multinuclear coordinated organometallic complexes.\cite{hoffman,khosla2017,jacko2017,merino2016,powell2016} The elementary building blocks of multinuclear complexes are molecular clusters 
containing transition metal ions whose d-orbitals are hybridized with molecular orbitals where each of the 
hybrids is typically described by a single Wannier orbital\cite{jacko2015}. The coupling of the spin with the electron currents around 
the Wannier orbitals describing each molecule gives rise to a spin molecular-orbital coupling (SMOC) \cite{khosla2017,jacko2017}.

\begin{figure}
	\centering 
	\includegraphics[width=0.9\columnwidth]{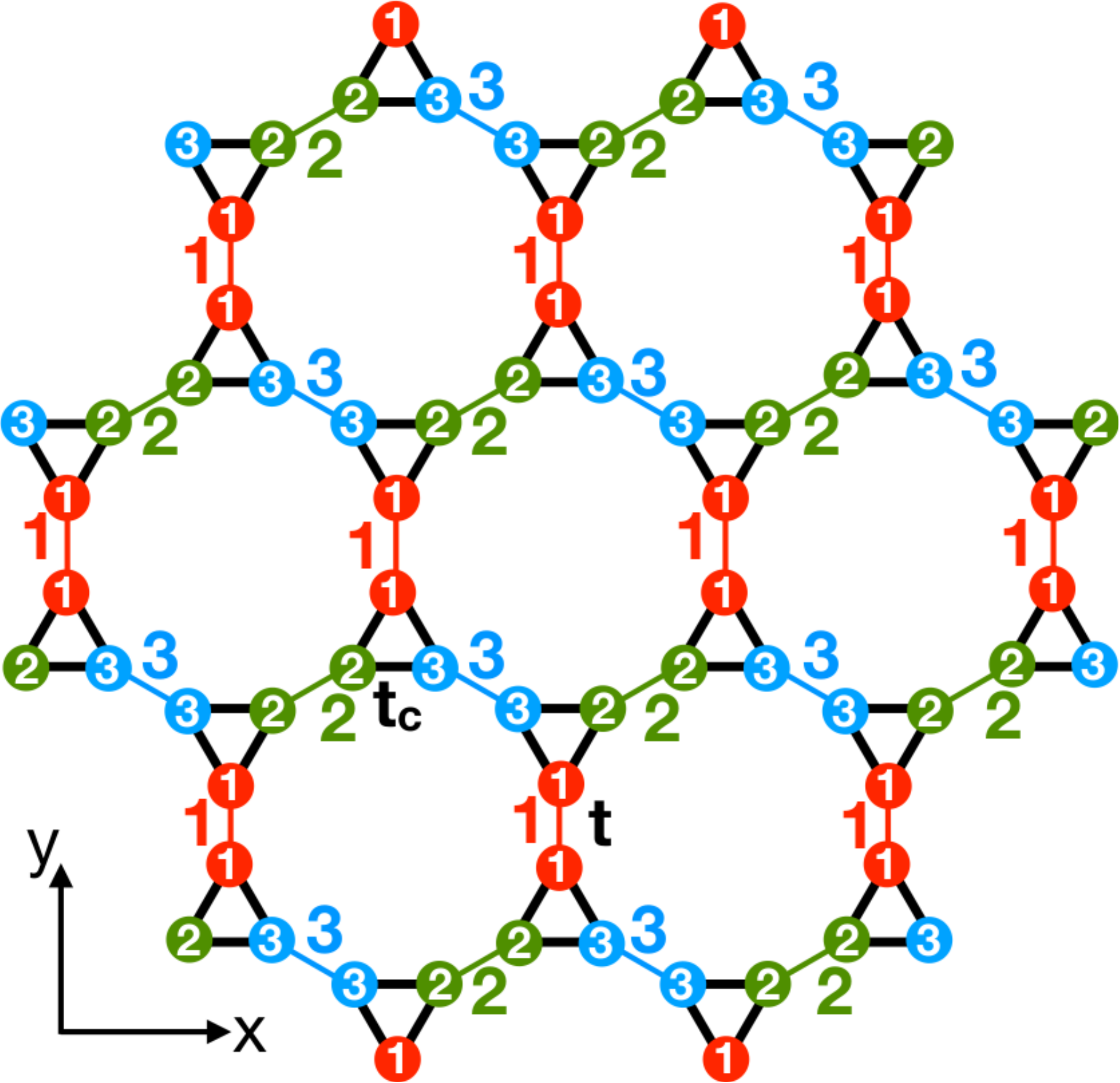}
	\caption{The decorated honeycomb lattice realized in the $a$-$b$ planes of Mo$_3$S$_7$(dmit)$_3$ crystals.  The small triangles represent the organometallic trinuclear 
	clusters located at sites of the honeycomb lattice. The intracluster hopping, $t_c$, and the intercluster hopping, $t$, entering our model (\ref{eq:genham}) are also shown. Note 
	the labeling (color coding) of both the sites within the trinuclear clusters and the intracluster $t$-bonds. The 
		full crystal consists of these decorated honeycomb layers stacked along the $c$-direction, see Fig. \ref{fig:tubes}b.}
	\label{fig:decorated}
\end{figure}

A typical multinuclear complex is Mo$_3$S$_7$(dmit)$_3$. Here the honeycomb networks of Mo$_3$S$_7$(dmit)$_3$ molecules are stacked on 
top of each other along the $c$-direction of the crystal.  
Mo$_3$S$_7$(dmit)$_3$ molecules can be described by three Wannier orbitals \cite{jacko2015}, and their packing on a honeycomb lattice within the layers leads to a decorated honeycomb lattices, 
as shown in Fig. \ref{fig:decorated}. The electronic and magnetic properties of the decorated honeycomb lattice
are interesting both in the weakly and strongly interacting limit. At weak coupling, a tight-binding model on such 
a lattice leads to topological insulating phases when SOC is turned on which display the quantum spin 
hall  effect\cite{ruegg2010} as predicted in graphene\cite{kane2005}.  
At strong coupling, the exact ground state of the Kitaev model on the decorated honeycomb lattice \cite{kivelson2007}, 
is a chiral spin liquid. Therefore, it is interesting
to find possible realizations of the decorated honeycomb lattice in actual materials to probe such rich physics.
Furthermore, similar models arise naturally in a number of other organic\cite{elise2017} and organometallic materials\cite{hoffman,Silveira,Wang1,Wang2,Wang3} and 
inorganic compounds with decorated lattices.\cite{Sheckelton,Bao,Chen}

\begin{figure}
	\centering 
	\includegraphics[width=0.6\columnwidth]{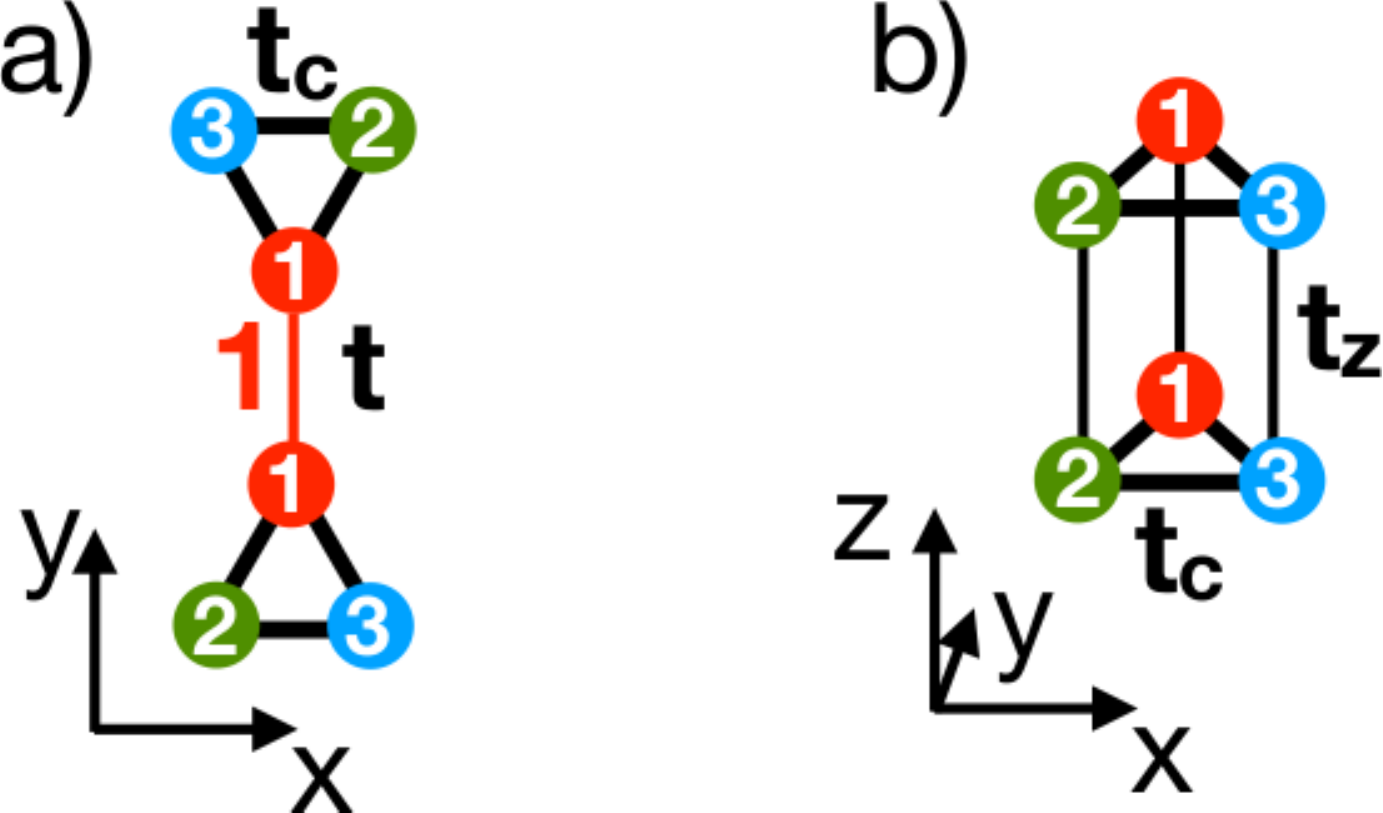} 
	\caption{The two arrangements of two neighbor trimers relevant to Mo$_3$S$_7$(dmit)$_3$ 
		crystals. In (a) we show two neighbor trimers in the $a$-$b$ plane whereas in (b) the two trimers are stacked 
		along the $c$-direction. In the dumbbell arrangement (a) the two molecules are related by inversion symmetry through 
		the midpoint of the bond while in the tube arrangement (b) they are related by translational symmetry. }
	\label{fig:tubes}
\end{figure}

Organometallic complexes have intrinsic structural properties which make them  different to transition metal 
oxides.  A crucial difference comes from the fact that isolated molecular clusters break the spherical 
symmetry present in isolated transition metal ions. While the total angular momentum of the ion is 
conserved, it is only the component perpendicular to the molecular plane that 
is conserved in cyclic molecular clusters. Hence, in these systems, anisotropies are intrinsic to the 
molecules constituting the material, whereas in transition metal oxides anisotropies can only be achieved via the 
environment surrounding the ions in the crystal. This suggests that anisotropic spin exchange interactions
may be easily generated in organometallic complexes due their intrinsic structure. These anisotropies may be
further enhanced by the anisotropic SMOC typically found in these systems. \cite{jacko2017,powell2016}
SMOC is an emergent coupling between electron currents circulating around the cyclic molecules and the electron spin. 
Also by tuning the relative orientation between molecules in the crystal a Dzyaloshinskii-Moriya interaction
can be generated\cite{powell2016}. All the above  suggests that these materials are ideal playgrounds 
for the realization of quantum compass models.\cite{nussinov2015}

Recently \cite{merino2016,powell2016} we  derived an effective super-exchange Hamiltonian that captures the  
magnetic properties of trinuclear coordinated complexes at strong coupling. The onsite Coulomb 
repulsion, $U$, leads to $S=1$ moments localized at each triangular cluster whence SMOC, $\lambda$,  induces 
a single-spin anisotropy, $D$. The $S=1$ moments behave as weakly coupled chains due
 to the decorated lattice structure of trinuclear organometallic complexes. \cite{merino2016,powell2016} The lattice structure is such
 that three hopping amplitudes connect two nearest-neighbor molecules along the $c$-direction while  
 only one hopping amplitude connects nearest-neighbor molecules in the $a$-$b$ planes, cf. Fig. \ref{fig:tubes}.  As $U$ is increased 
exchange of electrons between nearest-neighbor molecules in the $a$-$b$ plane is suppressed as compared to
exchange between molecules along the $c$-direction. This leads to a quasi-one-dimensional effective spin exchange model
of $S=1$ localized moments which is in the Haldane phase.\cite{powell2016}

Here we extend our previous work, which focused on \Mo, by studying the more general problem of trinuclear organometallic complexes with strong 
correlations and   strong anisotropic SMOC.
After introducing our general combined analytical and numerical approach to extract exchange coupling
parameters in these systems we show how anisotropy in SMOC plays a crucial role in determining the level of anisotropy of the 
effective spin Hamiltonian. We show that the effective spin exchange Hamiltonian 
for two-thirds filled trinuclear coordination crystals is  
	\begin{widetext}
		\begin{eqnarray}
		H_\text{eff}&=& D^* \sum_{\ell} (\mathcal{S}_{\bm r_\ell}^{z})^2 
		+J^c \sum_{\ell} \left( \mathcal{S}_{\bm r_\ell}^{x} \mathcal{S}_{\bm r_\ell+\bm \delta_z}^{x} + \mathcal{S}_{\bm r_\ell}^{y} \mathcal{S}_{\bm r_\ell+\bm \delta_z}^{y} + \Delta^c\mathcal{S}_{\bm r_\ell}^{z} \mathcal{S}_{\bm r_\ell+\bm \delta_z}^{z} \right)
		+ \sum_{\ell\alpha\beta} P_{\alpha\beta} \mathcal{S}_{\bm r_\ell}^{\alpha} \mathcal{S}_{\bm r_\ell}^{\beta} \mathcal{S}_{\bm r_\ell+\bm \delta_z}^{\alpha} \mathcal{S}_{\bm r_\ell+\bm \delta_z}^{\beta}
		\label{eq:finalH}
		\\
		\notag 
		&&+ J^{ab} \sum_{ \ell \in \bigtriangledown } \sum_{j=1}^3\left(
		{\mathcal S}^x_{\bm r_\ell} {\mathcal S}^x_{\bm r_\ell+\bm \delta_j}
		+ {\mathcal S}^y_{\bm r_\ell} {\mathcal S}^y_{\bm r_\ell+\bm \delta_j}
		+ \Delta^{ab} {\mathcal S}^z_{\bm r_\ell} {\mathcal S}^z_{\bm r_\ell+\bm \delta_j} 
		\right) 
		+ Q \sum_{ \ell \in \bigtriangledown} \sum_{j=1}^3 \left({\mathcal S}^y_{\bm r_\ell} {\mathcal S}^y_{\bm r_\ell+\bm \delta_j} \cos^2\phi_j + {\mathcal S}^x_{\bm r_\ell} {\mathcal S}^x_{\bm r_\ell+\bm \delta_j} \sin^2\phi_j\right),
		\end{eqnarray}
	\end{widetext}
where ${\mathcal S}^\alpha_{\bm r}$ is the $\alpha$th component ($\alpha=x,y,z$) of the pseudospin-one operator at position $\bm r$,
$\bm r_\ell$ is the position of site $\ell$,
$\bm \delta_z=(0,0,c)$, c is the interlayer spacing, 
$j=1,2,3$ labels the nearest neighbour bonds as marked in Fig. \ref{fig:decorated},
$\phi_j={2\pi(j-1)/3}$,
$\bm \delta_j=(\sin\phi_j,\cos\phi_j,0)a_g$ is the vector, of length $a_g$, connecting one sublattice to its three nearest neighbours in the plane,  
and 
$\sum_{ \ell \in \bigtriangledown }$ indicates that the sum runs over only the sublattice of triangles that point down in Fig. \ref{fig:decorated}.

For large $U$, the magnitude of the antiferromagnetic exchange coupling between nearest neighbor clusters in the $c$-direction, $J^c$, is much larger than 
 the exchange coupling between nearest-neighbor clusters in the $a$-$b$ plane, $J^{ab}$, we conclude that the magnetic properties of 
 two-thirds filled trinuclear coordination crystals can be effectively described by $S=1$ XXZ 
chains with a local single-spin anisotropy, $D^*$ and anisotropic biquadratic terms, $P_{\alpha\beta}$. We explore the effect of anisotropic SMOC, $\lambda_{xy} \neq \lambda_z$, finding that the largest 
 anisotropic spin exchange couplings and single-spin anisotropies emerge when $\lambda_{xy}/\lambda_z <1$, which is 
 the  relevant parameter regime for Mo$_3$S$_7$(dmit)$_3$.  
 
\begin{figure}
\centering 
\includegraphics[width=9cm,clip=]{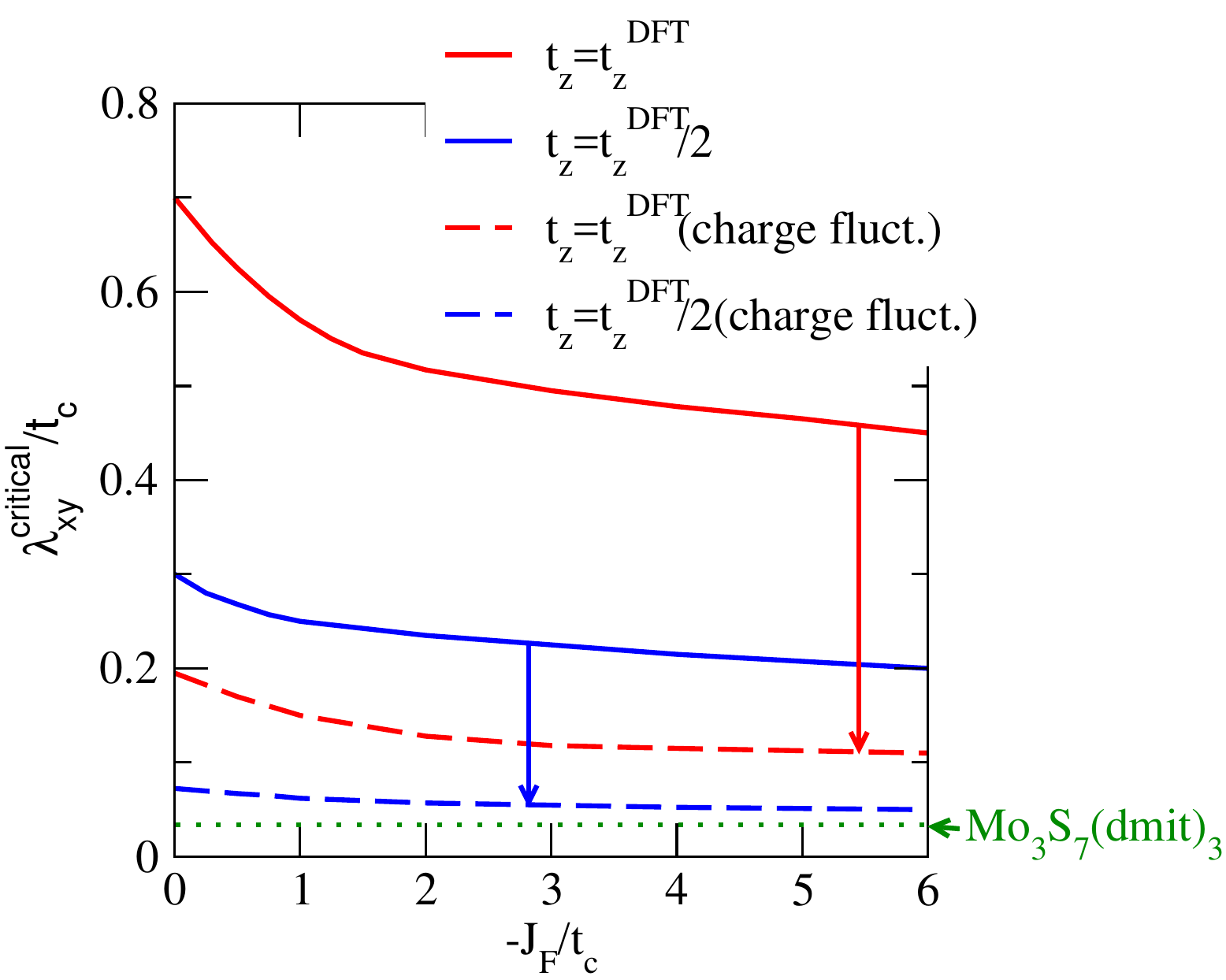}
\caption{Critical SMOC coupling for the transition from the Haldane to the $D$-phase.
At strong coupling, our effective spin exchange model consists of weakly coupled $S=1$ antiferromagnetic
chains in the presence of single-spin anisotropy, $D^*$. When the transverse SMOC, $\lambda_{xy} > \lambda^{critical}_{xy}$
the $D$-phase is stabilized whereas for $\lambda_{xy} < \lambda^{critical}_{xy}$ the Haldane phase occurs. The 
dependence of $\lambda^{critical}_{xy}$ with ferromagnetic intracluster exchange, $-J_F$, is 
shown for model parameters: $\lambda_z=\lambda_{xy}/2$, $U=10t_c$, and $t=0.785 t_c$ for two different hopping amplitudes along 
the chain: $t_z=t_z^{DFT}=0.683t_c$ and $t_z=t^{DFT}_z/2$, where $t_z^{DFT}$ is the value obtained from DFT calculations on the Mo$_3$S$_7$(dmit)$_3$ crystal.
The full lines are obtained from the condition $D^*(\lambda_{xy}) \sim J^c$  assuming a Haldane spin gap opens
in our spin model, while the dashed lines are obtained from: $D^*(\lambda_{xy}) \sim 0.066 J^c$ which 
includes renormalization effects due to charge fluctuations (not contained in our effective spin model) 
which strongly suppress the spin gap. The dotted horizontal line corresponds to the 
$\lambda_{xy}$ obtained from DFT calculations on Mo$_3$S$_7$(dmit)$_3$ crystals. This figure shows that by reducing $t_z$ and
increasing $-J_F$, Mo$_3$S$_7$(dmit)$_3$ can be effectively driven close to the $D$-phase. Varying the SMOC anisotropy also leads to significant changes in this curve, see particular Eq. (\ref{eq:Dapp})}
\label{fig:fig0}
\end{figure}

 For \Mo \textit{ab initio} estimates of SMOC\cite{jacko2017} indicate that 
 $\lambda_{xy} \approx 0.042 t_c$, and $\lambda_{xy} \approx \lambda_z/2$. This, suggests that single-spin anisotropies 
 are smaller than the exchange coupling along the $c$-direction, $D^* < J^c$, so that Mo$_3$S$_7$(dmit)$_3$ is in
 the Haldane phase rather than in the topologically 
 trivial `$D$-phase', {\it i.e.}, the tensor product of the $j=0$ singlets (where $j$ is the $z$-component of the total 
 angular momentum) at each cluster, which is expected for $D^*>J^c$. In spite of the small SMOC values 
 found in Mo$_3$S$_7$(dmit)$_3$ (see Fig. \ref{fig:fig0}), the chemical flexibility of molecular crystals
 can significantly enhance $\lambda_{xy}$ and $\lambda_z$, and suppress $t_z$. Together this could  drive other 
 related systems into the $D$-phase and enhance  anisotropies in the exchange interactions. 
 
 In Fig. \ref{fig:fig0} we show how the critical SMOC, $\lambda^{critical}_{xy}$, at which the 
 transition from the Haldane to the $D$-phase occurs {\it i. e.} when $ D^*(\lambda_{xy}) \sim J^c $, is strongly 
 suppressed by reducing  $t_z$ and/or by a ferromagnetic intracluster exchange, $-J_F$. Variations in the SMOC anisotropy (not shown) 
 can also significantly vary $D^*$ [see Appendix \ref{app:B} and particularly, Eq. (\ref{eq:Dapp})].
On the other hand, increasing $U$ by, say, a factor of two does not change $\lambda^{critical}_{xy}$ 
since $J^c$ is moderately influenced by $U$ when $U \rightarrow \infty$. Intracluster charge fluctuations not captured 
by our spin model but present in the original Hubbard model 
are found to strongly suppress the spin gap \cite{nourse2016}. For the microscopic parameters found from density functional theory (DFT)
\cite{jacko2015,jacko2017} for \Mo the transition line is given by $D^* \sim 0.066 J^c $. The charge fluctuation effect suppresses $\lambda^{critical}_{xy}$ even further
becoming comparable to the SMOC in Mo$_3$S$_7$(dmit)$_3$ crystals. Hence, even though SMOC is small
in Mo$_3$S$_7$(dmit)$_3$ it may be possible to drive it from the Haldane to the $D$-phase 
by modifying crystal parameters, in particular, by suppressing $t_z$. 
This may be achieved by applying negative uniaxial pressure along the $c$-direction of the crystal
which increases the interlayer distance. Alternatively, an expansion along the $c$-direction can be achieved 
by applying uniaxial (positive) pressure on the $a-b$ directions through the Poisson effect. However, this 
procedure can lead to changes in the in-plane arrangement of the molecules distorting the physics of 
the honeycomb lattice discussed here.

We analyze the possible magnetic anisotropies arising in the decorated honeycomb lattice of Fig. \ref{fig:decorated},
which can be realized by isolating the $a$-$b$ planes of trinuclear clusters. More specifically, we analyze the role played by 
the interplay of Coulomb repulsion, intracluster exchange and SMOC in producing anisotropic exchange couplings. We study 
the role played by SMOC anisotropy, $\lambda_{xy} \neq \lambda_z$, which is generically the case in these systems and has 
not been considered in previously.   
We find that the effective exchange couplings within the $a$-$b$ planes are anisotropic {\it only} when {\it both} SMOC {\it and} 
intracluster exchange, $J_F$, are present.  These magnetic anisotropies  lead to a spin-one XXZ + 120$^\circ$ degree honeycomb 
quantum compass model with single spin anisotropy. In the limit of $J_F \rightarrow 0 $, our effective 
spin exchange model reduces to the conventional isotropic $S=1$ antiferromagnetic Heisenberg model on a honeycomb lattice.

We predict that under a sufficiently large external magnetic field, the Haldane
phase can be destroyed giving way to a three-dimensional ordered magnet. This occurs at a critical 
magnetic field, $h_c \sim \Delta_s$,  where $\Delta_s$ is the zero-field Haldane gap of the 
$S=1$ chain.

The present paper is organized as follows. In Section \ref{sec:sec2} we introduce the minimal strongly correlated 
model for describing the electronic properties of isolated triangular  molecules in
 the presence of SMOC. The physics of a single molecule described by this model is discussed in the Appendix \ref{app:A}. In Section \ref{sec:sec3} we analyze the electronic structure of two coupled trimers
 arranged as two nearest-neighbor molecules in the $a$-$b$ plane and also as two nearest-neighbor molecules 
 along the $c$-direction. The energy level spectra of two coupled trimers is obtained exactly and compared to 
 second order perturbation theory. In Section \ref{sec:sec4} the combination of the numerical perturbative approach 
 with an analytical canonical transformation (see also Appendix B), used to extract the exchange interactions between the nearest neighbor 
 pseudospins, is  detailed.  In Section \ref{sec:sec5}, we discuss the qualitative phase diagram 
  expected for the quasi-one-dimensional spin model arising from our approach. Finally, in Section \ref{sec:sec6}, we 
  conclude providing an outlook of our work. 
  
 \section{Model of isolated trimers in the presence of SMOC}
 \label{sec:sec2}
 
Here we condisder crystals formed of triangular tri-nuclear molecules. 
In order to understand the effects of SMOC  on the electronic and magnetic properties  
of these systems we first discuss the relevant model for isolated triangular clusters.
The simplest strongly correlated model is a Hubbard model on a triangle in the presence 
of SMOC \cite{janani2014a,janani2014b,jacko2015}: 
\begin{equation}
H=H_0+H_{SMOC}+H_{U-J_F}.
\label{eq:genham}
\end{equation}
In general all operators should also have a molecular label but this is suppressed throughout the current Section as we deal only with a single complex.

The tight-binding part reads
\begin{equation}
H_0=-t_c \sum_{\langle ij\rangle \sigma} \left(a^\dagger_{i\sigma} a_{j\sigma} + H. c. \right),
\label{eq:tb}
\end{equation}
where $t_c$ is the hopping between the hybrid metal-ligand orbitals at nearest-neighbor sites in the cluster and
$a^\dagger_{i\sigma}$ creates an electron at the $i$th Wannier orbital with spin $\sigma$.  

The general SOC contribution is\cite{khosla2017}
\begin{equation}
H_{SOC}={\bf K} \cdot {\bf S},
\end{equation}
where ${\bf S}$ is the electron spin and ${\bf K}$ is a pseudovector operator,
\begin{equation}
{\bf K}={\hbar \over 4m^2 c^2}[ {\bf p} \times \bm\nabla V({\bf r})].
\end{equation}
We project onto a basis of one Wannier orbital per site of the model illustrated in Fig. \ref{fig:decorated}. The two spin states of the Wannier orbital are a Kramers pair thus 
this projection removes all non-trivial effects of the atomic SOC. For example, in \Mo the Mo atoms are in a $C_1$ environment. 
Consider an atom with ${\bf L}\cdot {\bf S}$ atomic SOC in a $C_1$ environment with time reversal symmetry. 
The most general coupling between  two  states is ${\bf B^*} \cdot{\bf S} + C^* \mathbb{1}$ (the most general $2\times 2$ Hamiltonian). However, 
we require that these states remain degenerate to maintain time-reversal symmetry. Thus only the $C^*$ term remains, providing a constant energy shift as the only effect of atomic SOC in the subspace of the Krammers pair. 
Note that the ${\bf B^*}$ term is a Zeeman splitting term; if we had projected out more states, this term could be non-zero. It has been argued that this is relevant to some transition metal oxides \cite{khaliullin05,kim2012} where this 
projection induces an effective anisotropy on the atomic SOC.
Thus the only SOC term possible in our model is the direct coupling of the spin to currents running around the plane of the molecule (SMOC), which has no analogue in 
transition metal oxides.

For $C_3$ symmetric molecules it can be shown\cite{khosla2017} the SMOC is
\begin{eqnarray}
H_{SMOC} &=& \lambda_{xy} (L_x S_x + L_y  S_y)+\lambda_z  L_z S_z 
\nonumber \\
&=& \lambda_{xy}\left( \frac{ L^+  S^-+L^-S^+}{2}  \right)+ \lambda_z L_z S_z,
\label{eq:hsoc}
\end{eqnarray}
where $L$ is the molecular orbital angular momentum of electrons in the cluster,
$\lambda_{xy}$ describes the transverse SMOC while $\lambda_z$ describes the longitudinal contribution.
 

Finally, the Hubbard-Heisenberg term reads 
\begin{eqnarray}
H_{U-J_F}&=&U\sum_i n_{i\uparrow}n_{i\downarrow}+J_F\sum_{\langle ij \rangle} \left( {\bf S}_i\cdot {\bf S}_j-\frac{n_i n_j}{4} \right),
\end{eqnarray}
where $U$ is the onsite Hubbard interaction, $J_F$ is an intracluster exchange interaction, 
and $n_{i\sigma}=a^\dagger_{i\sigma}a_{i\sigma}$ the number operator. The {\it direct} exchange, $J_F$,
between electrons at nearest-neighbor sites is generically non-zero and  favors ferromagnetic tendencies, \textit{i.e.}, it is
expected to be negative, $J_F < 0$. We will see below that, even if it is much smaller than the direct Coulomb interaction,  $J_F$ plays a crucial role in 
generating magnetic anisotropies. It plays a similar role as the Hunds coupling 
in transition metal oxides\cite{jackeli2009}, which also generates magnetic exchange anisotropies between spins in the lattice.

The non-interacting part (\ref{eq:tb}) can be readily diagonalized: 
\begin{equation}
H_0= \sum_{k\sigma} \epsilon_{k} b^\dagger_{k\sigma}b_{k\sigma},
\end{equation}
using Bloch operators: 
\begin{equation}
b^\dagger_{k\sigma}=\frac{1}{\sqrt{3}} \sum_{j=1}^3 e^{i k \phi (j-1)} a^\dagger_{j\sigma}, 
\label{eq:site}
\end{equation}
with $\phi={2\pi/3}$.  $k=0,\pm 1$ correspond to the allowed $0,\pm{\frac{2\pi}{3}}$ 
momenta in the first Brillouin zone of the triangular cluster with energies $\epsilon_0=-2t_c$ and $\epsilon_{1}=\epsilon_{-1}=t_c$.     

The SMOC contribution to $H$ is most naturally described using `Condon-Shortley' states
which are eigenstates of the $z$-component of the angular momentum, $L_z$, of the cluster\cite{khosla2017,powell2015},
\begin{equation}
c^\dagger_{k\sigma}=\text{sgn}^k(-k) \frac{1}{\sqrt 3 } \sum_{j=1}^3e^{ik\phi(j-1)} a^\dagger_{j\sigma}.
\end{equation}
More explicitly we have
\begin{eqnarray}
c^\dagger_{0,\sigma}=b^\dagger_{0\sigma}\nonumber 
\\
c^\dagger_{1\sigma}=-b^\dagger_{1,\sigma} \nonumber 
\\
c^\dagger_{-1\sigma}=b^\dagger_{-1,\sigma}.
\end{eqnarray}
Note that as Bloch's theorem applies to the cluster, the $z$-component of 
angular momentum is defined up to $3n$ (in units of $\phi$) with $n$ an integer, 
{\it i.e.}, Bloch states with momentum $k'$ satisfying $k=k' \pm 3 n$ are equivalent to 
the $k=0,\pm1$ states.

Hence, the tight-binding part of the Hamiltonian $H_0$ can be expressed either in terms of the Condon-Shortley or Bloch operators as  
\begin{eqnarray}
H_0&=&-2 t_c\sum_{\sigma,k=-1}^1 \cos(\phi k)c^\dagger_{k\sigma}c_{k\sigma} \notag\\
&=&-2 t_c\sum_{\sigma,k=-1}^1\cos(\phi k) b^\dagger_{k\sigma}b_{k\sigma}.
\label{eq:h0}
\end{eqnarray}

\begin{table}
	\caption{List of parameters entering our microscopic model for Mo$_3$S$_7$(dmit)$_3$. The 
		exchange couplings of our derived effective spin exchange model (\ref{eq:fullmodel}) using the actual 
		DFT parameters \cite{jacko2017} obtained for the crystal are also tabulated. The exchange couplings are isotropic so $\alpha$ can be $x,y, z$. 
		Parameters of the effective model that are smaller than $10^{-4}$ are not included. All energy units are in eV. }
	\label{table1}
	\begin{tabular}{lllllll}
		t$_c$ &  t &  t$_z$ & $ \lambda_{xy}$ & $ \lambda_z$  & J$^{ab}$ &J$^c$  \\
		\hline
		0.06 & 0.047 &  0.041   &  0.0025   &   0.005    & 0.0024 &  0.01296 \\
		\hline
	\end{tabular}
\end{table}

Similarly, from the expressions of the angular momentum in terms of the Bloch states:
\begin{eqnarray}
L^+&=&\sqrt{2}\sum_\sigma ( b^\dagger_{0\sigma} b_{-1\sigma} - b^\dagger_{1\sigma} b_{0\sigma} )
\nonumber \\
L^-&=&\sqrt{2}\sum_\sigma ( -b^\dagger_{0\sigma} b_{1\sigma} +b^\dagger_{-1\sigma} b_{0\sigma} )
\nonumber \\
L_z&=& \sum_{k\sigma} k b^\dagger_{k \sigma} b_{k \sigma}, 
\end{eqnarray}
the SMOC contribution to the Hamiltonian of the isolated cluster reads: 
\begin{widetext}
\begin{eqnarray}
H_{SMOC}
= \frac{\lambda_{xy} }{\sqrt{2}} (b^\dagger_{0\downarrow} b_{-1\uparrow} -b^\dagger_{1\downarrow} b_{0\uparrow}
-b^\dagger_{0\uparrow} b_{1\downarrow} +b^\dagger_{-1\uparrow} b_{0\downarrow}) 
+{\lambda_z  \over 2} (b^\dagger_{1\uparrow} b_{1\uparrow} -b^\dagger_{1\downarrow} b_{1\downarrow}
-b^\dagger_{-1\uparrow} b_{-1\uparrow} + b^\dagger_{-1\downarrow} b_{-1\downarrow}). 
\label{eq:ls}
\end{eqnarray}
We may also express $H$ in the site basis, ${|i\sigma\rangle}$, using the transformation of Eq. (\ref{eq:site}) which leads to:
\begin{eqnarray}
H&=& \sum_\sigma \left( ( -t_c +\sigma \lambda_z B^*)a^\dagger_{1\sigma} a_{2 \sigma} + (-t_c+\sigma \lambda_z B) a^\dagger_{1\sigma} a_{3 \sigma}
+(-t_c +\sigma \lambda_z B^*) a^\dagger_{2\sigma} a_{3\sigma} + H.c  \nonumber \right) 
\nonumber \\
&+& \lambda_{xy} \sqrt{2} \left( A  a^\dagger_{1 \downarrow} a _{2 \uparrow} + A^* a^\dagger_{1 \downarrow} a_{3 \uparrow} 
-A a^\dagger_{2 \downarrow} a_{1\uparrow} +B^* a^\dagger_{2\downarrow} a_{3\uparrow}-A^* a^\dagger_{3\downarrow} a_{1 \uparrow}
+B a^\dagger_{3\downarrow} a_{2 \uparrow}  \right)  +H.c. 
\nonumber \\
&+&U\sum_i n_{i\uparrow}n_{i\downarrow}+J_F\sum_{\langle ij \rangle} \left( {\bf S}_i\cdot {\bf S}_j-{n_i n_j \over 4} \right),
\label{eq:hamsite}
\end{eqnarray} 
\end{widetext}
with $A={(e^{i\phi}-1) \over 6} $, $B={i  \over 3}\sin(\phi)$, and  $\sigma=\pm1$. It is evident from the above Hamiltonian
that SMOC can be understood as a spin-dependent hopping between nearest-neighbor sites of the trimers.

Four-component relativistic {\it ab initio} calculations \cite{jacko2017} for Mo$_3$S$_7$(dmit)$_3$ have found anisotropic SMOC:  
$\lambda_{xy}  \approx \lambda_z /2>0$, cf. Table \ref{table1}. 
Below we will fix $t_c>0$ as the unit of energy and explore different
values of SMOC and different $\lambda_{xy}/ \lambda_z$ ratios. Note that the electronic properties of the model
are invariant under the particle-hole transformation $a^\dagger_i \rightarrow h_i,a_i \rightarrow h^\dagger_i$, where $h^\dagger_i$ and $h_i$ are 
hole operators together with the transformation $t_c \rightarrow -t_c$, $\lambda_{xy} \rightarrow -\lambda_{xy}$, 
$\lambda_z \rightarrow -\lambda_z$. 
The onsite Coulomb repulsion within each Wannier orbital, $U$,
is comparable to or even larger than the bandwidth of the relevant Mo$_3$S$_7$(dmit)$_3$ bands crossing 
the Fermi energy. We will assume $U=10t_c$ as a reasonable estimate.
Since the Mo$_3$S$_7$(dmit)$_3$ crystal is at $2/3$-filling there are $N=4$ electrons per triangular cluster in the crystal.
In order to fully characterize the electronic structure of two coupled clusters through perturbation theory techniques we have analyzed
triangular clusters with $N=3,4, 5$ electrons and the parameters $t_c, \lambda_{xy}, \lambda_z>0$, relevant 
to Mo$_3$S$_7$(dmit)$_3$ crystals. Through the particle-hole transformation we can also obtain the electronic structure of triangular clusters with $N=1$ ($N=2$) electrons
from the $N=5$ ($N=4$) solutions by switching the sign of $\lambda_{xy}, \lambda_z, t_c$.

Since $J_z=L_z+S_z$ is a conserved quantity: $[J_z,H]=0$, it is convenient to use the $(k,\sigma)$
representation instead of the site representation to classify the basis states according to their 
quantum number:  $j=k+\sigma$.  We have already expressed $H_0+H_{SMOC}$ in the $(k,\sigma)$ basis 
through Eq. (\ref{eq:h0})-(\ref{eq:ls}). The Hubbard-Heisenberg contribution is expressed in the $(k,\sigma)$ basis as
\begin{widetext}
\begin{eqnarray}
H_{U-J_F}&=&{1\over 3}\sum_k (U -2 J_F) n_{k\uparrow}n_{k\downarrow}+{1 \over 3}\sum_{k,k',k\ne k'} (U -J_F \cos((k-k') \phi)-J_F) n_{k\uparrow}n_{k'\downarrow}
\nonumber \\
&+&{1 \over 3 } \sum_{k,k',q \ne 0} (U-J_F \cos((k'-k-q)\phi)-J_F \cos(q\phi) ) b^\dagger_{k\uparrow} b^\dagger_{k'\downarrow} b_{k'-q\downarrow} b_{k+q\uparrow}.
\label{eq:hUJFk}
\end{eqnarray}
\end{widetext}
For the triangular clusters studied here ${1 \over 3}\sum_{k,k',k\ne k'} (U -J_F \cos((k-k') \phi)-J_F) n_{k\uparrow}n_{k'\downarrow}=
(U/3-J_F/6)\sum_{k,k',k\ne k'} n_{k\uparrow}n_{k'\downarrow}$.
Note that while for the Hubbard-Heisenberg model the effective Coulomb repulsion 
between electrons is different for electrons in different orbitals, in a pure Hubbard model ($J_F=0$),  all 
Coulomb interactions are equal to $U/3$. This has been shown to be important for finding spin exchange anisotropies 
in the context of transition metal oxides\cite{aharony,perkins2014a}. 

Hence, the full Hamiltonian can be explicitly expressed in the $(k,\sigma)$ basis using
the expressions for $H_0$, $H_{SMOC}$ and $H_{U-J_F}$ in Eq. (\ref{eq:h0}), (\ref{eq:ls}) and (\ref{eq:hUJFk}), respectively.

\begin{figure}
  \centering
    \includegraphics[width=5cm]{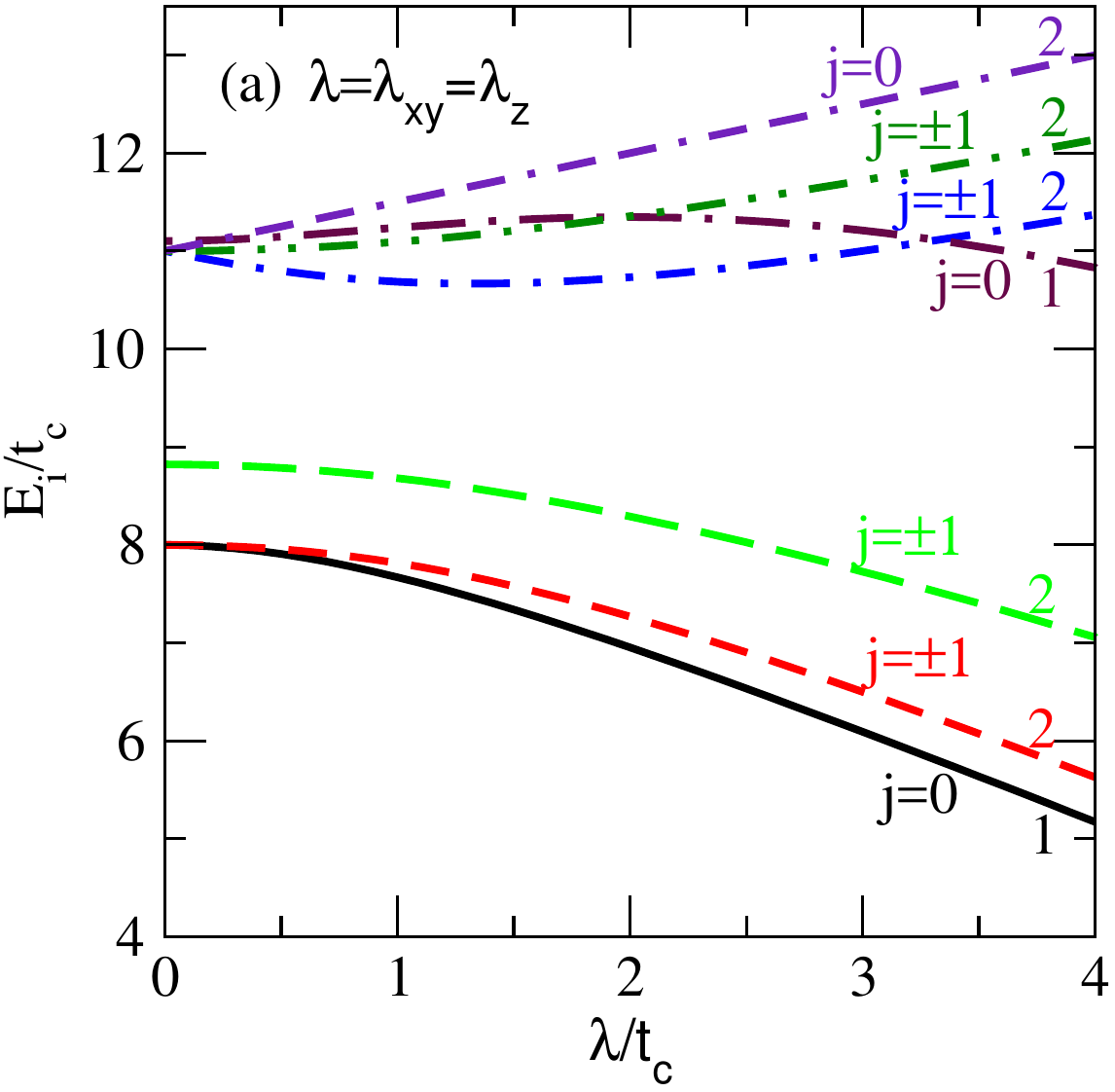}
     \includegraphics[width=5cm]{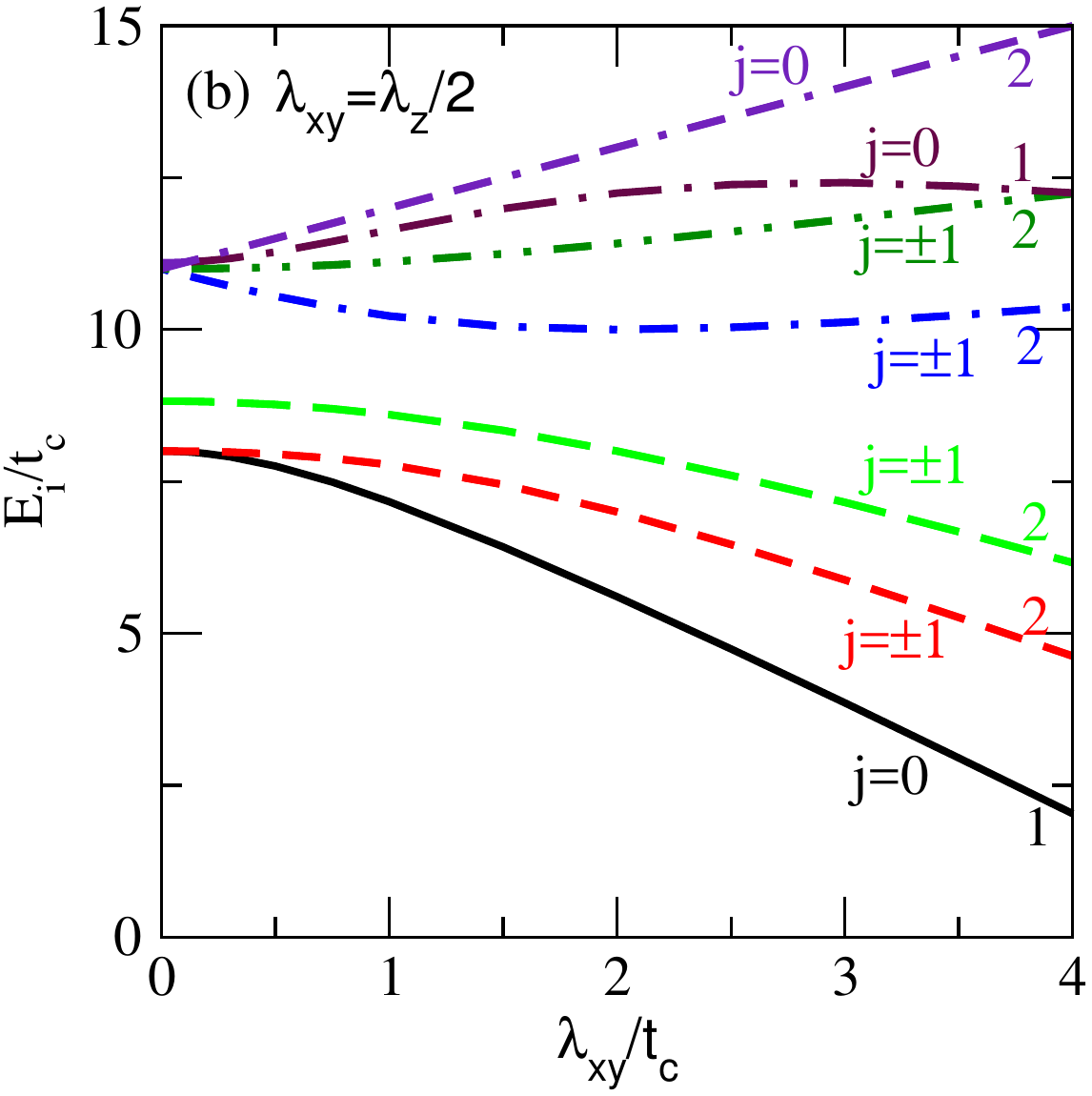}
       \includegraphics[width=5cm]{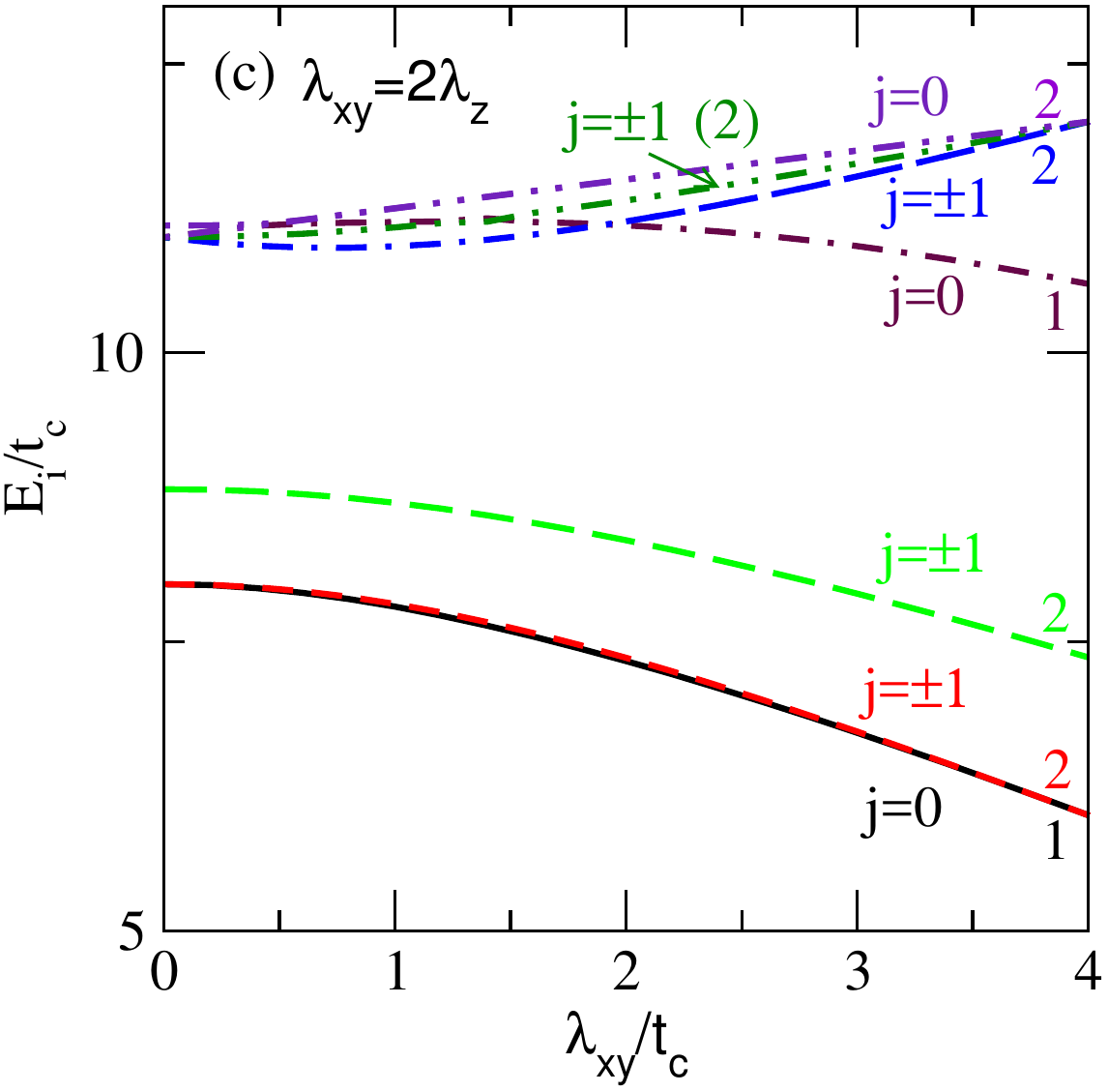}
   \caption{Dependence of electronic structure of isolated triangular clusters on the strength and anisotropy of the SMOC. 
   We plot the eigenvalues of Hamiltonian (\ref{eq:hamsite}) with $N=4$ electrons for $U=10t_c$ and $J_F=0$.  We compare (a) the isotropic SMOC 
   case, $\lambda=\lambda_{xy}=\lambda_z$, with anisotropic SMOC in (b) $\lambda_{xy}=\lambda_{z}/2$ and  in (c) $\lambda_{xy}=2\lambda_{z}$. 
   The eigenstates are classified according to the $z$-component of total angular momentum $j=k+\sigma$. The numbers
  denote energy level degeneracies. For $J_F\neq 0$ the electronic structure of the isolated cluster remains very similar, and in particular conserves 
  the energy level degeneracies shown here.}
 \label{fig:fig1}
 \end{figure}

\begin{figure}
  \centering
    \includegraphics[width=5.5cm]{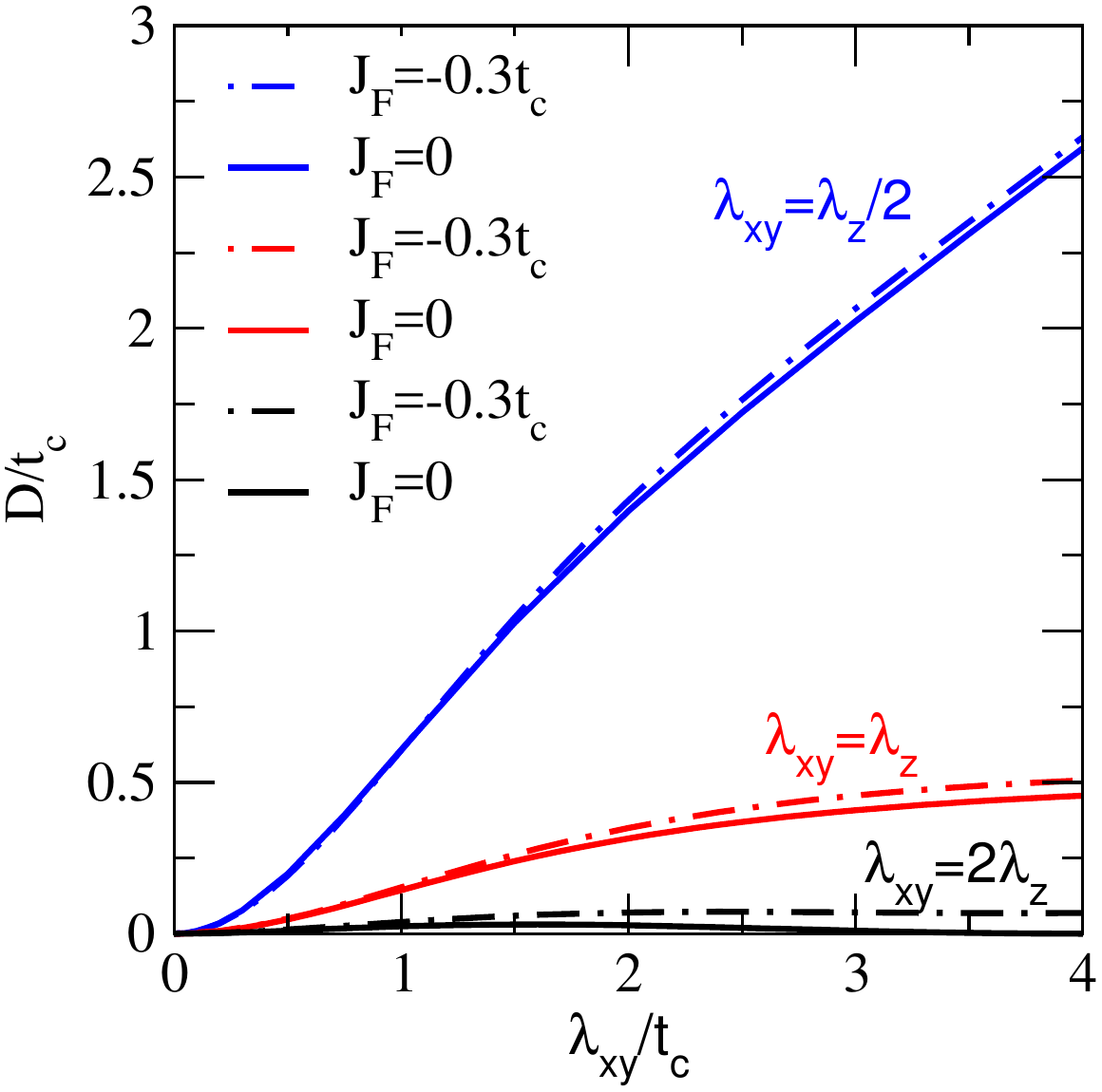}
      \caption{Dependence of the single-spin anisotropy, $D$, of isolated triangular clusters on the strength and anisotropy of the SMOC. 
 The energy difference between the lowest $j=\pm 1$ doublet and the ground state $j=0$ singlet in Fig. \ref{fig:fig1}, which defines
 $D$, cf. Eq. (\ref{eq:trig}), is plotted as a function of $\lambda_{xy}$ for different $\lambda_{xy}/\lambda_z$ ratios. A 
 large enhancement of $D$ is found when $\lambda_{xy} < \lambda_z$, which is relevant to Mo$_3$S$_7$(dmit)$_3$.
 Here, we have used $U=10t_c$ and $J_F=0, -0.3 t_c$.}
 \label{fig:fig1bis}
 \end{figure}

In  Appendix \ref{app:A} we present results for the electronic structure of trimers with $N=3,4,5$ electrons expressed in this basis.
From this analysis, we conclude that
isolated trimers with $N=4$ electrons in the presence of SMOC effectively behave as pseudospin-one localized
moments.  In Fig. \ref{fig:fig1} we show that under SMOC the lowest energy triplet splits into a non-degenerate 
singlet ($j=0$) and a doublet ($j =\pm 1 $), where $j$ is the $z$-component of total angular momentum.  Higher energy excitations are doublets or
non-degenerate under SMOC. Note that since we have an even number of electrons in the cluster, Kramers theorem 
does not apply and non-degenerate states are possible. 
Hence, SMOC induces a single-spin anisotropy at each cluster  
so that the effective spin model for $N=4$ electrons in the $m$-th Mo$_3$S$_7$(dmit)$_3$ molecule
in the crystal reads:
\begin{equation}
H_m^\text{eff}= D ({{\mathcal{ S}}}_{\bm r_m}^z)^2.
\label{eq:trig}
\end{equation} 
As shown in Fig. \ref{fig:fig1} the overall energy level structure 
of the cluster {\it i.e.}, level splittings and degeneracies remain unaffected by anisotropies in SMOC, 
$\lambda_{xy} \neq \lambda_z$ and/or intracluster exchange $J_F \ne 0$. However, the 
absolute value of $D$ is strongly enhanced when $\lambda_{xy}/\lambda_z <1 $ as shown 
in Fig. \ref{fig:fig1bis}.  This is directly relevant to  Mo$_3$S$_7$(dmit)$_3$ crystals in which $\lambda_{xy}/\lambda_z \approx 1/2$.

\section{Two coupled triangular clusters} 
\label{sec:sec3}

We now consider two triangular coupled clusters. We analyze 
the electronic structure of two nearest neighbor triangular clusters as arranged in
Mo$_3$S$_7$(dmit)$_3$ crystals and shown in Fig. \ref{fig:tubes}.  In Fig. \ref{fig:tubes}(a) 
we show two nearest-neighbor clusters in the $a$-$b$ plane, whereas in 
\ref{fig:tubes}(b)  we show two nearest-neighbor clusters along the $c$-direction.  
The molecules in the ``dumbbell"  configuration of Fig. \ref{fig:tubes}(a) are related by 
inversion symmetry as in Mo$_3$S$_7$(dmit)$_3$. 
Molecules in the ``tube" configuration of Fig. \ref{fig:tubes}(b) are  related by a rigid translation along the $c$-axis
but no inversion symmetry is present. We  first report exact results for the
energy level structure. This gives key information about the type of spin exchange acting between the 
effective pseudospins localized at each trimer. These exact results are also used to benchmark perturbation theory 
calculations discussed in Section \ref{sec:pert}.

\subsection{Electronic structure}

Consider a model of two trimers, $\ell$ and $m$ coupled by $H_{kin}$:
\begin{equation}
H=H_\ell+H_m+H_{kin},
\label{eq:coupled}
\end{equation}
where $H_\ell$ is the Hubbard-Heisenberg model of an isolated trimer, $\ell$, in the presence of SMOC as 
introduced previously, Eq. (\ref{eq:hamsite}) in Sec. \ref{sec:sec2}B.  The hopping between two neighbor  clusters is 
described through, $H_{kin}$.

As shown in Fig. \ref{fig:tubes}a, in the coplanar dumbbell arrangement, there is only one hopping 
amplitude connecting the trimers, so $H_{kin}$ reads
\begin{equation}
H^{dumbbell}_{kin}=-t \sum_\sigma \left( a^\dagger_{\ell 1\sigma}a_{m1\sigma}+a^\dagger_{m1\sigma}a_{\ell 1\sigma} \right),
\label{eq:dumb}
\end{equation}
which connects, say, site 1 of the $\ell$-cluster with site 1 of the $m$-cluster. Here $a^{(\dagger)}_{mi\sigma}$ annihilates (creates) an electron with spin $\sigma$ in the $i$th Wannier orbital on molecule $m$.  
The kinetic energy contains off-diagonal hopping matrix elements in the Bloch basis:
\begin{equation}
H^{dumbbell}_{kin}=-{t \over 3}\sum_{k_1k_2\sigma}\left( b^\dagger_{\ell k_1\sigma}b_{m k_2\sigma}+b^\dagger_{m k_2\sigma}b_{\ell k_1\sigma} \right),
\label{eq:kindumb}
\end{equation}
showing that the orbital momentum is not conserved in this case due to the breaking of
trigonal symmetry.  

In the tube arrangement, Fig. \ref{fig:tubes}b the three vertices of the two clusters are connected by a hopping, $t_z$, 
and $H_{kin}$, reads
\begin{equation}
H^{tube}_{kin}=-t_z\sum_{i\sigma} \left( a^\dagger_{\ell i\sigma}a_{mi\sigma}+a^\dagger_{mi\sigma}a_{\ell i\sigma} \right).
\label{eq:tubeHa}
\end{equation}
As the tubes respect the trigonal symmetry of the isolated trimers, the angular momentum about the $C_3$ axis is conserved. Hence, the kinetic energy between
two trimers in the tube arrangement is diagonal when expressed in the Bloch basis: 
\begin{equation}
H^{tube}_{kin}=-t_z \sum_{k\sigma}\left( b^\dagger_{\ell k\sigma}b_{m k\sigma}+b^\dagger_{m k\sigma}b_{\ell k\sigma} \right),
\label{eq:kintube}
\end{equation}
where $k=0,\pm 1$, are the allowed momenta at each trimer. 
of isolated trimers.

\begin{figure}
	\centering 
	\includegraphics[width=4.cm,clip=]{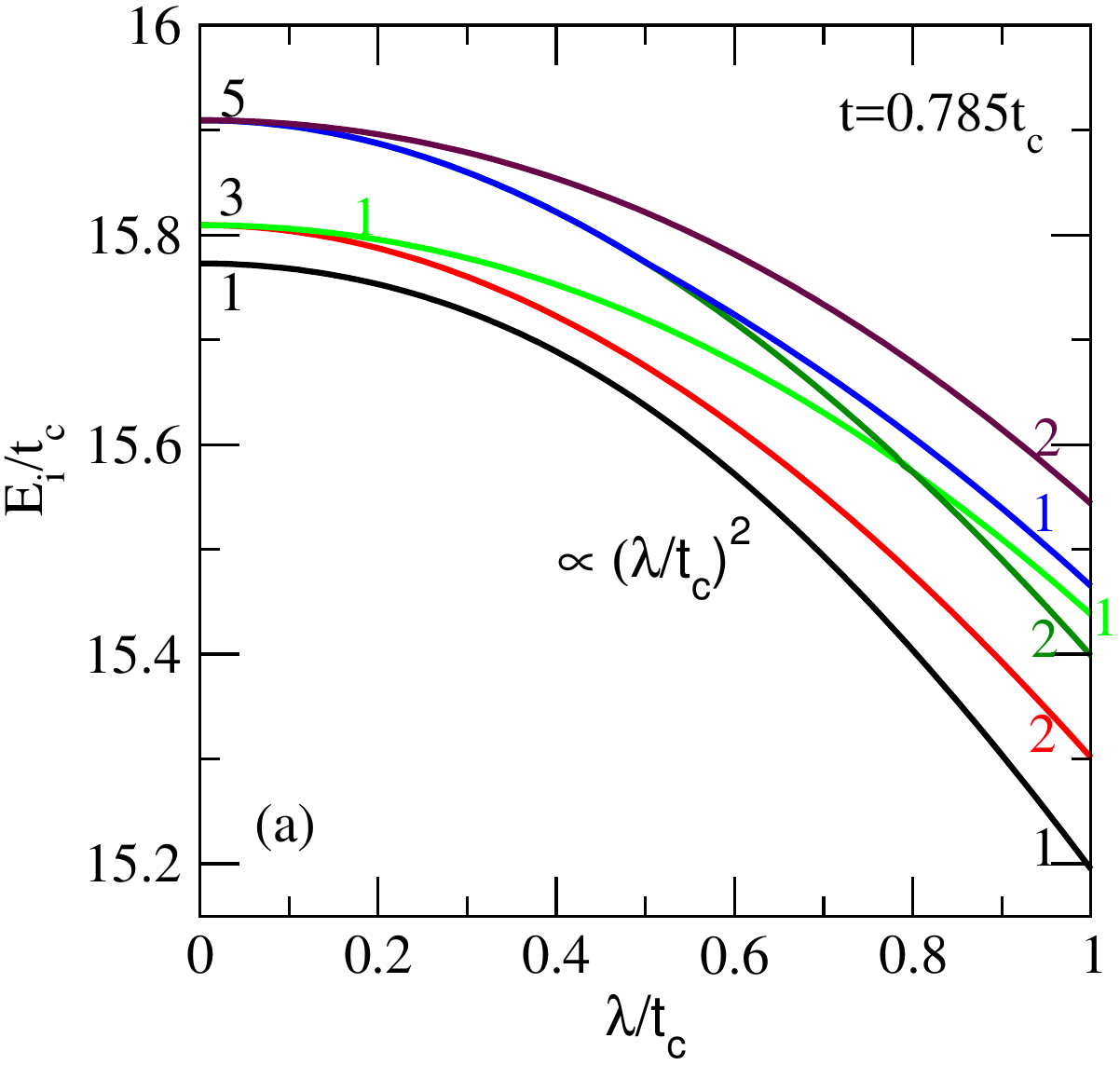}
	\includegraphics[width=4.cm,clip=]{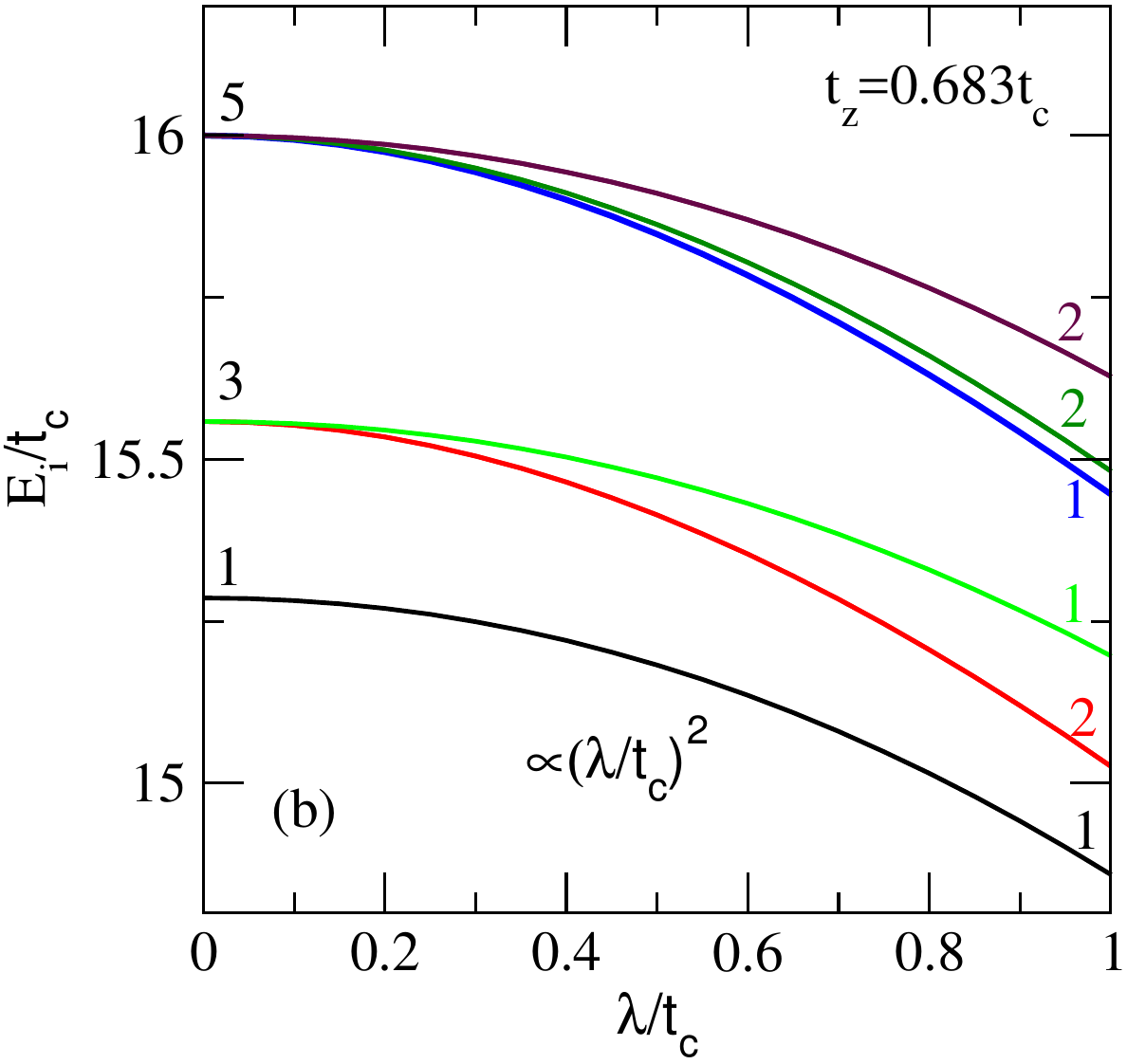}  
	\includegraphics[width=4.cm,clip=]{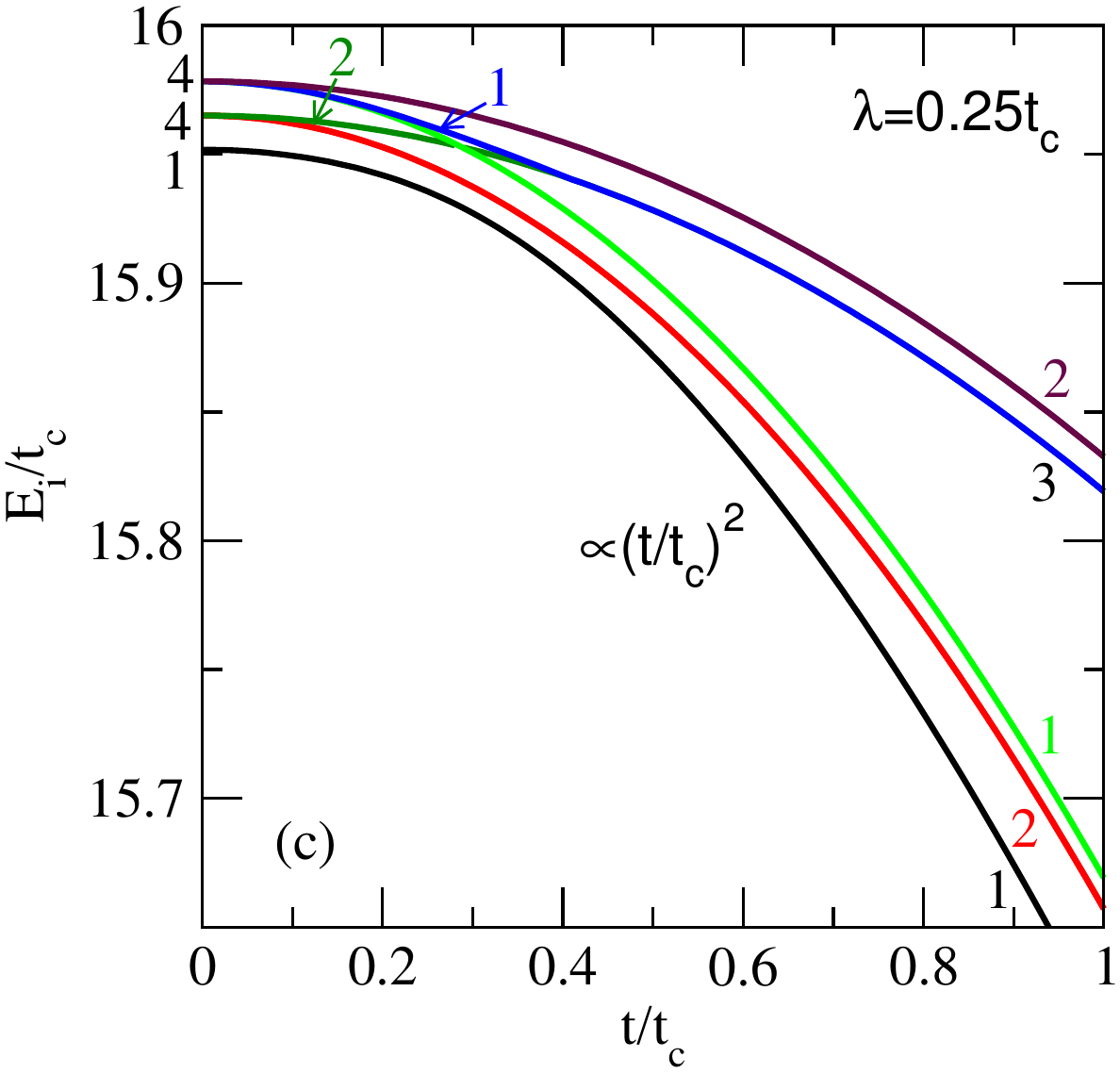}
	\includegraphics[width=4.cm,clip=]{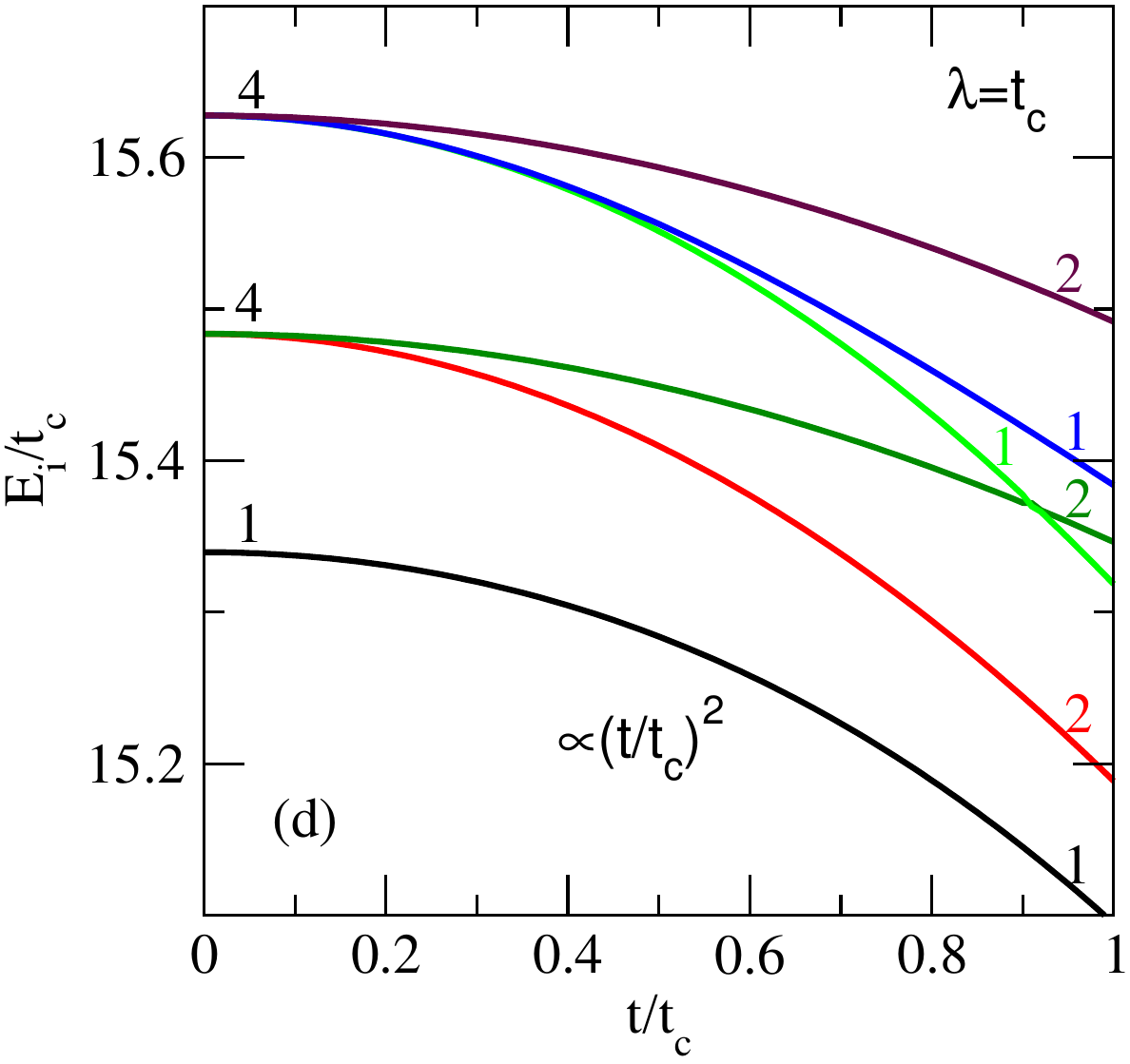}  
	\caption{Exact energy level spectra of two coupled trimers. The plots show exact eigenenergies of  model (\ref{eq:coupled}) 
		for $U=10t_c$ and isotropic SMOC, $\lambda=\lambda_{xy}=\lambda_z$ and $J_F=0$. The dependence of eigenstates, $E_i$, 
		with $\lambda$ for $t=0.785t_c$ are shown for the dumbbell (a) and tube arrangement (b). 
		These plots show an $E_i \propto \lambda^2$ dependence. In (c) and (d) we fix $\lambda$ and analyze the 
		dependence of $E_i$ on the hopping, $t$, in the dumbbell configuration. A quadratic dependence, $E_i \propto t^2$,
		is found for both weak SMOC, $\lambda=0.25 t_c$ in (a) and strong SMOC, $\lambda=t_c$ in (d).  The numbers denote the 
		energy level degeneracies. }
	\label{fig:fig4}
\end{figure}

We have  exactly diagonalized model (\ref{eq:coupled}) for two coupled triangular clusters 
in the presence of SMOC. We consider the case in which each cluster is filled with $N=4$ electrons which is the relevant case 
for Mo$_3$S$_7$(dmit)$_3$ crystals. 
In Fig. \ref{fig:fig4}(a) and (b) we  show the dependence of the eigenenergies, $E_i$, on $\lambda=\lambda_{xy}=\lambda_z$ (isotropic SMOC)
for $U=10t_c$, $t=0.785t_c$ and $J_F=0$ in the dumbbell (a) and tube (b) arrangements.  For $\lambda=0$ we find that 
the eigenspectrum of the coupled trimers consists of a ground state non-degenerate singlet, a triplet and a pentuplet. This is
the eigenspectrum expected for an {\it isotropic} antiferromagnetic exchange interaction between two localized $S=1$ moments 
\cite{janani2014b}. As $\lambda$ is increased the energy levels are split partially removing 
$\lambda=0$ degeneracies. The ground state of the coupled trimers is found to be non-degenerate for any 
value of $\lambda$. 

In Figs. \ref{fig:fig4}(c) and (d) we show the dependence of $E_i$ on $t$ for fixed SMOC, $\lambda=0.25t$ 
and $\lambda=t$. In both cases the eigenenergies depend quadratically on $t$, $E_i \propto t^2$ up to 
large values of $t/t_c \sim 1$ indicating that second order perturbation theory  ($O(t^2)$) is 
reliable. Below, we will further analyze the accuracy of the $O(t^2)$ calculation 
for the model parameters that are  relevant to Mo$_3$S$_7$(dmit)$_3$ crystals.

In order to understand  these spectra, it is important to understand the symmetries of the models. This can be a little subtle when SMOC is included.

In the absence of SMOC the dumbbell model is $D_{2h}$ symmetric as it also contains three mutually perpendicular two-fold rotation axes (cf. Fig. \ref{fig:tubes}a).
If two molecules, $\ell$ and $m$ are related to one another by inversion symmetry then the pseudovectorial nature of angular momenta requires that the SMOC is equal on both molecules: $\lambda_{\ell,xy}=\lambda_{m,xy}$ and $\lambda_{\ell,z}=\lambda_{m,z}$. On the other hand if two molecules are related by a $\pi$-rotation about, say, the $z$-axis this yields $\lambda_{\ell,z}=\lambda_{m,z}$, but $\lambda_{\ell,xy}=-\lambda_{m,xy}$. 
This leads to significant changes in the effective interactions between the molecular spins, which we have discussed elsewhere.\cite{merino2016,powell2016} Thus the case $\lambda_{\ell,xy}=\lambda_{m,xy}$ 
and $\lambda_{\ell,z}=\lambda_{m,z}$, which we consider here, lowers the symmetry to $C_i$ (triclinic).

In the absence of SMOC the tube model is D$_{3h}$ symmetric. This is lowered to C$_{3v}$ in the presence of SMOC, 
which can be understood as follows. 
In our model $\lambda_{\ell,xy}=\lambda_{m,xy}$  and $\lambda_{\ell,z}=\lambda_{m,z}$. 
Under a mirror reflection with respect to a plane perpendicular to the $z$-axis passing through the middle of the
tube,  {\it i.e.}, a $\sigma_h$ operation, there is a change in sign of the transverse SMOC contribution: 
$\lambda_{\ell,xy}= -\lambda_{m,xy}$, which would be inconsistent
with our  model, except for $\lambda_{\ell,xy}=0$.  We find that our model is symmetric under $C_3$ rotations and only has three $\sigma_v$ reflection planes.
Hence, we conclude that the point group symmetry for the tube in the presence of SMOC is C$_{3v}$.

In both the dumbbell (Fig. \ref{fig:fig4}a) and tube  (Fig. \ref{fig:fig4}b) configurations the $\lambda=0$ triplet  is split into a singlet and a doublet while the pentuplet is split into two 
doublets and a singlet. The energy levels are found to depend quadratically on $\lambda$: $E_ i \propto \lambda^2 $,  indicating the absence of the 
linear DM antisymmetric exchange.
In both cases this is expected on symmetry grounds. For the dumbbell this is straightforward, since there is an inversion  center at the midpoint between the two triangular clusters\cite{powell2016,DM2}. 
For the tube
the C$_3$ rotation symmetry implies that ${\bm D}^{DM} \parallel  z$-axis (Moriya's rule 5; Ref. \onlinecite{DM2})
and the $\sigma_v$ reflection symmetry implies that ${\bm D}^{DM} \parallel xy$-plane (Moriya's rule 3). Both  conditions
taken together lead to ${\bm D}^{DM}={\bm 0}$, and there is no DM coupling between the two spins in the tube 
arrangement. 

In \Mo the symmetry of the tube is lowered from $C_{3v}$ to $C_3$ by small intermolecular interactions neglected in the current model \cite{jacko2017}. This allows for a non-zero DM coupling parallel to the $C_3$ axis, which points along the crystallographic c-axis \cite{powell2016}. 

The level degeneracies for both pairs of coupled  clusters (Fig. \ref{fig:fig4}) are those expected for  an {\it isotropic} antiferromagnetic Heisenberg model with a trigonal single ion anisotropy  described by Eq. (\ref{eq:trig}), which we have seen arises for non-zero $\lambda$.  
This is expected for the tube, as in $C_{3v}$ symmetry there are two-fold degenerate states corresponding to the $E$ irreducible representation.  

\begin{figure*}
	\centering 
	\includegraphics[width=5.5cm,clip=]{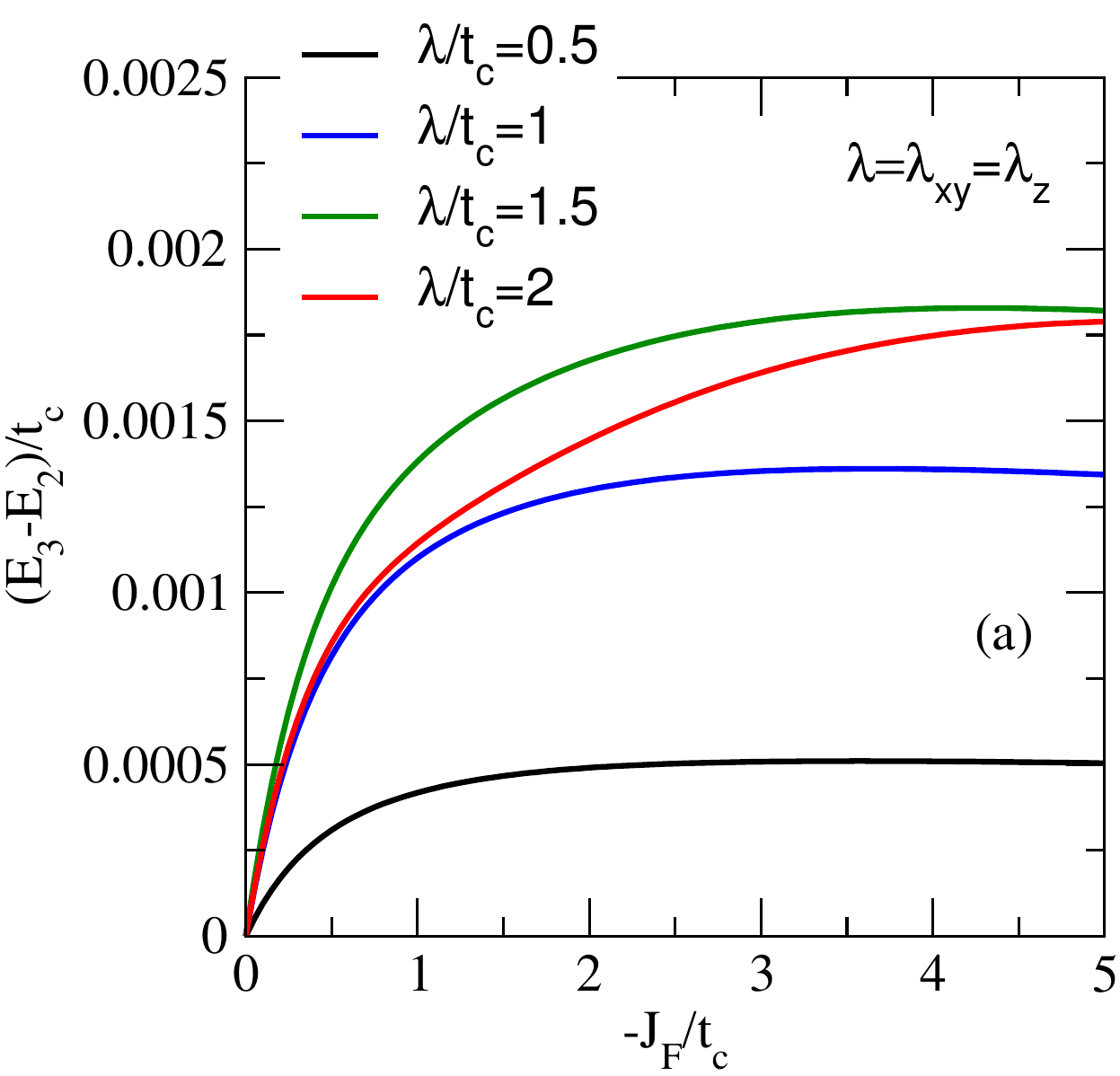}
	\includegraphics[width=5.5cm,clip=]{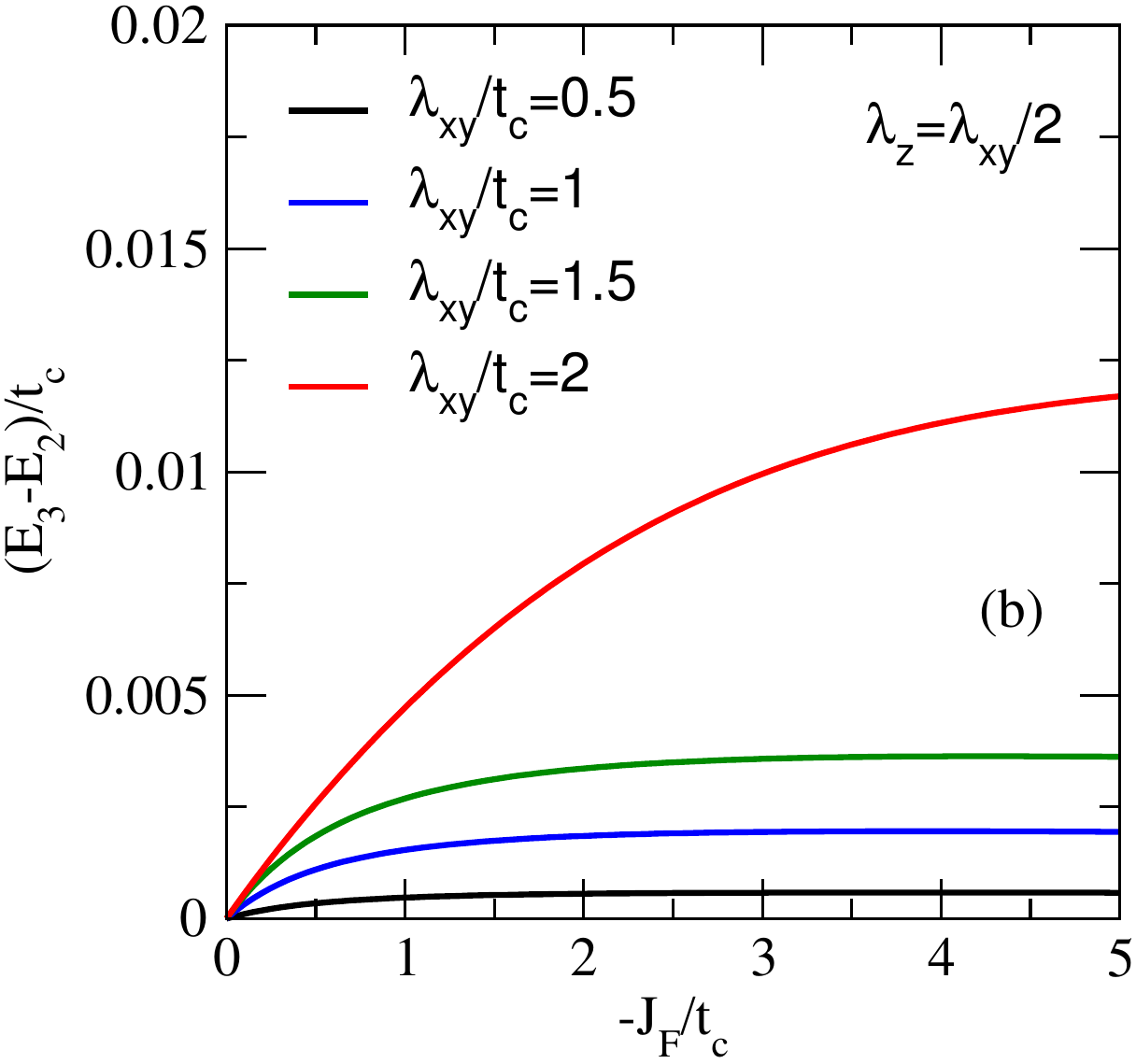}
	\includegraphics[width=5.5cm,clip=]{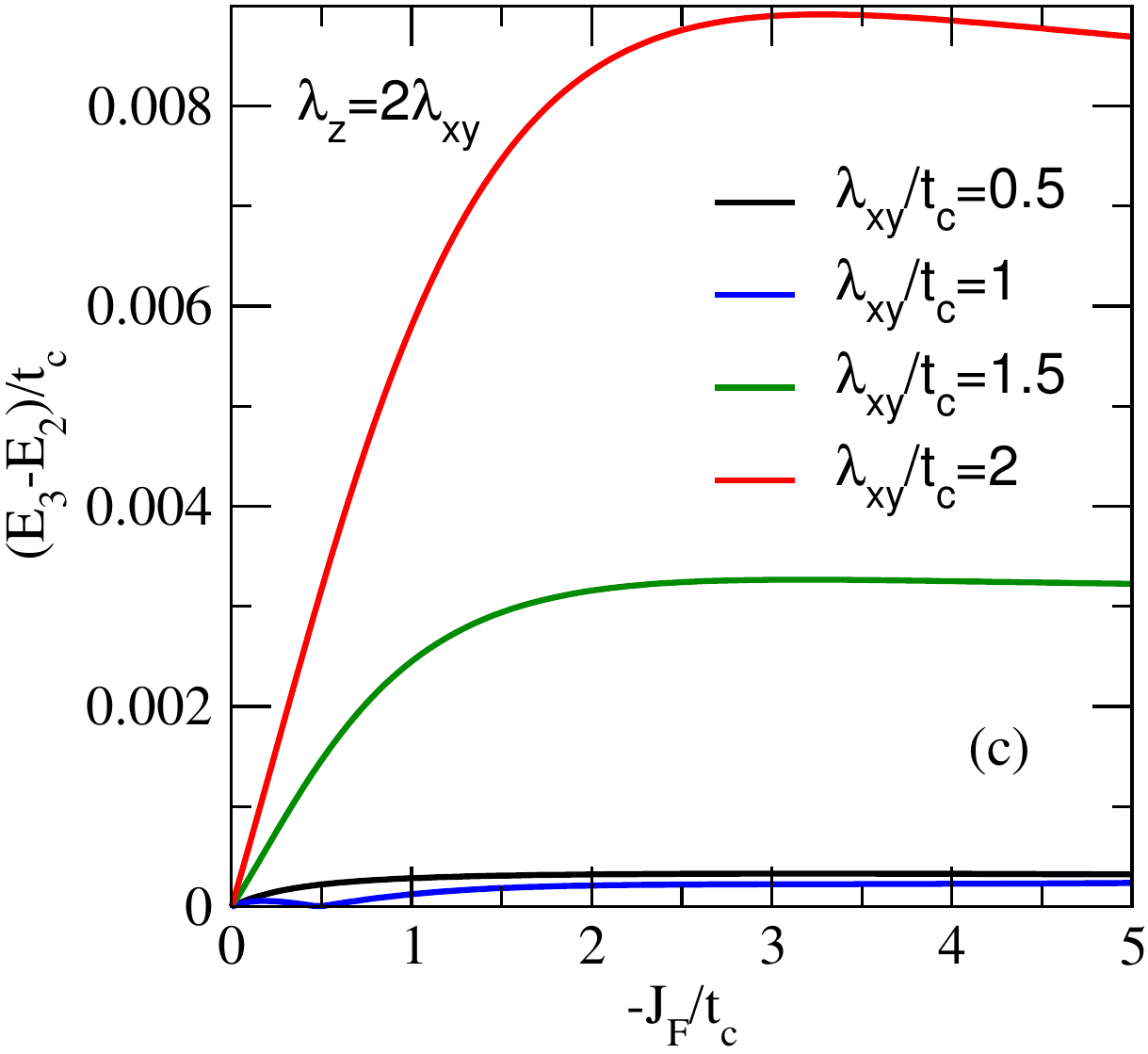}
	\caption{Triclinic anisotropies induced by the Heisenberg intracluster exchange, $J_F$, and SMOC, $\lambda_{xy}, \lambda_z$, for 
		different ratios and strengths of SMOC. The energy difference between the exact third and second lowest energy levels for the dumbbell configuration
		is shown as a function of $-J_F$ (ferromagnetic direct exchange).  We have fixed $U=10t_c$, $t=0.785t_c$ in all figures. In (a) we show results for the 
		Hubbard-Heisenberg model with isotropic SMOC, $\lambda=\lambda_{xy}=\lambda_z$, (b) $\lambda_z=\lambda_{xy}/2$ and 
		in (c) $\lambda_z=2 \lambda_{xy}$.  In contrast, in the case of the tube arrangement the energy splittings are zero: $E_3-E_2=0$,
		for any  value of $J_F$ and $\lambda_{xy}/\lambda_z$ ratio due to the trigonal ($C_3$) symmetry in that case.}
	\label{fig:fig5}
\end{figure*}

However, the $C_i$ symmetry of the dumbbell configuration admits only one-dimensional irreducible representations. Thus, one expects  the level
degeneracies associated with the trigonal symmetry to be fully lifted in the presence of SMOC.  We will  denote these level splittings as triclinic splittings. 
The absence of such triclinic splittings for $J_F=0$  in the dumbbell arrangement therefore indicates a hidden symmetry in the model. This is broken for $J_F\ne0$.
To quantify the  degree of hidden symmetry breaking we plot
the difference in energy between the second and third eigenstates, $E_3-E_2$ in Fig. \ref{fig:fig5}.  
 For $J_F=0$ no level splitting is present for any $\lambda_{xy}/\lambda_z$ 
ratio.  However, a triclinic splitting arises as $-J_F$ is increased,  saturating at sufficiently large $-J_F$. The largest 
splittings are found when SMOC is anisotropic, particularly when $\lambda_{xy}/\lambda_z >1$.

Thus, it is apparent that hidden symmetry is related to Coulomb matrix and is  present in the absence of direct exchange interaction. For $J_F=0$ the symmetric and antisymmetric spin exchange tensors are proportional, but this is lifted for $J_F\ne0$. This  hidden symmetry  plays a similar role in controlling the anisotropy of effective spin models of transition metal oxides.\cite{sea}

We stress that the C$_3$ rotation symmetry of the tube conformation forbids trigonal level splittings, even for $J_F\ne0$. Consistent with this expectation, no triclinic level splittings are observed in our calculations for the tube configuration.

\subsection{Second order perturbation theory in the intercluster hopping}\label{sec:pert}

In order to derive a low energy effective Hamiltonian for the two coupled clusters we now perform perturbation 
theory calculations to $O(t_{conf}^2)$, where $conf=dumbbell$, $tube$ and $t_{dumbbell}=t$ and $t_{tube}=t_z$. The effective Hamiltonian for two coupled clusters 
with $N$ electrons in each cluster is given by
\begin{eqnarray}
H^{(2),conf}_{eff} &=& E_0(N,j_{\ell z}) |N,j_{\ell z}\rangle \langle N,j_{\ell z}|
\nonumber \\
&+& E_0(N,j_{mz}) |N,j_{mz}\rangle \langle N,j_{mz}|
\nonumber \\
&+& \sum_{| m_0 \rangle} { H^{conf}_{kin} |m_0\rangle \langle m_0 | H^{conf}_{kin} \over 2 E_0(N,0) - \langle m_0|H_0+H_U+H_{SMOC}  |m_0 \rangle },
\nonumber \\
\label{eq:hameff}
\end{eqnarray}   
where $E_0(N,j_{iz})$, is the energy of the isolated trimer, $i$, with $j_{iz}=0,\pm1$  with $N$ electrons ($N=4$ in the case of 
interest here), with corresponding eigenstate $|N,j_{iz}\rangle$. In the expression above we are implicitly assuming that the ground state of isolated uncoupled trimers is three-fold 
degenerate even for non-zero SMOC. From a comparison to exact results and the canonical transformation, discussed below, we find that this approximation is very accurate 
for the parameter regime analyzed.  The $\{ |m_0\rangle \}$ are the complete set of virtual excitations in which an electron 
is transferred from one cluster to the other and may be written as
\begin{widetext}
\begin{eqnarray}
|m_0 \rangle &=& |N-1,\gamma_\ell \rangle | N+1, \gamma_m \rangle
= \sum_{\mu_\ell,\mu_m} A_{\gamma_\ell} (N-1,\mu_\ell)A_{\gamma_m}(N+1,\mu_m)  |N-1,\mu_\ell \rangle |N+1,\mu_m \rangle,
\end{eqnarray}  
where $A_{\gamma_i} (N \pm 1 ,\mu_i)=\langle N \pm 1,\mu_i | N\pm 1, \gamma_i \rangle$, $\gamma_i$ denotes the excitations and $\mu_i$ runs over the the Hilbert state configurations with
$N\pm1$ electrons on trimer $i=\ell,m$.

Introducing these states in Eq. (\ref{eq:hameff}) we find for a given configuration of the coupled clusters
\begin{eqnarray}
H^{(2),conf}_{eff} &=& E_0(N,j_{\ell z}) |N,j_{\ell z}\rangle \langle N,j_{\ell z}| +E_0(N,j_{mz}) |N,j_{mz}\rangle \langle N,j_{mz}|
\nonumber \\
&&+ t_{conf}^2 \sum_{\gamma_\ell,\gamma_m} \sum_{\sigma,\sigma'} \sum_{\mu_\ell,\nu_m,\mu'_\ell,\nu'_m} A_{\gamma_\ell}(N-1,\mu_\ell) A_{\gamma_m}(N+1,\nu_m) A^*_{\gamma_\ell}(N-1,\mu'_\ell) A^*_{\gamma_m}(N+1,\nu'_m) \times
\nonumber \\
&&{c^\dagger_{\ell1\sigma}}c_{m1\sigma}|N-1, \mu_\ell \rangle  |N+1,\nu_m \rangle \langle N+1,\nu'_m | \langle N-1,\mu'_\ell|
c^\dagger_{m1\sigma'}c_{\ell 1\sigma'} \over {\Delta \epsilon(N-1,\gamma_\ell ;N+1, \gamma_m) }
\nonumber \\
&&+t_{conf}^2 \sum_{\gamma_\ell,\gamma_m} \sum_{\sigma,\sigma'} \sum_{\mu_m,\nu_\ell,\mu'_m,\nu'_\ell} A_{\gamma_\ell}(N+1,\nu_\ell) A_{\gamma_m}(N-1,\mu_m) A^*_{\gamma_m}(N-1,\mu'_m) A^*_{\gamma_\ell}(N+1,\nu'_\ell) \times
\nonumber \\
&&\frac{{c^\dagger_{m1\sigma}c_{\ell1\sigma}}|N-1, \mu_m \rangle  |N+1,\nu_\ell \rangle \langle N+1,\nu'_\ell | \langle N-1,\mu'_m|
c^\dagger_{\ell1\sigma'}c_{m1\sigma'} } {\Delta \epsilon(N-1,\gamma_m;N+1, \gamma_\ell) },
\label{eq:2nd}
\end{eqnarray}
\end{widetext}
where  
the excitation energies are $\Delta \epsilon(N-1,\gamma_\ell;N+1,\gamma_m)=2 E_0(N)-(E_{\gamma_\ell}(N-1)+E_{\gamma_m}(N+1))$.

It is important to test the reliability of the present second order perturbative calculation for the values of the 
inter cluster hopping amplitudes relevant to  Mo$_3$S$_7$(dmit)$_3$ crystals.
We have checked the accuracy of the second order perturbation theory calculations by comparing the 
nine lowest  energy eigenstates with the exact eigenspectrum in our previous work\cite{merino2016}. 
From Fig. 3 of [\onlinecite{merino2016}]
we concluded that the second order, $O(t^2)$, calculation is very accurate in the dumbbell arrangement with $U=10t_c$, even for the 
large inter-molecular hopping amplitude, $t=0.785 t_c$ relevant to Mo$_3$S$_7$(dmit)$_3$ crystals. 

In the tube arrangement, comparable accuracies
can only be achieved  at larger $U$. The  poorer accuracy at intermediate $U$ in the tube configuration is due the stronger charge fluctuations in this configuration\cite{nourse2016,janani2014a}. In the tube particles can be exchanged between the two clusters 
through $\sim t_z ^2/t_c$ processes without paying energy cost $\sim U$.\cite{powell2016} In contrast, in the dumbbell case, since particles
can only be exchanged through the single hopping connecting the two vertices there is always an energy cost $\sim U$ inherent
to the exchange process $\sim 4t^2/U$.  In spite of this, at sufficiently large values of $U$ we find that the second order perturbation theory is sufficiently accurate
for both the dumbbell and tube arrangements even for the  large values of $t=0.785t_c$ and $t_z=0.683t_c$ extracted from DFT for \Mo.\cite{jacko2015,jacko2017}

\section{Effective magnetic spin exchange model}
\label{sec:sec4}
 
In order to determine the  analytical form of the pseudospin exchange Hamiltonian, we 
have performed a canonical transformation. Analytical expressions of the pseudospin
model valid to $O(\lambda^2)$ and $O(t^2)$, are obtained assuming a $t$-$J$ model for 
the triangular clusters, specified in Appendix \ref{app:B}. By equating the matrix elements of the effective pseudospin 
exchange Hamiltonian obtained from the canonical transformation to the matrix elements of $H_{eff}^{(2)}$ 
evaluated in the low energy subspace, $\{ |j_\ell, j_m \rangle \}$, with $j_\ell,j_m=0,\pm 1$, we are able to 
extract the parameters entering the pseudospin exchange model.

\subsection{Canonical transformation for a nearly degenerate low-energy subspace}\label{sec:canon}

Consider an arbitrary Hamiltonian, $H=H_0+H_1$ where $H_0=\sum_{\nu}P_\nu H P_\nu$, $H_1=\sum_{\mu\ne\nu}P_\nu H P_\mu$, and 
$P_\nu$ is a projector onto the $\nu$th subspace. Now define $H(\varepsilon)=H_0+\varepsilon H_1$. Let
\begin{eqnarray}
\overline{H}(\varepsilon) &\equiv& e^{-i\varepsilon S} H(\varepsilon)  e^{i\varepsilon S} 
\nonumber \\
&=& H_0 + \varepsilon\left(H_1+i\left[H_0,S\right]\right) 
\nonumber \\
&+& \frac{\varepsilon^2}{2}\left( 2i\left[H_1,S\right] - \big[\left[H_0,S\right],S\big]  \right) + \dots
\end{eqnarray}

We choose $S$ so that the linear term vanishes, i.e., such that
$
iH_1=\left[H_0,S\right]. 
$
This implies that
\begin{equation}
 P_\mu H P_\nu (1-\delta_{\mu\nu}) + iP_\mu  H P_{\mu}(P_{\mu}SP_\nu) -i(P_\mu S P_{\nu})P_{\nu}H P_\nu =0.
 \\
\end{equation}
because $P_\mu P_\nu=P_\mu\delta_{\mu\nu}$ and $\sum_\mu P_\mu=1$.
For $\mu=\nu$ this yields
$P_{\mu}SP_\mu = \gamma P_\mu$
for  $\gamma\in\mathbb{C}$.
While, for $\mu\ne\nu$ we find
\begin{eqnarray}
iP_\mu H P_\nu &=& P_\mu  H P_{\mu}(P_{\mu}SP_\nu) -(P_\mu S P_{\nu})P_{\nu}H P_\nu 
\end{eqnarray}
If we choose the projectors such that they project onto \textit{strictly} degenerate subspaces then 
\begin{eqnarray}
P_\mu S P_{\nu} &=& \frac{iP_\mu H P_\nu}{\langle P_\mu  H P_{\mu} \rangle - \langle P_{\nu}H P_\nu\rangle}. \label{expect}
\end{eqnarray}
Therefore, keeping only second order $O(\varepsilon^2)$ terms, we find that
\begin{widetext}
\begin{eqnarray}
\overline{H}\equiv\overline{H}(1) &=& H_0 +  \frac{i}{2}\left[H_1,S\right]  
\nonumber \\
&=&\sum_{\mu}P_\mu H P_{\mu} 
-  \frac{1}{2} \sum_{\mu\ne\nu}\sum_{\mu'\ne\nu}
P_{\mu} H  P_{\nu} H P_{\mu'}
\left( \frac{1}
{\langle P_{\nu}  H P_{\nu} \rangle - \langle P_{\mu'}H P_{\mu'}\rangle}  
+  \frac{1}
{ \langle P_{\nu}H P_{\nu}\rangle - \langle P_{\mu}  H P_{\mu} \rangle} \right).
\end{eqnarray}

Finally, we find the effective low-energy Hamiltonian by projecting onto the low-energy subspace, henceforth denoted $\cal L$. Here it is convenient 
to associate all of the subspaces with the states chosen so that the low energy subspace is diagonal, i.e., $P_\mu H P_{\nu}=0$ if $\mu\ne\nu$ and both $\mu$ and $\nu\in\cal L$. 
(This is always possible provided we can solve the problem restricted purely to $\cal L$, as in elementary degenerate perturbation theory.) We then find that 
\begin{eqnarray}
H_\text{eff} &\equiv& P_{\cal L} \overline{H} P_{\cal L} \notag\\
&=& \sum_{\mu\in\cal L}P_{\mu} H P_{\mu} 
-  \frac{1}{2} \sum_{\mu,\mu'\in\cal L}\sum_{\nu\notin\cal L}\left( \frac{P_\mu H P_{\nu}  H P_{\mu'}}
{\langle P_{\nu}  H P_{\nu} \rangle - \langle P_{\mu'}H P_{\mu'}\rangle}  
+ \frac{P_{\mu} H  P_{\nu} H P_{\mu'}}
{\langle P_{\nu}H P_{\nu}\rangle - \langle P_{\mu}  H P_{\mu} \rangle} \right)
, \label{eq:Heff_CT}
\end{eqnarray}
\end{widetext}
where 
$
P_{\cal L}=\sum_{\mu\in\cal L}P_\mu.
$
In the case that $\cal L$ is strictly degenerate this reduces to the standard result. In the case where there is a small  spread of energies
in $\cal L$ and these are treated as a single subspace, as in the derivation of the $t$-$J$ model, a similar result holds but is approximate because the
 replacement of $P_\mu  H P_{\mu}$ by its expectation value in Eq. (\ref{expect}) is no longer exact. We note that this is precisely the approximation made in Eq. (\ref{eq:hameff}) where we neglected the single-ion splitting of the ground state triplet in the denominator.

The effective Hamiltonian derived from this canonical transformation describing the coupling between two isolated nearest-neighbor trimers, $\ell$ and $m$, in the 
tube arrangement of Fig. \ref{fig:tubes}b is
\begin{eqnarray}
H^{c}_{\ell m}&=&D^c[({\mathcal{S}}_{\bm r_\ell}^z)^2  +({\mathcal{S}}_{\bm r_m}^z)^2]
+\sum_{\alpha\beta}J^c_{\alpha\beta} \mathcal{S}_{\bm r_\ell}^{\alpha} \mathcal{S}_{\bm r_m}^{\beta} 
\notag \\ &&
+ \sum_{\alpha\beta} P_{\alpha\beta} \mathcal{S}_{\bm r_\ell}^{\alpha} \mathcal{S}_{\bm r_\ell}^{\beta} \mathcal{S}_{\bm r_m}^{\alpha} \mathcal{S}_{\bm r_m}^{\beta},
\label{eq:can}
\end{eqnarray}
where $J^c_{\alpha\beta}$ is diagonal and $J^c_{xx}=J^c_{yy}\ne J^c_{zz}$, and the anisotropic biquadratic couplings, $P_{\alpha\beta}=P_{\beta\alpha}$, obey $P_{xx}=P_{yy}=P_{xy}$ and $P_{zx}=P_{zy}=(P_{zz}+P_{xx})/2$. Both numerically and analytically we find $P_{xx}\ll P_{zz}$, indeed we find numerically that $P_{xx}$ is negligibly small and thus do not discuss it further below. 
$D^{c}=D + \Delta D^{c}$ is the single-spin anisotropy
including corrections, $\Delta D^{c}$, due to hopping processes between the clusters. The  perturbative expressions for these parameters are given in Appendix \ref{app:B}.  
Thus, one can recast the bilinear terms of $H^{c}_{\ell m}$ in the familiar XXZ form. Doing so, one finds that the Hamiltonian for a single chain is
\begin{eqnarray}
H^{c}&=&\sum_{\ell} D^c({\mathcal{S}}_{\bm r_\ell}^z)^2  
+ \sum_{\ell\alpha\beta} P_{\alpha\beta} \mathcal{S}_{\bm r_\ell}^{\alpha} \mathcal{S}_{\bm r_\ell}^{\beta} \mathcal{S}_{\bm r_\ell+\bm\delta_z}^{\alpha} \mathcal{S}_{\bm r_\ell+\bm\delta_z}^{\beta}
\label{eq:c}
\\\notag  &&
+J^c \sum_{\ell} \left( \mathcal{S}_{\bm r_\ell}^{x} \mathcal{S}_{\bm r_\ell+\bm\delta_z}^{x} + \mathcal{S}_{\bm r_\ell}^{y} \mathcal{S}_{\bm r_\ell+\bm\delta_z}^{y} + \Delta^c\mathcal{S}_{\bm r_\ell}^{z} \mathcal{S}_{\bm r_\ell+\bm\delta_z}^{z} \right),
\end{eqnarray}
where $J^c=J_{xx}^c$ and $\Delta^c=J_{zz}^c/J_{xx}^c$.

For two isolated nearest-neighbor trimers in the dumbbell arrangement with the $t$ bond connecting the two sites labeled `1' (cf. Figs. \ref{fig:decorated} and \ref{fig:tubes}a), the exchange Hamiltonian   is
 \begin{eqnarray}
H^{ab}_{1}&=&  D^{ab} \left[ ({\mathcal S}^z_{\bm r_\ell})^2 + ({\mathcal S}^z_{\bm r_\ell+\bm\delta_1})^2\right]\notag
\\
&+&  K_{\pm\pm} \big[ {\mathcal S}_{\bm r_\ell}^+{\mathcal S}_{\bm r_\ell}^+
+ {\mathcal S}_{\bm r_\ell+\bm\delta_1}^+{\mathcal S}_{\bm r_\ell+\bm\delta_1}^+
 +H.c. \big] \notag
\\
&+& K_{z\pm}\big[ {\mathcal S}_{\bm r_\ell}^z {\mathcal S}_{\bm r_\ell}^x 
+ {\mathcal S}_{\bm r_\ell+\bm\delta_1}^z {\mathcal S}_{\bm r_\ell+\bm\delta_1}^x  +H.c. \big] \notag 
\\
&+& \sum_{\alpha\beta}J^{ab}_{\alpha\beta} {\mathcal S}^\alpha_{\bm r_\ell} {\mathcal S}^\beta_{\bm r_\ell+\bm\delta_1}. 
 \label{eq:ab}
\end{eqnarray}
$D^{ab}=D + \Delta D^{ab}$ is the single-spin anisotropy
including corrections, $\Delta D^{ab}$, due to hopping processes between the clusters and is plotted in Figs. \ref{fig:fig7} and \ref{fig:fig7bis}. We find that $\Delta D^{ab}$
is very small so that $D^{ab} \sim D$.

To derive the effective Hamiltonian for the full crystal we know need to note that we have, so far, only considered the $t$-bonds between Wannier orbitals labeled `1', cf. Figs.  \ref{fig:decorated} and \ref{fig:tubes}, and Eq. \ref{eq:dumb}. Rather than repeating the derivation for `2' and `3' bonds we can simply use the $C_3$ symmetry of the molecules and note that the $\bm{\mathcal S}_{\bm r_m}$ operators transform as vectors under rotation. Hence we can replace 
	\begin{subequations}
	\begin{eqnarray}
	{\mathcal S}_{\bm r_m}^x &\rightarrow& {\mathcal S}_{\bm r_m}^x \cos\phi_j - {\mathcal S}_{\bm r_m}^y \sin\phi_j \\
	{\mathcal S}_{\bm r_m}^y &\rightarrow& {\mathcal S}_{\bm r_m}^y \cos\phi_j + {\mathcal S}_{\bm r_m}^x \sin\phi_j
	\end{eqnarray} 
	\end{subequations}
in Eq. (\ref{eq:ab}), where  $j$ labels the bond, as shown in Fig. \ref{fig:decorated}.

	Firstly, one finds that the $K_{\pm\pm}$ and $K_{z\pm}$ terms vanish in the full crystal due to
	cancellation among the contributions from the three nearest-neighbor bonds. Transforming the other terms, one can rewrite that Hamiltonian as 
\begin{widetext}
		\begin{eqnarray}
	H^{ab}&=&  \sum_\ell D^{ab} ({\mathcal S}^z_{\bm r_\ell})^2  \notag 
	+ J^{ab} \sum_{\ell\in \bigtriangledown}\sum_{j=1}^3\left(
		{\mathcal S}^x_{\bm r_\ell} {\mathcal S}^x_{\bm r_\ell+\bm\delta_j}
		+ {\mathcal S}^y_{\bm r_\ell} {\mathcal S}^y_{\bm r_\ell+\bm\delta_j}
		+ \Delta^{ab} {\mathcal S}^z_{\bm r_\ell} {\mathcal S}^z_{\bm r_\ell+\bm\delta_j} 
	\right) 
	\notag \\
	&& 
	+ Q \sum_{\ell\in \bigtriangledown}\sum_{j=1}^3 \left({\mathcal S}^y_{\bm r_\ell} {\mathcal S}^y_{{\bm r_\ell}+\bm\delta_j} \cos^2\phi_j + {\mathcal S}^x_{\bm r_\ell} {\mathcal S}^x_{{\bm r_\ell}+\bm\delta_j} \sin^2\phi_j\right) \notag \\
&& + J_{xz}^{ab} \sum_{\ell\in \bigtriangledown}\sum_{j=1}^3 \left[ \left( {\mathcal S}_{\bm r_\ell}^x \cos\phi_j - {\mathcal S}_{\bm r_\ell}^y \sin\phi_j \right) {\mathcal S}^z_{{\bm r_\ell}+\bm\delta_j} 
+ {\mathcal S}^z_{\bm r_\ell} \left( {\mathcal S}_{{\bm r_\ell}+\bm\delta_j}^x \cos\phi_j - {\mathcal S}_{{\bm r_\ell}+\bm\delta_j}^y \sin\phi_j \right) \right], 
	\label{eq:ab-compassXXZ}
	\end{eqnarray}
	where $J^{ab}=(J_{xx}^{ab}+J_{yy}^{ab})/2$, $\Delta^{ab}=J_{zz}^{ab}/J^{ab}$ and $Q=(J_{xx}^{ab}-J_{yy}^{ab})/2$. The perturbative expressions for these parameters are given in Appendix \ref{app:B}. Thus, we see that the second term (proportional to $J^{ab}$) is simply the XXZ model and the third term (proportional to $Q$) is the honeycomb 120$^\circ$ compass model.\cite{nussinov2015}

Finally, combining the results obtained above we obtain the full effective spin exchange model for the crystal, which reads:
\begin{eqnarray}
H_\text{eff}&=& D^* \sum_{\ell} (\mathcal{S}_{\bm r_\ell}^{z})^2 
+J^c \sum_{\ell} \left( \mathcal{S}_{\bm r_\ell}^{x} \mathcal{S}_{\bm r_\ell+\bm\delta_z}^{x} + \mathcal{S}_{\bm r_\ell}^{y} \mathcal{S}_{\bm r_\ell+\bm\delta_z}^{y} + \Delta^c\mathcal{S}_{\bm r_\ell}^{z} \mathcal{S}_{\bm r_\ell+\bm\delta_z}^{z} \right)
+ \sum_{\ell\alpha\beta} P_{\alpha\beta} \mathcal{S}_{\bm r_\ell}^{\alpha} \mathcal{S}_{\bm r_\ell}^{\beta} \mathcal{S}_{\bm r_\ell+\bm\delta_z}^{\alpha} \mathcal{S}_{\bm r_\ell+\bm\delta_z}^{\beta}
\notag 
\\
&& + J^{ab} \sum_{\ell\in \bigtriangledown}\sum_{j=1}^3\left(
{\mathcal S}^x_{\bm r_\ell} {\mathcal S}^x_{\bm r_\ell+\bm\delta_j}
+ {\mathcal S}^y_{\bm r_\ell} {\mathcal S}^y_{\bm r_\ell+\bm\delta_j}
+ \Delta^{ab} {\mathcal S}^z_{\bm r_\ell} {\mathcal S}^z_{\bm r_\ell+\bm\delta_j} 
\right) 
+ Q \sum_{\ell\in \bigtriangledown}\sum_{j=1}^3 \left({\mathcal S}^y_{\bm r_\ell} {\mathcal S}^y_{{\bm r_\ell}+\bm\delta_j} \cos^2\phi_j + {\mathcal S}^x_{\bm r_\ell} {\mathcal S}^x_{{\bm r_\ell}+\bm\delta_j} \sin^2\phi_j\right) \notag \\
&& + J_{xz}^{ab} \sum_{\ell\in \bigtriangledown}\sum_{j=1}^3 \left[ \left( {\mathcal S}_{\bm r_\ell}^x \cos\phi_j - {\mathcal S}_{\bm r_\ell}^y \sin\phi_j \right) {\mathcal S}^z_{{\bm r_\ell}+\bm\delta_j} 
+ {\mathcal S}^z_{\bm r_\ell} \left( {\mathcal S}_{{\bm r_\ell}+\bm\delta_j}^x \cos\phi_j - {\mathcal S}_{{\bm r_\ell}+\bm\delta_j}^y \sin\phi_j \right) \right],
\label{eq:fullmodel}
\end{eqnarray}
\end{widetext}
where $D^*=D + \Delta D^c +\Delta D^{ab}$. This expression neglects `three molecule' terms analogous to the `three site'  terms neglected in the 
usual formulation of the $t$-$J$ model.\cite{Chao,Harris} We will see below that $J_{xz}^{ab}$ is extremely small. On neglecting this term 
one finds that the effective Hamiltonian is given by Eq. (\ref{eq:finalH}).
 
The parameters governing the spin exchange between molecules $\ell$ and $m$ in our spin exchange Hamiltonian, $H_\text{eff}$, 
are obtained by comparing the canonical transformation with our numerical second order perturbation theory
\begin{eqnarray}
\langle j_\ell,j_m|H^{ab}_{lm} | j_\ell,
j_m \rangle &=& \langle j_\ell,j_m|H^{(2),dumbbell}_{eff} | j_\ell,j_m \rangle,
\nonumber \\
\langle j_\ell,j_m|H^{c}_{lm} | j_\ell,j_m \rangle &=& \langle j_\ell,j_m|H^{(2),tube}_{eff} | j_\ell,j_m \rangle,
\end{eqnarray}
recall $H^{(2),conf}_{eff}$ is defined in  Eq. (\ref{eq:2nd}).
The above equations are solved for a given set of parameters: $U$, $J_F$, $t_c$,  $t$, $t_z$, $\lambda_{xy}$, 
and $\lambda_z$ entering our original microscopic model (\ref{eq:genham}).

\subsection{Anisotropic exchange in the $ab$-plane}

We have explored anisotropies arising in the exchange couplings of the effective exchange
model, Eq. (\ref{eq:ab}) for the two clusters coupled as in Fig. \ref{fig:tubes}(a). Since
the non-pseudospin-conserving $K_{\alpha\beta}$ terms exactly cancel in the crystal they will
not be discussed any further. We find that when $J_F=0$ the exchange coupling tensor  is diagonal  and 
 isotropic, $J^{ab}_{\alpha\beta}=J^{ab}\delta_{\alpha\beta}$. This is consistent with our previous results  (see Fig. 4(a) of Ref. [\onlinecite{merino2016}]) and the lack of triclinic splittings 
observed in the energy level spectrum for two clusters in the dumbbell configuration shown in Fig. \ref{fig:fig5}.

\begin{figure*} 
	\includegraphics[width=4.5cm,clip=]{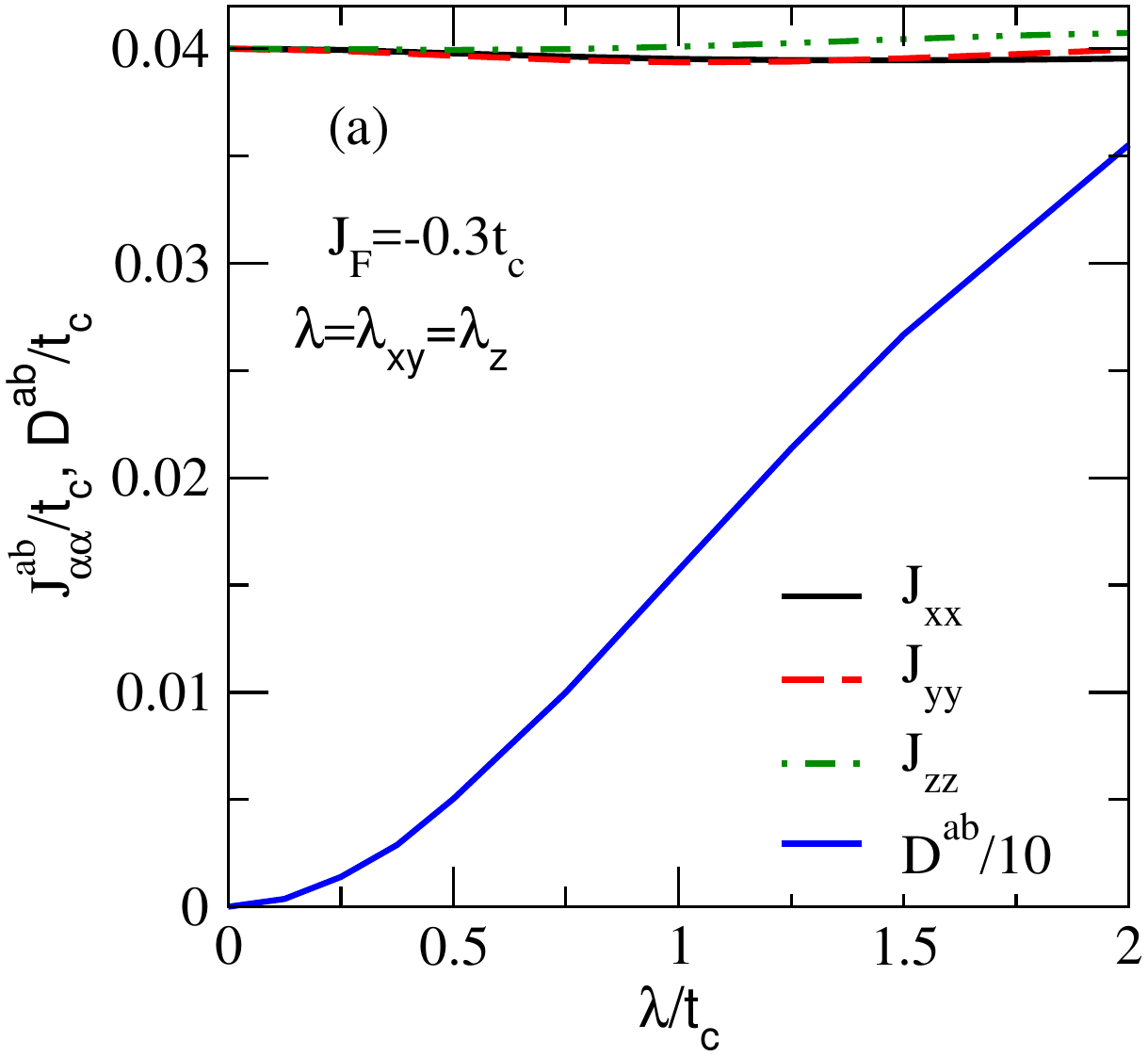}
	\includegraphics[width=4.5cm,clip=]{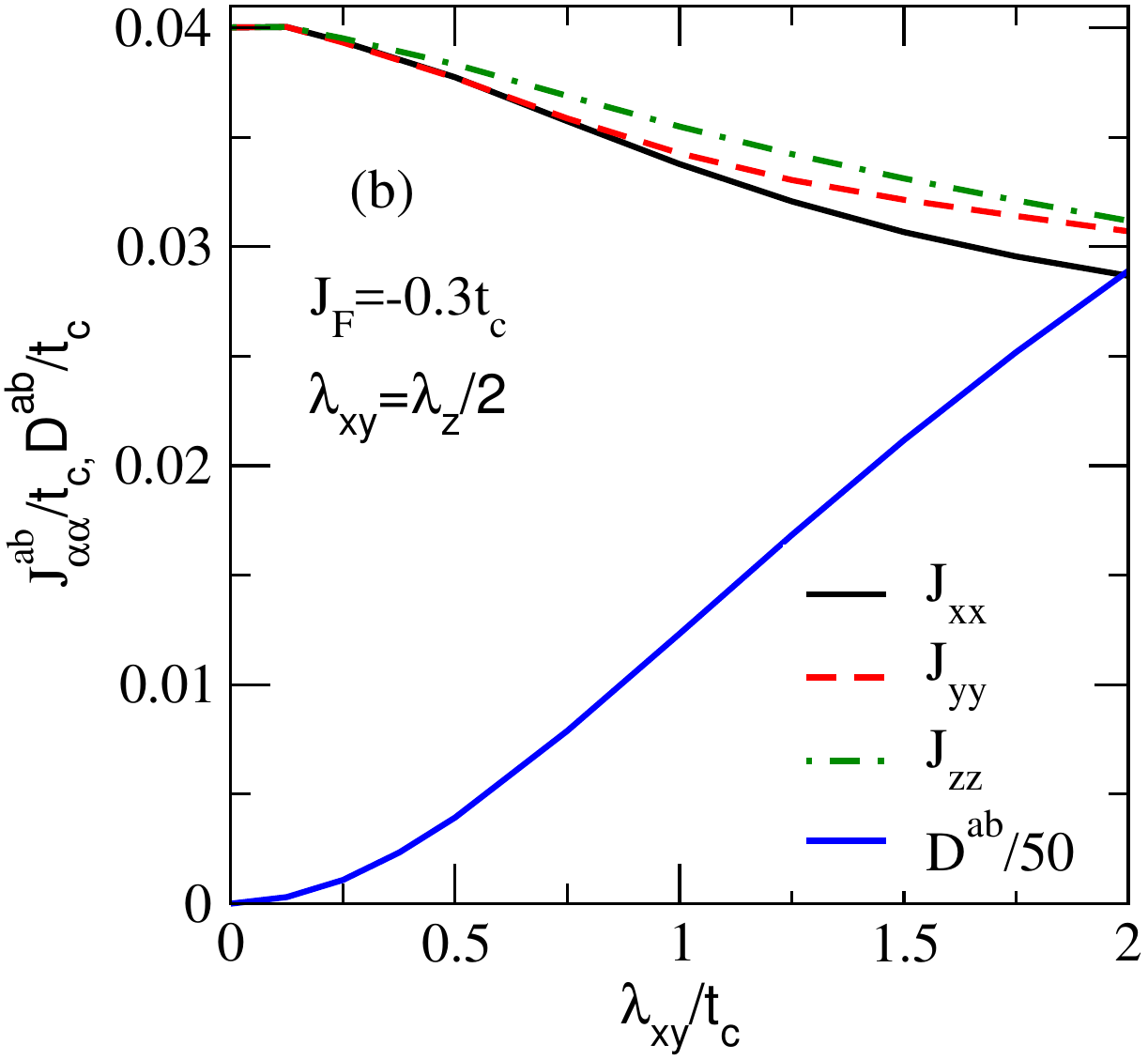}
	\includegraphics[width=4.5cm,clip=]{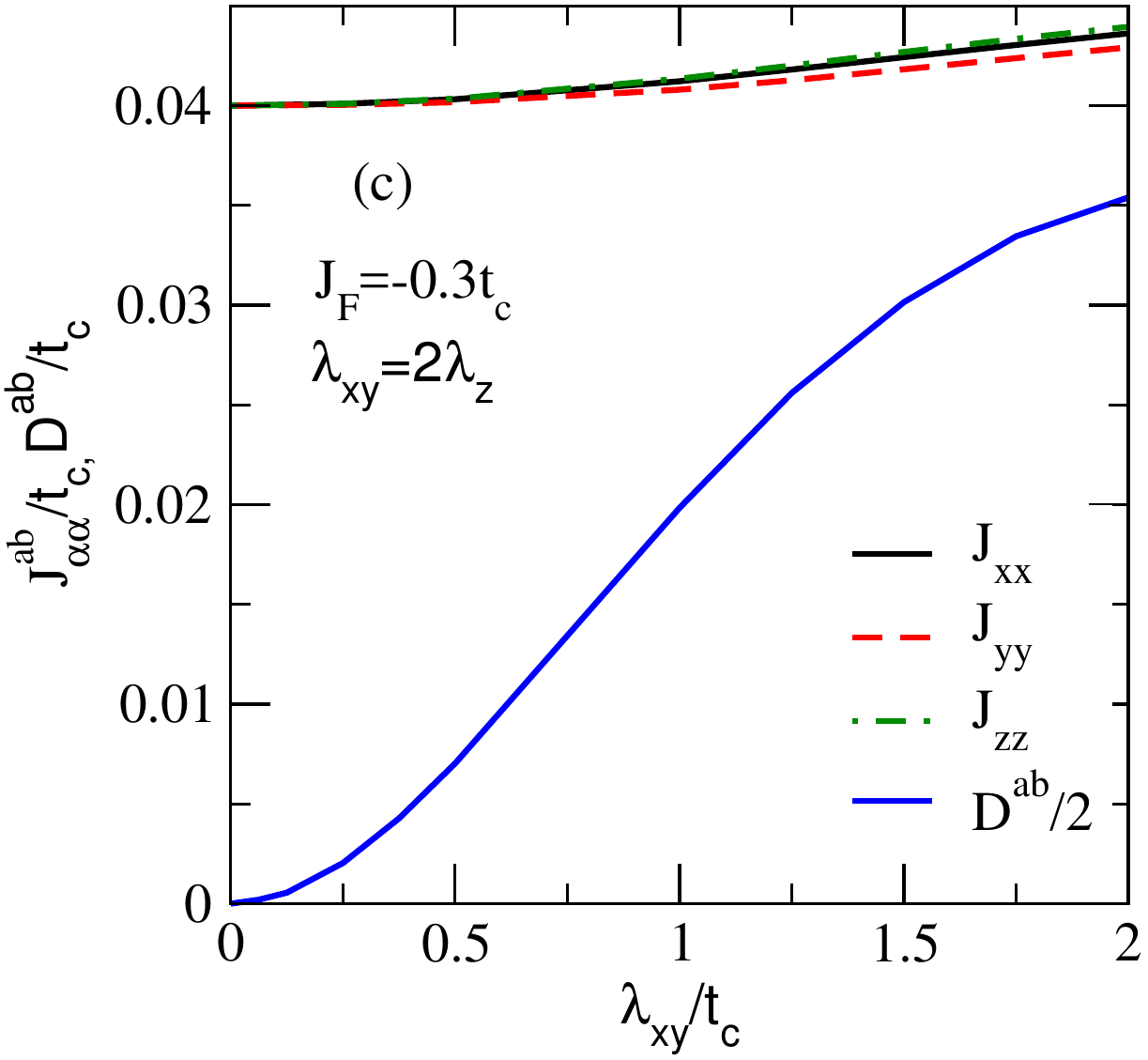}
	\includegraphics[width=4.5cm,clip=]{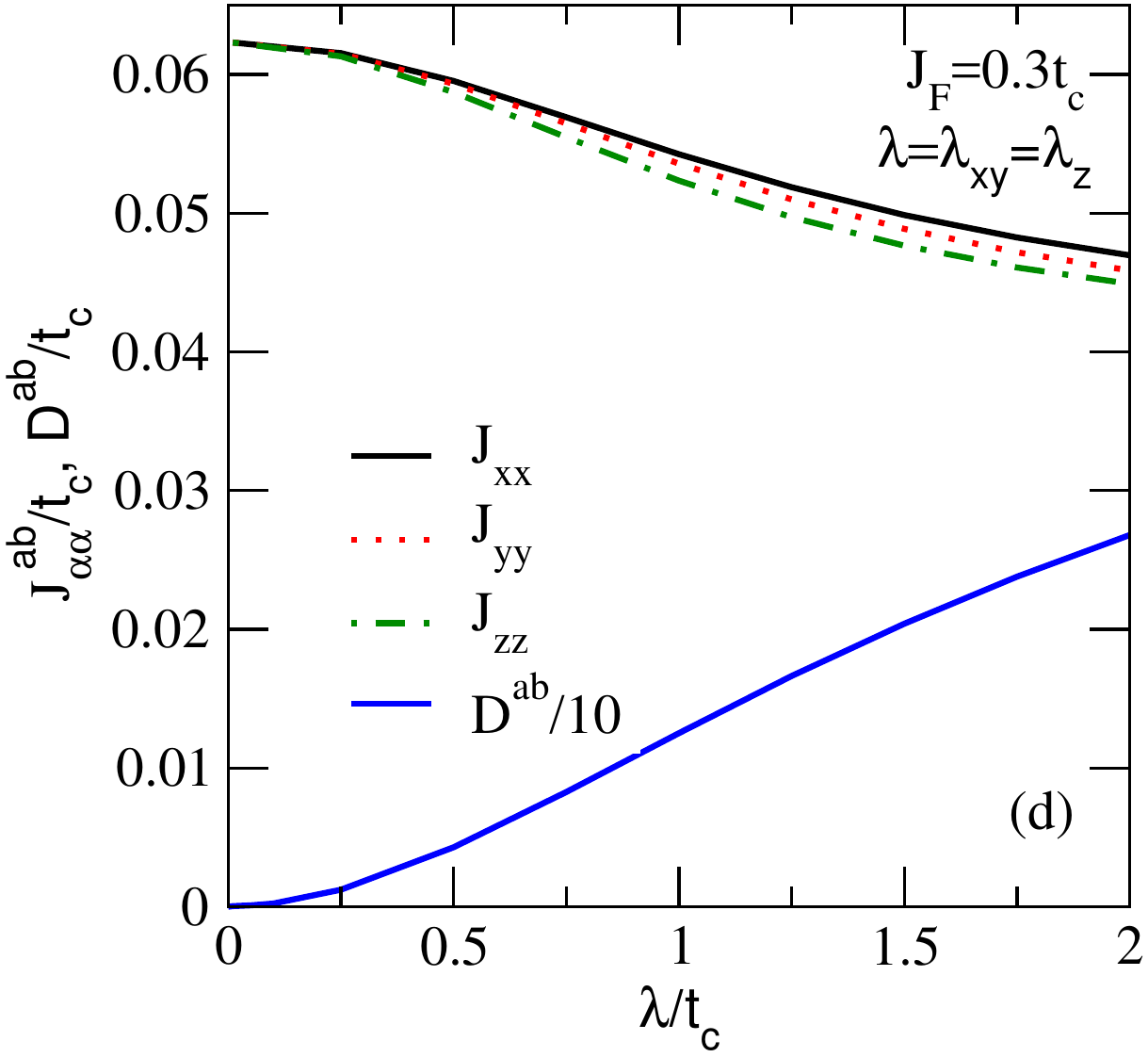}
	\includegraphics[width=4.5cm,clip=]{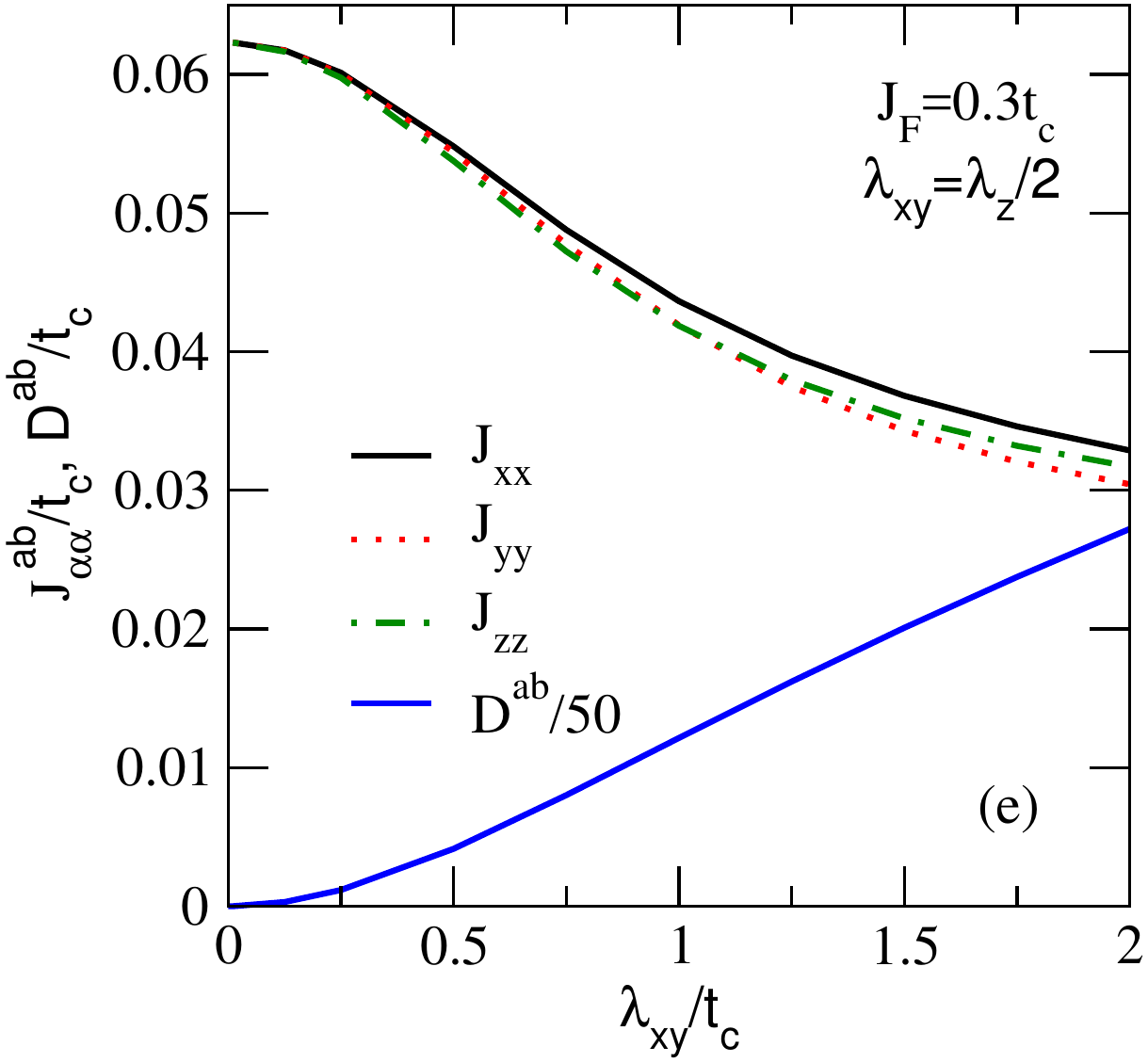}
	\includegraphics[width=4.5cm,clip=]{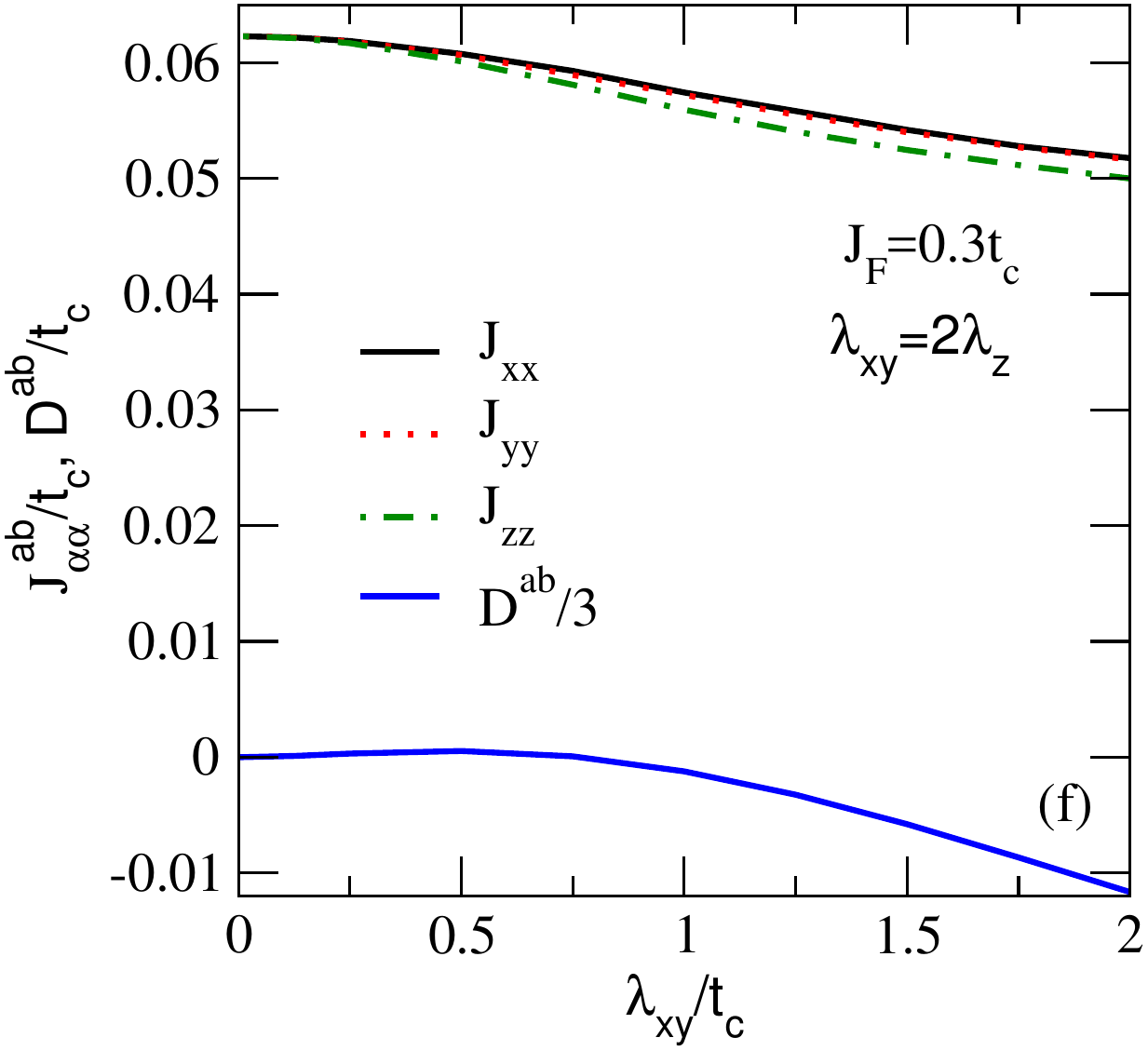}
	\caption{Anisotropic exchange couplings and single-spin anisotropy in the $a$-$b$ plane of trinuclear complexes.  
		The dependence of the parameters entering  model (\ref{eq:ab}) on SMOC are shown for $U=10t_c$.  
		The hopping between the trimers is $t=0.785t_c$.
		In the upper row panels we show the dependence on SMOC of the exchange couplings, $J_{\alpha\alpha}$ and the single-spin anisotropy, $D^{ab}=D+\Delta D^{ab}$,
		for different $\lambda_{xy}/\lambda_z$ ratios in the presence of an intracluster ferromagnetic coupling, $J_F=-0.3t_c$:  (a) $\lambda_{xy}/\lambda_z=1$, (b) $\lambda_{xy}/\lambda_z=1/2$ and  (c) $\lambda_{xy}/\lambda_z=2$.
		In the lower row panels [(d), (e) and (f)] we show the same cases but with an intracluster antiferromagnetic exchange: $J_F=0.3t_c$. The only non-zero off-diagonal exchange coupling, 
		$J^{ab}_{xz}$, is at most $\sim -4.4 \times 10^{-4}$; too small to be appreciable in the scale of the figure.}
	\label{fig:fig7}
\end{figure*}

 As shown in Fig. \ref{fig:fig7} anisotropic exchange couplings
$J^{ab}_{xx}\neq J^{ab}_{yy} \neq J^{ab}_{zz}$ arise when $J_F\neq 0$,  which are consistent with the triclinic 
splittings found in the exact level spectrum of Fig. \ref{fig:fig5}. Also we find off-diagonal 
exchange couplings: $J_{xz}^{ab} \neq 0$ to all orders of SMOC consistent with the analytical expression for $J_{xz}^{ab}$ 
derived in our previous work \cite{merino2016} valid to O($\lambda_{xy}^2, \lambda_z^2$). However, we typically find small
values of $J_{xz}^{ab} \sim -0.00044 t_c  ~(\lambda_{xy}=\lambda_z/2)$ and $J_{xz}^{ab} \sim -0.0003 t_c ~(\lambda_{xy}=\lambda_z)$
and so this parameter is not displayed in Fig. \ref{fig:fig7}. Therefore, to an excellent approximation, the in-plane Hamiltonian is an XXZ + 120$^\circ$ honeycomb model with single ion anisotropy.

Comparing the results shown in Fig. \ref{fig:fig7} for different $\lambda_{xy}/\lambda_z$ ratios,  
we observe that the anisotropies in the exchange couplings are enhanced for  $\lambda_{xy}/\lambda_z \neq 1$. 
In fact, larger anisotropies are found to occur for $\lambda_{xy}=\lambda_z/2$,  which is the parameter regime relevant to Mo$_3$S$_7$(dmit)$_3$ crystals.\cite{jacko2017} 
Also note from Fig. \ref{fig:fig7} the strong dependence of the magnitude of $D$ on the SMOC anisotropy. 
The single-spin anisotropy increases rapidly with SMOC, becoming equal to the exchange
couplings, $D \sim J^{ab}$ at $\lambda \approx 0.45 t_c$ 
($\lambda=\lambda_{xy}=\lambda_z$),  at $\lambda_{xy} \approx  0.22 t_c$ ($\lambda_{xy}=\lambda_z/2$) and 
at $\lambda_{xy} \approx 1.045 t_c $ ($\lambda_{xy}=2\lambda_z$).  At sufficiently large $D \gtrsim J^{ab}$ we 
expect the $D$-phase,  {\it i.e.}, a tensor product of $j=0$ states located at each cluster of the crystal.  Hence, a $D$-phase 
is favored by anisotropic SMOC with $\lambda_{xy}<\lambda_z$.  

In Fig. \ref{fig:fig7} we also show results for an antiferromagnetic exchange coupling 
inside the cluster, $J_F>0$. This could arise in, say, Mo$_3$S$_7$(dmit)$_3$ due to superexchange via the sulphur atoms in the core.  We find similar spin exchange 
anisotropies for both  ferromagnetic and antiferromagnetic $J_F$. In the antiferromagnetic case we find that $D^{ab}$ becomes negative for 
sufficiently large SMOC and $\lambda_{xy}=2\lambda_z$, consistent with our perturbative results for the $t$-$J$ model [cf. Eqs. (\ref{eq:Dapp}), (\ref{eq:Dappc}), and (\ref{Delta_g_pp})]. This 
signifies a switch of the ground state of the isolated cluster from the $j=0$ singlet to the $j=\pm 1$ doublet. In contrast, in
the ferromagnetic cases, $J_F<0$, we have explored a large parameter set and we always find $D^{ab}>0$.

\begin{figure} 
	\includegraphics[width=4cm,clip=]{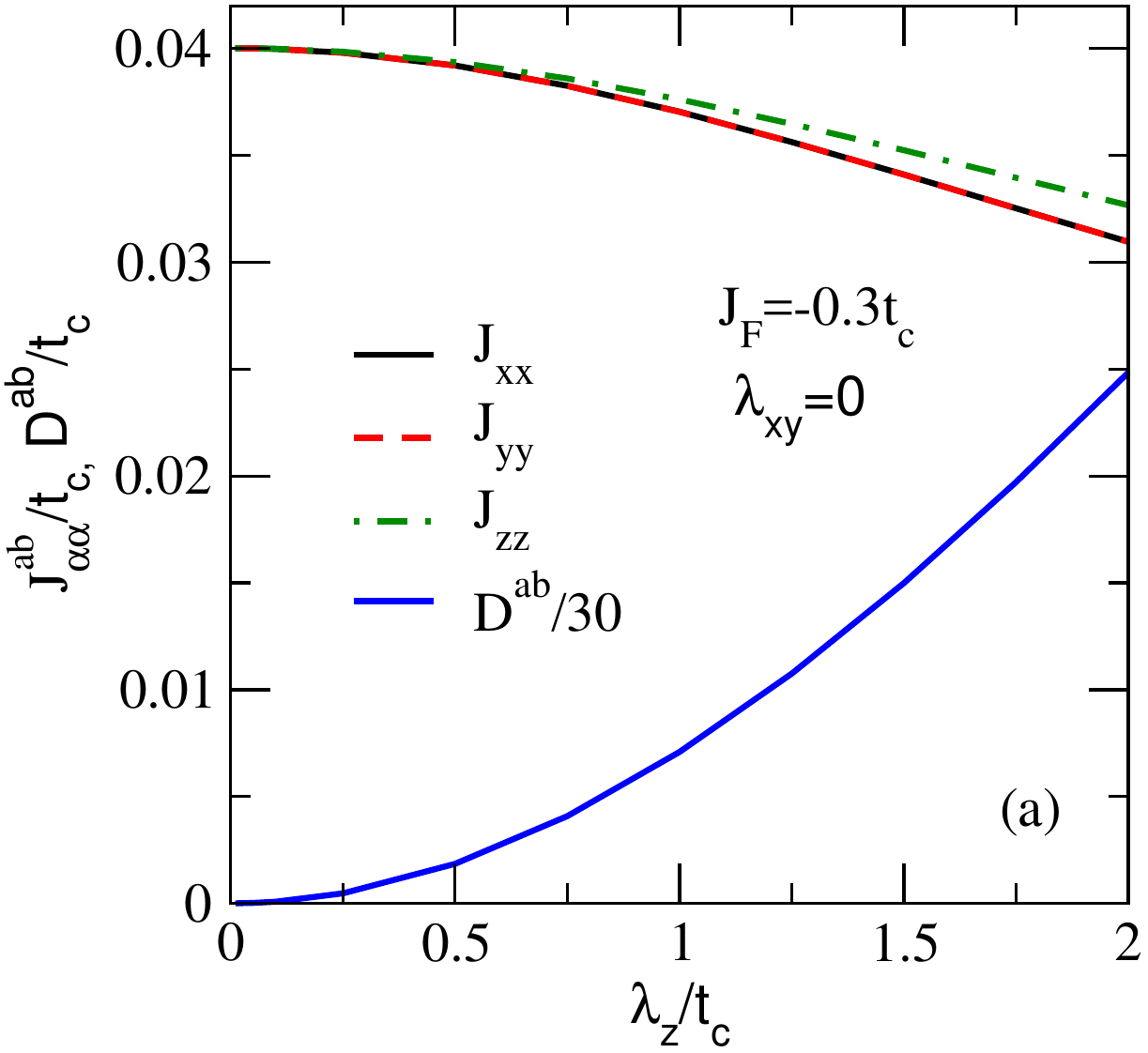}
	\includegraphics[width=4cm,clip=]{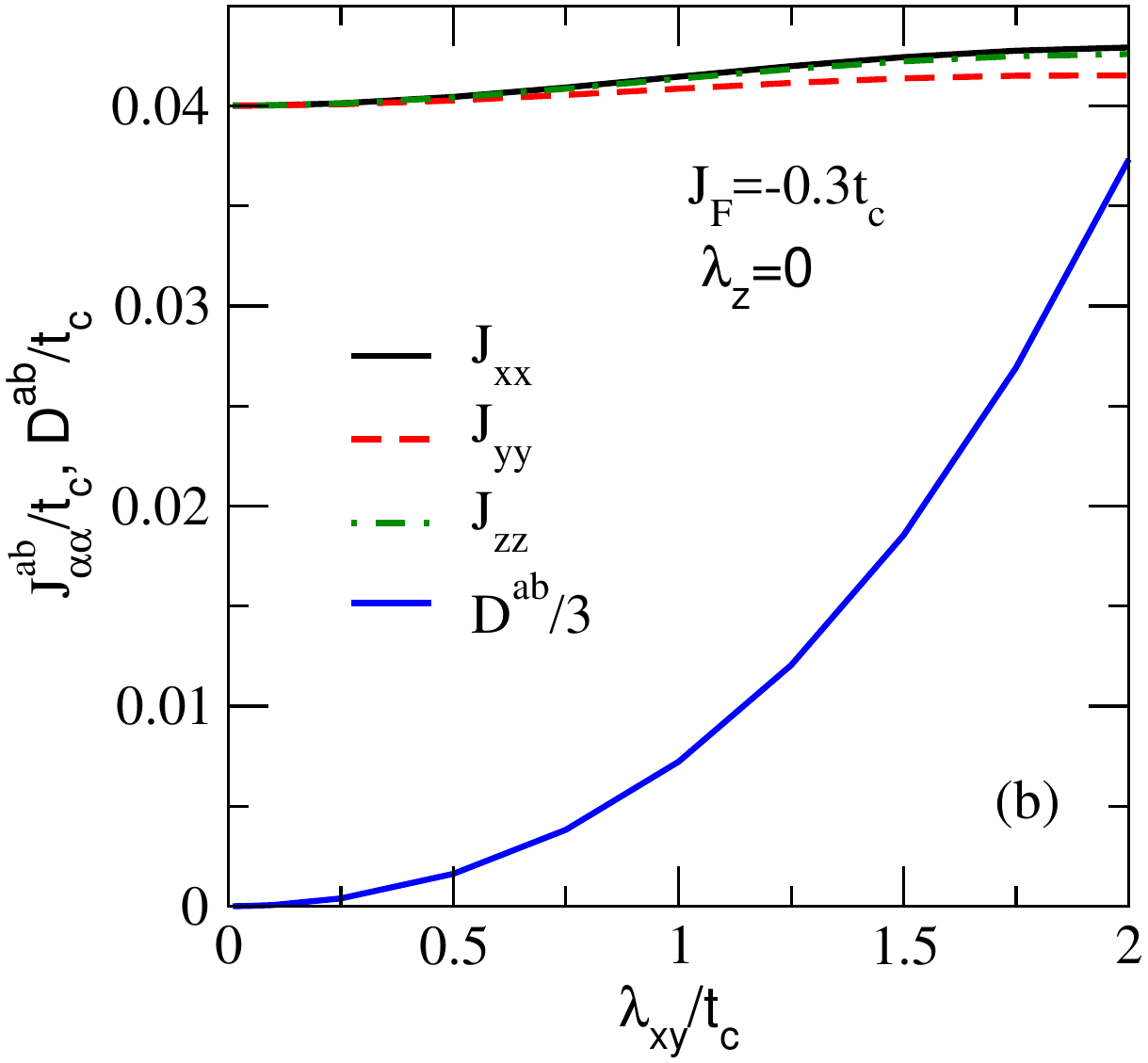}
	\caption{Anisotropic exchange couplings in the $a$-$b$ plane of trinuclear complexes in the limit of extreme SMOC anisotropies. 
		The dependence on SMOC of the parameters entering  model (\ref{eq:ab}) are shown for $U=10t_c$ and  $J_F=-0.3t_c$.  The hopping between the trimers is $t=0.785t_c$.
		We compare different ratios of the SMOC: (a) $\lambda_{xy}=0$ and  (b) $\lambda_z=0$.}
	\label{fig:fig7bis}
\end{figure}

In order to understand the effect of
exchange couplings with SMOC anisotropy, we show in Fig. \ref{fig:fig7bis} exchange couplings, $J^{ab}_{\alpha\alpha}$,
and $D$ in two extreme cases: $\lambda_{xy}=0$ and $\lambda_z=0$ with $J_F=-0.3t_c$. The $J^{ab}_{\alpha\alpha}$ are suppressed (enhanced) with SMOC for 
$\lambda_{xy}=0$ ($\lambda_z=0$),
 consistent with the results shown in Fig. \ref{fig:fig7}. 
Only when $\lambda_{xy}$ is turned on, does one find that the
transverse couplings become different {\it i.e.},  $J_{xx} \neq J_{yy}$. Furthermore, the single-spin anisotropy  is 
much more strongly enhanced by $\lambda_{xy}$ than by $\lambda_z$  (by more than an order of magnitude),
consistent with the analytical expressions [see Eqs. (\ref{eq:Dapp}), (\ref{eq:Dappc}), and (\ref{Delta_g_pp})].

\subsection{Anisotropies in the exchange interactions along the $c$-direction}

\begin{figure*} 
 	\includegraphics[width=4.5cm,clip=]{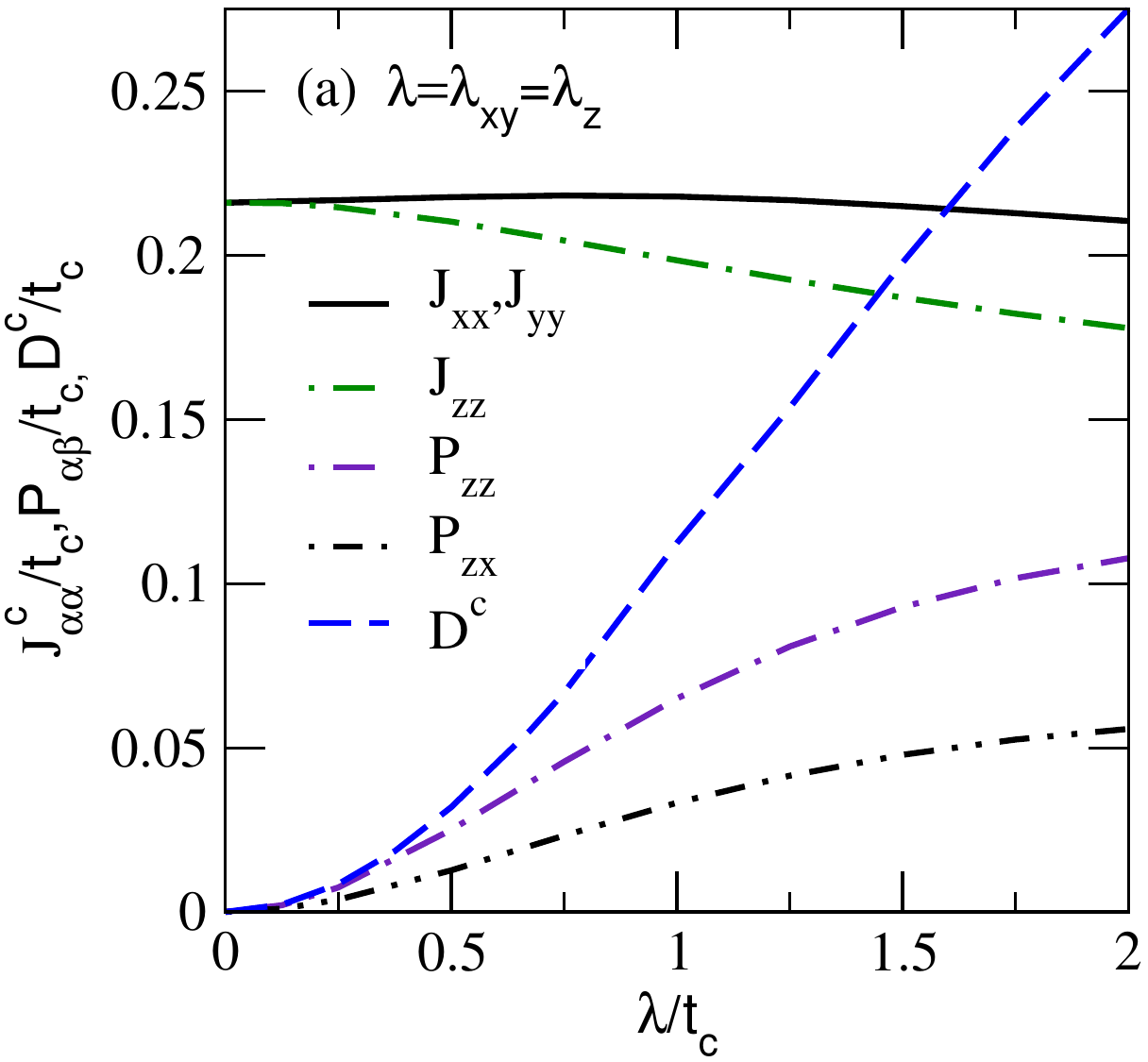}
 	\includegraphics[width=4.5cm,clip=]{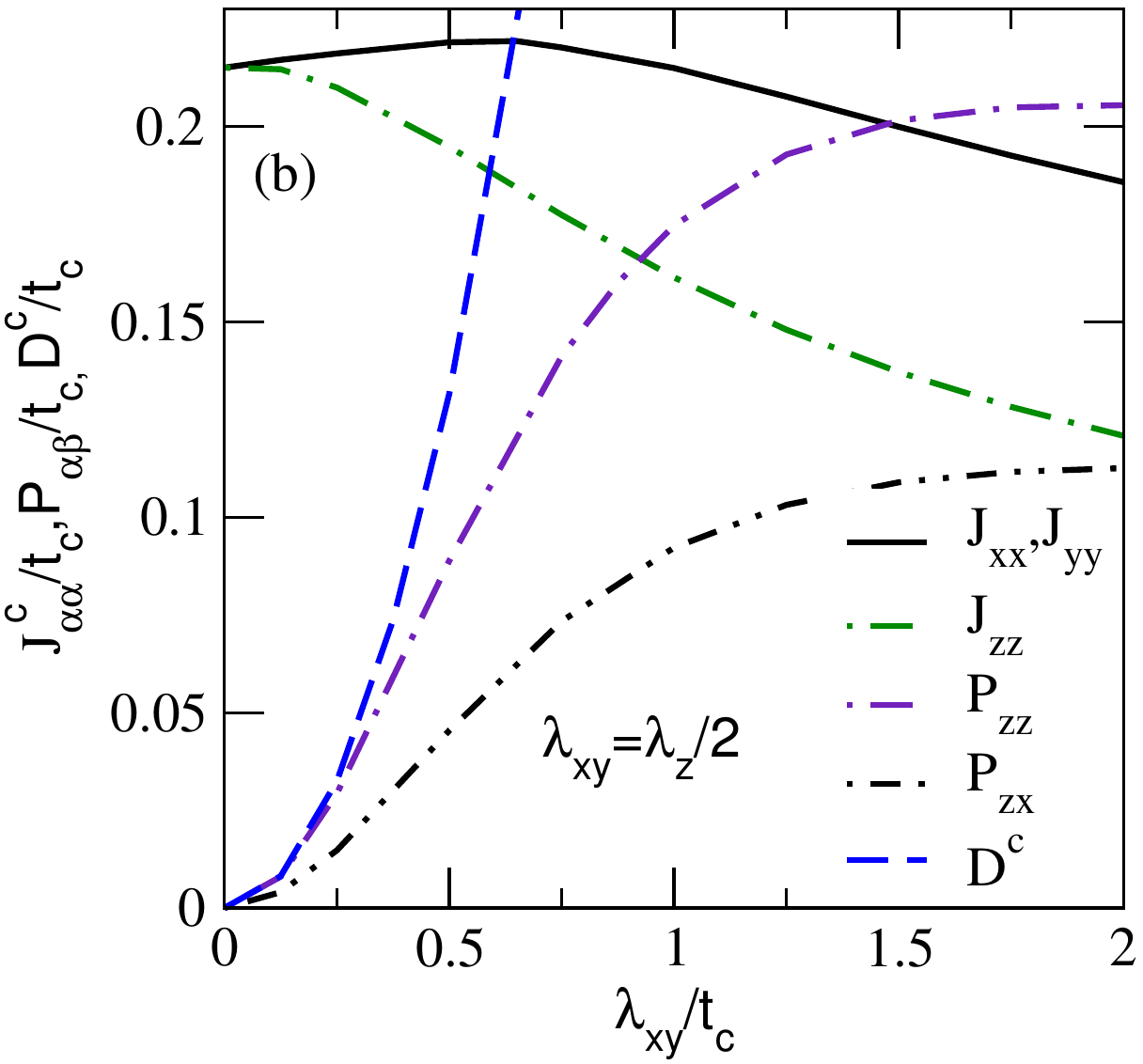}
 	\includegraphics[width=4.5cm,clip=]{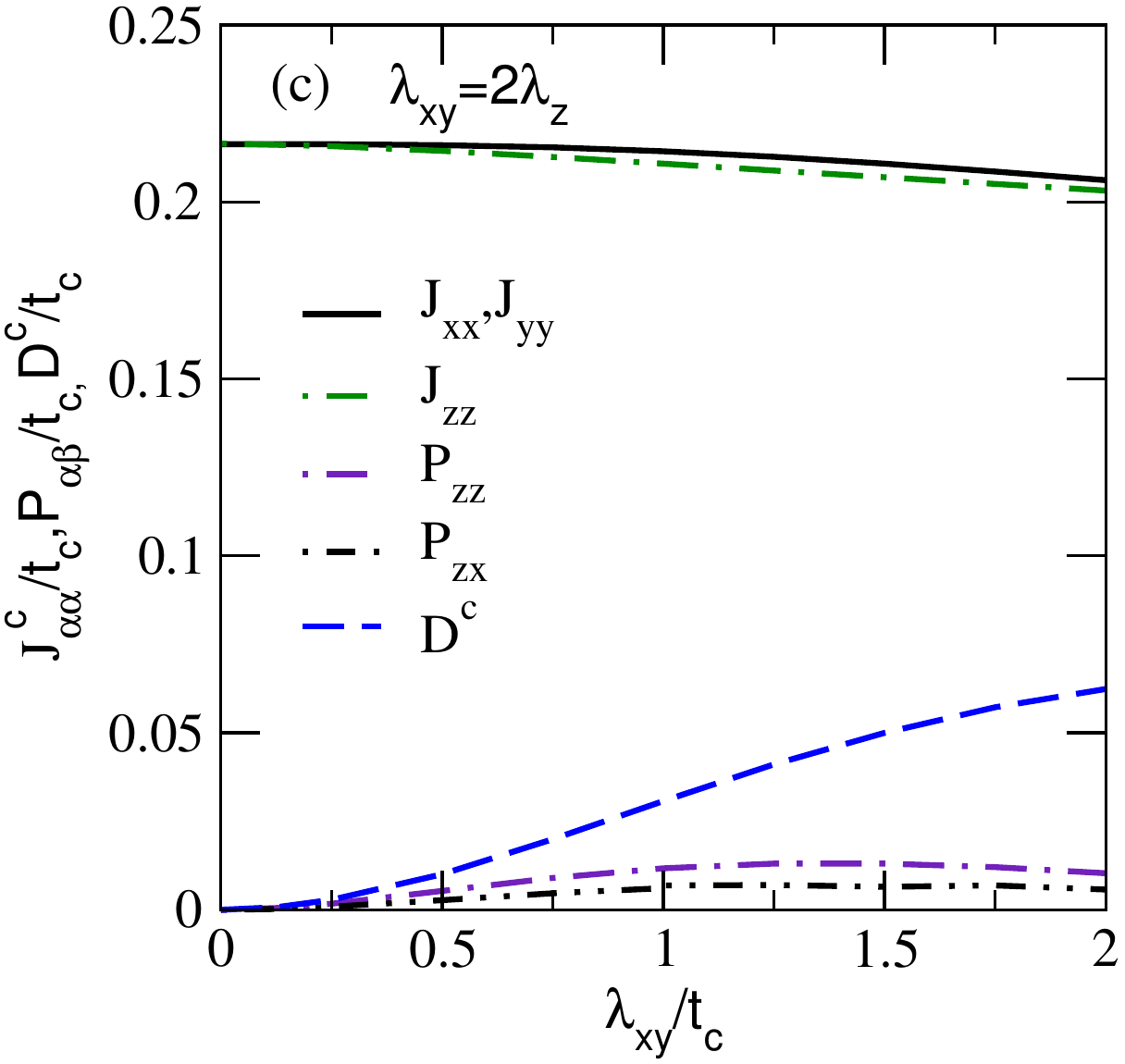}
 	\caption{Anisotropic exchange couplings in the $c$-direction of trinuclear complexes.  
 		The dependence of the parameters entering  model (\ref{eq:ab}) on SMOC are shown for $U=10t_c$ and  $J_F=-0.3t_c$.  
 		The hopping between the trimers is $t_z=0.683t_c$.
 		We compare different SMOC anisotropies: (a) $\lambda_{xy}/\lambda_z=1$, (b) $\lambda_{xy}/\lambda_z=1/2$ and  (c) $\lambda_{xy}/\lambda_z=2$.
 		Note the large enhancement of the single-spin anisotropy, $D^c=D + \Delta D^c$, for anisotropic SMOC becoming the largest for
 		$\lambda_{xy} < \lambda_z$. For $\lambda_{xy} = \lambda_z/2$, relevant to Mo$_3$S$_7$(dmit)$_3$ crystals,\cite{jacko2017} we have 
 		that $D^c \sim J^c_{zz}$ at about $\lambda_{xy}=0.65t_c$.}
 	\label{fig:fig8}
 \end{figure*}

The exchange couplings between two neighboring clusters in the $c$-direction
are shown in Fig. \ref{fig:fig8}. 
We find a diagonal exchange tensor:  
$J^c_{\alpha\beta}=J^c_{\alpha\alpha}\delta_{\alpha\beta}$, with $J^c_{xx}=J^c_{yy} \neq J^c_{zz}$  for any $J_F\ne0$ and $\lambda_{xy}/\lambda_z$ ratio.  
The higher symmetry than for a pair of  molecules in the $ab$ plane
 is due to  the $C_3$ rotational symmetry of the tube dimer, as discussed above.  
 
The largest anisotropies with $J^c_{xx}=J^c_{yy}>J^c_{zz}$ are seen in the case of anisotropic SMOC with $\lambda_{xy}=\lambda_z/2$
as shown in Fig. \ref{fig:fig8}(b). The only non-negligible biquadratic exchange terms, $P_{zz}>P_{zx}$,  increase rapidly with $\lambda_{xy}$
starting to saturate around $\lambda/t_c \sim  1-1.5$. The single-spin anisotropy equals 
the exchange coupling, $D = J^c$, at $\lambda_{xy}=0.65 t_c$ for $\lambda_{xy}/\lambda_z =1/2$ and
at $\lambda_{xy}=1.457 t_c $  for  $\lambda_{xy}/\lambda_z =1$, while for $\lambda_{xy}/\lambda_z =2$ there is no critical
$\lambda_{xy}$ at which  $D \sim J^c$ within the parameter range explored. Hence, anisotropic SMOC with $\lambda_{xy} < \lambda_z$ again 
favors the $D$-phase as in the dumbbell arrangement.

\begin{figure} 
	\includegraphics[width=4cm,clip=]{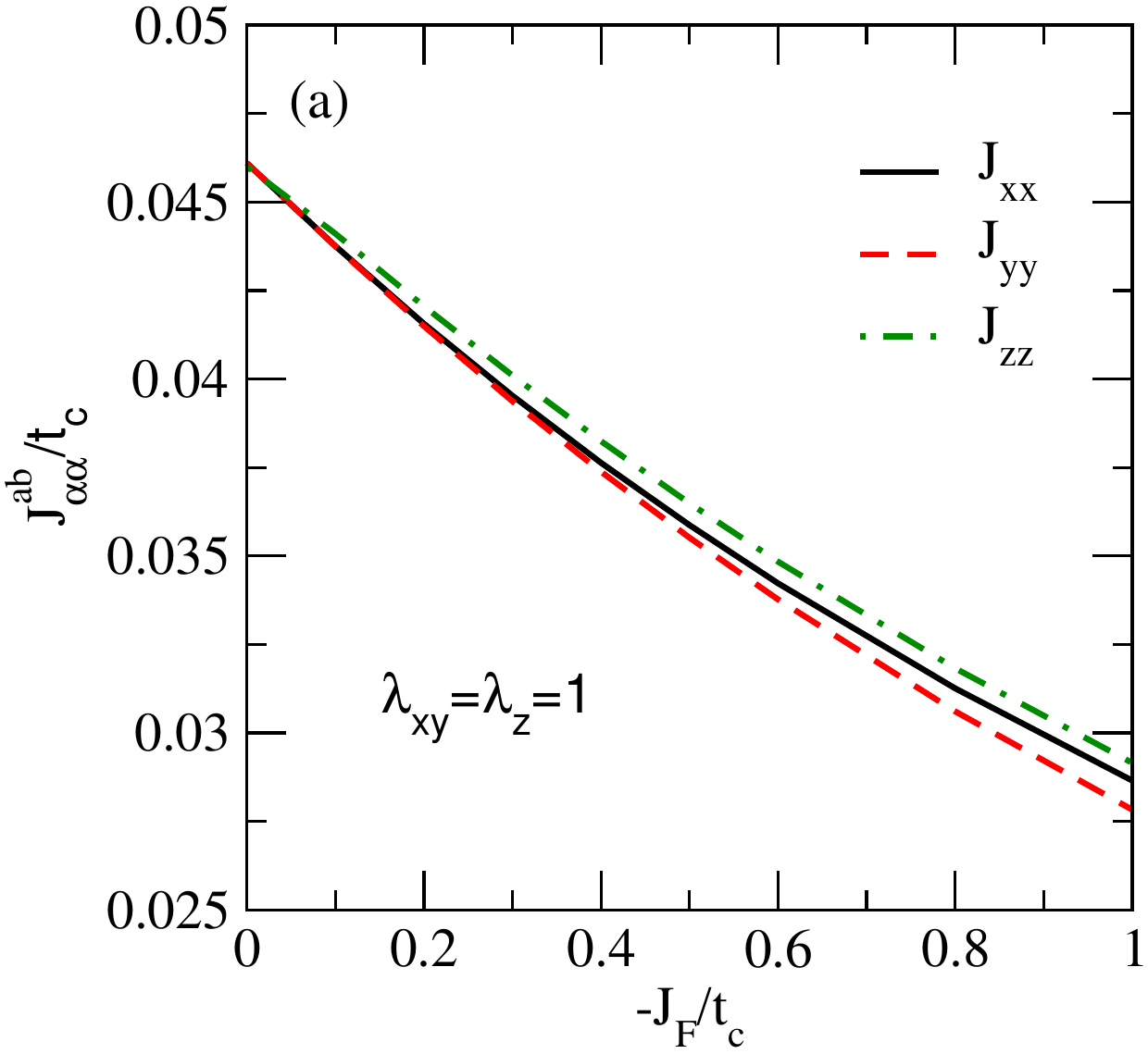}
	\includegraphics[width=4cm,clip=]{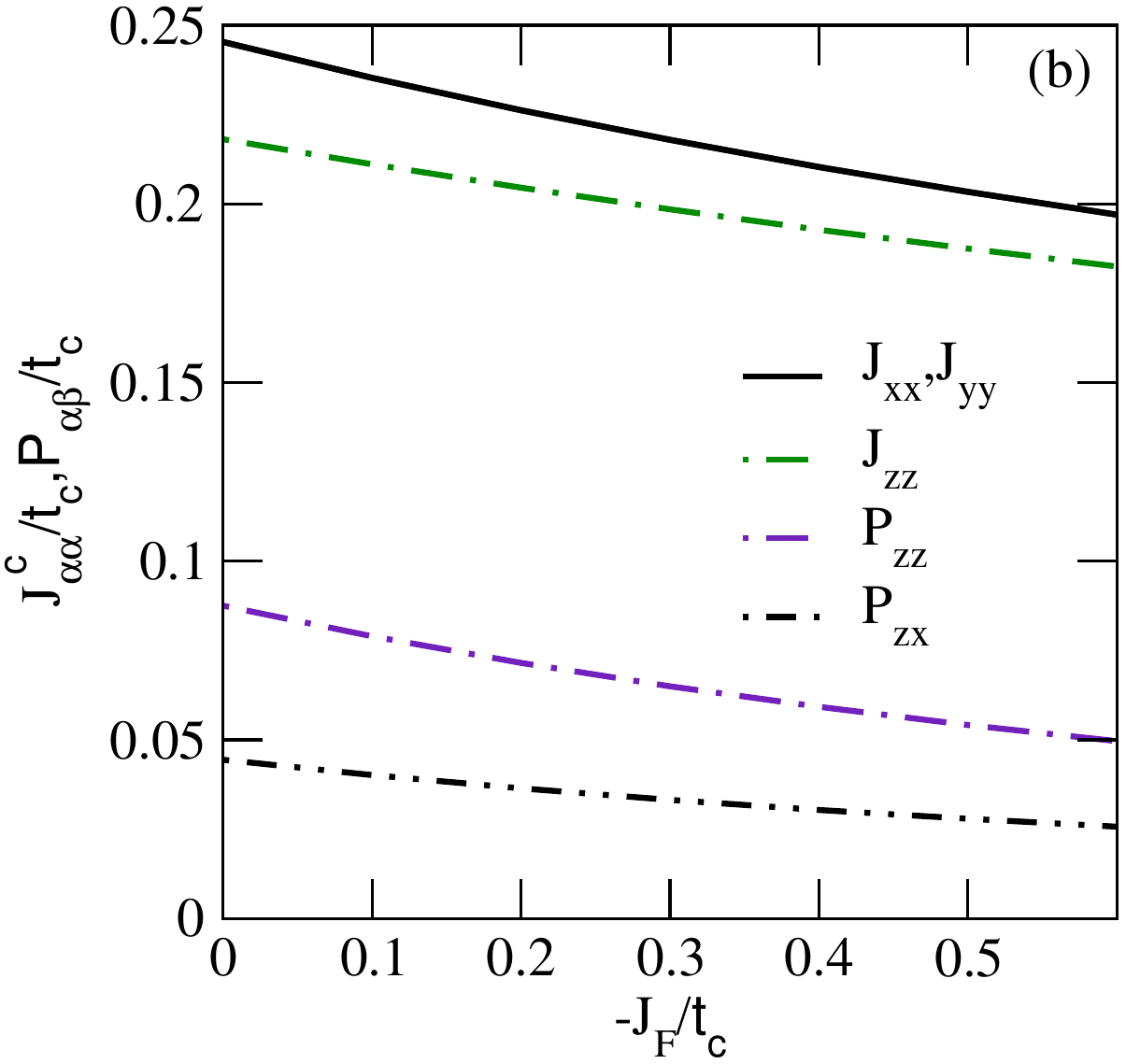} 
	\caption{Effect of the intracluster exchange, $J_F$, on the exchange couplings between trimers. In (a) we show the dependence 
		of the exchange couplings in the $ab$-plane, $J^{ab}$, on $-J_F$ while in (b) we show the dependence of exchange couplings in 
		the $c$-direction, $J^c$, on $-J_F$. We have used $U=10t_c$, $t=0.785$, $t_z=0.683$ and $\lambda_{xy}=\lambda_z=1$}
	\label{fig:fig9}
\end{figure} 

Finally, in Fig. \ref{fig:fig9} we compare the dependence of the exchange couplings on $J_F$ for $\lambda_{xy}=\lambda_z=1$.
The couplings in the $ab$ plane, $J^{ab}_{\alpha\alpha}$ are suppressed and become gradually anisotropic, $J^{ab}_{xx} \neq J^{ab}_{yy} \neq J^{ab}_{zz}$  
as  $J_F$ increases. This is in contrast to the exchange couplings in the $c$-direction which do not display larger
anisotropies but rather $J^c_{xx}=J^c_{yy} \neq J^c_{zz}$ for any $J_F$.

\section{Discussion of properties of the quasi-one-dimensional pseudospin-one model}
\label{sec:sec5}

Our analysis shows that the magnetic properties of layered decorated honeycomb lattice model  at strong coupling, $U\gg t_c,t, t_z, \lambda_{xy}, \lambda_z$,
are captured by  model  (\ref{eq:fullmodel}) with the exchange couplings obtained from our
combined approach described above. 
On comparing $J^{ab}$  in Fig. \ref{fig:fig7} with $J^{c}$ in Fig. \ref{fig:fig8} we find that 
$J^{c} \sim 5 J^{ab}$ for $U=10t_c$. This is related to the fact that
two clusters in the tube arrangement are connected by three hoppings so that they can 
exchange electrons without paying an energy cost\cite{merino2016,powell2016} $\sim U$. This mechanism is generic to decorated lattices and not 
specific to the model considered here.\cite{elise2017} In contrast,  neighboring clusters in the
dumbbell arrangement pay energy, $U$, since they can only exchange particles through a single hopping
connecting them. Hence,  $J^{ab}$ is strongly suppressed by $U$ in contrast to  $J^c$,
leading to an increase of the $J^c/J^{ab}$ ratio. Hence, 
at large $U$ the system becomes quasi-one-dimensional consisting on a set of weakly coupled 
pseudo spin-one antiferromagnetic chains. 

An isotropic version of model, (\ref{eq:c}) i.e., $J^c_{\alpha\beta}=J^c\delta_{\alpha\beta}$ $P_{\alpha\beta}=P\delta_{\alpha\beta}$ and $D^c = 0$ is 
just the bilinear-biquadratic model: $H = J^c \sum_\ell \bm{\mathcal{S}}_{\bm r_\ell} \cdot \bm{\mathcal{S}}_{\bm r_\ell+\bm \delta_z} +P \sum_\ell (\bm{\mathcal{S}}_{\bm r_\ell} \cdot \bm{\mathcal{S}}_{\bm r_\ell+\bm \delta_z})^2$, which becomes the Affleck-
Kennedy-Lieb-Tasaki (AKLT) model for $P/J^c=1/3$. The AKLT model can be solved exactly and has the valence bond solid ground state and is in
the Haldane phase\cite{AKLT}. 

We finally note that the next-nearest-neighbor exchange couplings 
between clusters in the $c$-direction can be neglected since recent estimates \cite{merino2016} suggest that they are about 20 times smaller than the nearest 
neighbor exchange coupling. This is because the small parameter in the  perturbation theory is $t_z/3$ so  fourth order terms (such as next-nearest-neighbor exchange couplings) 
must be at least an order of magnitude smaller than second order terms (such as nearest-neighbor exchange coupling).

\subsection{One-dimensional antiferromagnetic $S=1$ Heisenberg chains}

When no interchain coupling is present, $J^{ab}=0$, and $D^*<J^c$, the system consists on a set of 
uncoupled one-dimensional $S=1$ antiferromagnetic chains that are in the Haldane phase. The Haldane
phase is characterized by exponentially decaying spin correlations associated with \cite{dmrg} 
the Haldane spin gap $\Delta_s=0.4107(3)J^c$ to the lowest triplet state and string order. It is a 
symmetry-protected topological phase with nonlocal string
order and fractionalized edge states \cite{haldane1983a,haldane1983b,gomezsantos1989}. Topological protection can arise
from either (i) the dihedral group of $\pi$-rotations around the $x$ and $y$
axis, (ii) time-reversal symmetry or (iii) reflection through a plane perpendicular to the chain (or bond-center inversion symmetry, which is 
equivalent in one-dimension).\cite{pollmann2012} In the underlying fermionic model, charge fluctuations imply that topological protection 
can only come from reflection symmetry with respect to a plane perpendicular to the $c$-axis at the midpoint of a 
bond\cite{nourse2016}.

On the other hand, when $D^*\gg J^c$, the ground state
is adiabatically connected to a trivial state consisting on the 
tensor product of the ${\mathcal S}^z_{\bm r_\ell}=0$ at each cluster. The lowest energy 
excitations of the $D$-phase which reside in the ${\mathcal S}^z_{\bm r_\ell}=\pm 1$ sector, 
are gapped and consist of pairs  of excitons and antiexcitons which can be bound.  
Numerical studies \cite{oitmaa2009,normand2011,langari2013,tzeng2008} have established that in the pure spin model the 
quantum critical point	 separating the $D$-phase and Haldane phase 
occurs at $D^*/J^c \sim 0.96-0.971$. It has been found that in a pure spin model such as the one discussed here,
a quantum phase transition between the Haldane phase and the topologically trivial $D$-phase 
is signalled by the change in sign of an inversion-symmetry-based order parameter
\cite{langari2013} which is a non-local topological order parameter.  Hence,
a transition from a Haldane phase to a $D$-phase occurs when increasing SMOC until $D^* \sim J^c$.   

{From our analysis of Fig. \ref{fig:fig8} (b), which is the relevant SMOC ratio to Mo$_3$S$_7$(dmit)$_3$,
(assuming $J_F=-0.3 t_c$), we predict a transition from the Haldane to the $D$-phase at $\lambda_{xy} \sim 0.65 t_c$.
{\it Ab initio} estimates of SMOC\cite{jacko2017} in Mo$_3$S$_7$(dmit)$_3$ find that $\lambda_{xy}=\lambda_z/2=0.042t_c$,  
which would naively mean that the single-spin anisotropy is too small, $D^*\ll J^c$, to induce a $D$-phase 
in the crystal. By moving to suitable materials containing heavier elements \cite{khosla2017}, SMOC can be increased
by, at most, a factor of $4-5$ leading to $\lambda_{xy}  \approx 0.2 t_c\ll\lambda^{critical}_{xy}$ which means that 
the system is still in the Haldane phase. However, the critical $\lambda_{xy}$ for the transition can be reduced by suppressing 
$t_z$ and increasing $-J_F$ as shown in Fig. \ref{fig:fig0}. Also in the underlying 
fermionic model (neglecting SMOC), the Haldane gap is suppressed by more than an order of magnitude by 
charge fluctuations.\cite{nourse2016} More specifically, charge fluctuations renormalize the critical condition
$D^* \sim J^c$ to $D^* \sim 0.066J^c$ for the parameters relevant to \Mo. This  leads to a smaller $\lambda^{critical}_{xy}$ as shown in Fig. \ref{fig:fig0}. 
The above discussion indicates that  a series of materials related to \Mo with slight variations in model parameters could easily  effectively span the phase Haldane--to--$D$-phase transition. Furthermore, a material on the D-phase side of the transition could be driven into the Haldane phase by uniaxial pressure along the c-axis.} In particular, our results above suggest that the critical ratio $D^*/J^c$ could 
be exceeded by moving to suitable materials containing heavier elements.\cite{khosla2017}
Furthermore, one expects that the interlayer hopping $t_z$ will be extremely sensitive to chemical details. As $J^c\sim t_z^2$ 
structures with increased interlayer separation will strongly favor the $D$-phase.

\subsection{Effect of the interchain couplings}

When the quantum pseudospin-one chains are coupled through a sufficiently strong interchain coupling, $J^{ab}$,  the Haldane phase 
becomes unstable to 3D magnetic order. In previous numerical studies of weakly coupled $S=1$ antiferromagnetic Heisenberg chains   
(with $D^*=0$), it was estimated\cite{sengupta2014} that the critical value for the transition from the Haldane to the ordered 3D magnet occurs 
around $J^{ab}/J^{c} \geq (0.08-0.11) z \sim0.3$, where the  coordination number $z=3$ for the honeycomb lattice. Since we find that 
$J^{ab}/J^{c} \lesssim 0.2$, we expect that the ground state of our model is in the Haldane phase when $D^* =0$. This critical ratio, $J^{ab}/J^c$, 
for the onset of  3D magnetic order is suppressed  by $D^*$ as shown\cite{sakai1990} by mean-field treatments of the interchain 
coupling, $J^{ab}$. 

\subsection{Effect of an external magnetic field}

An external magnetic field suppresses the 1D quantum fluctuations and the Haldane gap, $\Delta_s$, closes\cite{affleck1990} at $h_c \sim \Delta_s$, whence a
transition to a 3D ordered magnet occurs. A quantum critical region with a V-shape emerges around $h_c$ in the 
temperature versus magnetic field, $T$-$h$, phase diagram\cite{sakai1990,giamarchi2007,bera2015}. The temperature, $T \sim J^{ab}$, sets the energy scale at which 3D quantum criticality for $T<J^{ab}$
crosses over to 1D behavior for $T>J^{ab}$. Similarly the three-dimensional magnetically ordered phase found for $h>h_c$ and $T=0$ crosses over to a gapless
Tomonaga Luttinger Liquid (TLL) at temperatures $T>J^{ab}$. We note that, strictly speaking, 
the TLL behavior should only occur\cite{capponi2016} in the range $J^{ab} < T < J^c$, since at too large temperatures, $T\gg J^c$, classical 
behavior sets in.  In the presence of a nonzero and small $D^*$, with $D^*\ll J^c$,  the lowest triplet state 
is split into a $j=\pm 1$ doublet with energy $\Delta_{\pm}$ above the ground state and a $j=0$ singlet at energy $\Delta_0$ with $\Delta_\pm < \Delta_0$.  
Hence, under an applied magnetic field,  $\Delta_{\pm}$, is suppressed and the transition from the Haldane phase to the 3D 
ordered phase occurs around $h_c=\Delta_{\pm}<\Delta_s$.  Apart from the downward shift of $h_c$, we can expect, qualitatively, a similar 
$T$-$h$ phase diagram as in the case with no single-spin anisotropy, $D^*=0$.  

\section{Conclusions}

We have analyzed the magnetic properties of the trinuclear organometallic materials, such as Mo$_3$S$_7$(dmit)$_3$.
These materials are potential candidates for realizing compass interactions in their  layers.  
In order to explore such possibilities we have derived an effective magnetic model describing the 
magnetic interactions between the pseudospin-one at each molecular cluster arising from strong Coulomb repulsion, 
lattice structure and SMOC. In spite of the crystals being nearly 
isotropic, we find that the exchange coupling between nearest-neighbor pseudospins along the $c$-direction is much larger
than between pseudospins within the hexagonal $a$-$b$ planes. Hence, the spin exchange model for these crystals is 
effectively quasi-one-dimensional. Magnetic anisotropies are found to arise 
under the simultaneous effect of spin orbit coupling and intra-cluster exchange interaction. These  
anisotropies are further enhanced by SMOC anisotropy, particularly when $\lambda_{xy} < \lambda_z$, which is
naturally present in organometallics.  Our analysis suggests that Mo$_3$S$_7$(dmit)$_3$ is most probably in
the Haldane phase since  the efffective model consists of weakly coupled $S=1$ antiferromagnetic chains in the presence
of small single-spin anisotropy induced by SMOC. However, by increasing the interlayer 
distances through changes in the chemistry of the material, increasing the anisotropy of magnitude of the SMOC  it should be possible to effectively drive it
into to the $D$-phase. A larger SMOC should be realised in complexes containing heavier metals.\cite{khosla2017}

The Haldane phase is strongly sensitive to an external magnetic field. Under applied magnetic fields larger than 
the Haldane gap, $h>h_c \sim \Delta_s $,  the Haldane phase is destroyed and a three-dimensional magnet may
be stabilized. We have estimated this critical field, $h_c$, based on our present analysis using  
DFT parameters\cite{jacko2017} for Mo$_3$S$_7$(dmit)$_3$ (Table \ref{table1})
with an onsite $U=10t_c$ and $J_F=-0.3t_c$. Using these parameters we extract  $J^c =0.0126$ eV from 
our Fig. \ref{fig:fig8}(b) which leads to a critical magnetic field $h_c \sim  \Delta_s \sim 41.4$ T assuming the Haldane spin gap, 
$\Delta_s = 0.414 J^c \approx  0.09 t_c$, in the pure Haldane chain. However, recent DMRG calculations on Hubbard tubes\cite{nourse2016}
have shown that charge fluctuations strongly suppress the spin gap when decreasing $U$. For the parameter range 
considered here, we would find: $\Delta_s \sim 0.006 t_c$, implying experimentally accessible critical fields: 
$h _c \sim 3 $ T.  A V-shaped quantum critical region in the $T-h$ phase diagram separating the Haldane phase from the three-dimensional
magnetically ordered phase should then emerge as observed in inorganic Haldane chain materials.\cite{bera2015}

Exfoliation or growth of a monolayer of trinuclear complexes arranged as in the ab-planes of Mo$_3$S$_7$(dmit)$_3$, 
would lead to the realization of a decorated hexagonal lattice which is known to contain rich physics. We have found that 
at large $U$ and no SMOC, the magnetic interactions between the pseudospin-one is that 
of a conventional nearest-neighbor antiferromagnetic Heisenberg model on an hexagonal lattice.\cite{footnote} The ground state of 
this model is a pure N\'eel antiferromagnet.  However, if crystal parameters are tuned so that 
magnetic exchange anisotropies are enhanced, disordered spin liquid phases \cite{sheng2015} may be achieved. For instance, 
if the relative orientation 
between the molecules in the crystal is modified so that inversion symmetry within the planes is broken, a DM interaction 
arises\cite{powell2016} which competes with the 
magnetic order,\cite{cepas2008} which can lead to interesting spin liquid phases\cite{messio2017}. 
All this illustrates how isolated layers of trinuclear organometallic complexes are ideal playgrounds to explore the
quantum many-body phases realized in a decorated honeycomb lattice.
\label{sec:sec6}

\section*{Acknowledgements.}
J. M. acknowledges financial support from (Grant No. MAT2015-66128-R) MINECO/FEDER, Uni\'on Europea..
Work at the University of Queensland was supported by the Australian Research Council (FT13010016 and DP160100060) and by computational resources provided by the Australian Government through Raijin under the National Computational Merit Allocation Scheme..

\begin{appendix}

\section{Electronic structure of isolated triangular clusters}\label{app:A}

Here, we provide the details of the electronic structure of isolated clusters with
different numbers of electrons. 

\subsection{Isolated triangular cluster with five electrons}

We start studying isolated trimers with $N=1$ electrons. This is due to its intrinsic importance and
due to the fact that the electronic structure of trimers with $N=5$ electrons and $t_c, \lambda_{xy}, \lambda_z>0$, relevant 
to Mo$_3$S$_7$(dmit)$_3$ can be obtained from the $N=1$ case by a particle-hole transformation  
switching the sign of the parameters: $t_c \rightarrow  -t_c, \lambda_{xy} \rightarrow -\lambda_{xy}, \lambda_z \rightarrow -\lambda_z$
apart from a rigid energy shift. 

For only one electron in the cluster, $N=1$,  the Hamiltonian is just $H=H_0+H_{SMOC}$. 
Since $[H,J_z]=0$, where $J_z=L_z+S_z$, then the projection of the total momentum along the $z$-axis is
a good quantum number.  In the following we denote the basis states for a fixed number of particles, $N$, as
$|N;j,n \rangle $ where $j=k+\sigma$ and $n$ numbers the different possible configurations for each $j$-sector. 
Hence, in this case the possible basis states are
\begin{eqnarray}
|1;1/2,1\rangle &=& b^\dagger_{0 \uparrow} |0 \rangle
\nonumber \\
|1;1/2,2 \rangle &=& b^\dagger_{1 \downarrow} |0 \rangle
\nonumber \\
|1; -1/2,1 \rangle &=& b^\dagger_{0 \downarrow} |0 \rangle
\nonumber \\
|1; -1/2,2 \rangle &=& b^\dagger_{-1 \uparrow} |0 \rangle
\nonumber \\
|1; 3/2,1 \rangle &=& b^\dagger_{1 \uparrow} |0 \rangle
\nonumber \\
|1; -3/2,1 \rangle &=& b^\dagger_{-1 \downarrow} |0 \rangle.
\end{eqnarray}
The eigenenergies, $E_n(N;j)$ of the Hamiltonian are
\begin{eqnarray}
E_2(1;j=\pm 3/2)&=&t_c+{\lambda_z \over 2}, \nonumber \\
E_1(1;j= \pm1/2)&=&-{\lambda_z \over 4}-{t_c \over 2}+\sqrt{ \left({\lambda_z-6 t_c \over 4} \right)^2+{\lambda_{xy}^2 \over 2} }, \nonumber \\
E_0(1;j= \pm 1/2)&=&-{\lambda_z \over 4}-{t_c \over 2}-\sqrt{ \left({\lambda_z-6 t_c \over 4} \right)^2+{\lambda_{xy}^2 \over 2} }. \nonumber \\
\label{eq:spect}
\end{eqnarray}
Hence the level spectra for $N=1$ consists of three doublets with the energies given above. 
The ground state of the system with one electron, $N=1$, is a doublet 
with energy, $E_0$. Time-reversal invariance of the Hamiltonian, $[T,H]=0$, and 
Kramers theorem ensures that all states should have a minimum degeneracy   
of two since the cluster has an odd number of electrons. Note that the level spectra of the
triangular cluster with $N=5$ electrons (one hole) would be the same as (\ref{eq:spect})   but with the signs reversed: $t_c \rightarrow -t_c$, $\lambda_{xy} \rightarrow -\lambda_{xy} ,
\lambda_z \rightarrow -\lambda_z$ and with an upward rigid shift of all energies by $+2U$.  

To make contact with previous work on transition metal oxides it is illustrative to consider our model Hamiltonian: $H=H_0+H_{SMOC}+H_{U-J_F}$,
with $H_0$, $H_{SMOC}$ and $H_{U-J_F}$  expressed in the $(k,\sigma)$ basis as given by Eq. (\ref{eq:h0}), (\ref{eq:ls}) and (\ref{eq:hUJFk}), respectively.
For $U,J_F=0$, this model is reminiscent of a model previously considered \cite{jackeli2009,perkins2014a,perkins2014b} for Ir$^{4+}$ ions in $A_2$IrO$_3$ ($A$=Na,Li) compounds.
In these systems, five electrons occupy the lowest $t_{2g}$ manifold of the Ir ions 
which is well separated from the high energy $e_g$ doublet. The low energy effective model for the hole in the $t_{2g}$ manifold 
of the {\it isolated } Ir-ions includes a trigonal crystal field resulting from the surrounding oxygen octahedra and a large
SOC contribution\cite{perkins2014b}: $H=\Delta (L^z)^2+\lambda {\bf L} \cdot {\bf S}$, with $\Delta >0$.  

Through the particle-hole transformation discussed above, the three-fold degenerate $t_{2g}$ manifold of the isolated Ir ion with one hole
is equivalent to our model of the isolated molecule with one electron, $N=1$, with the signs of $\lambda=\lambda_{xy}=\lambda_z$
 and $t_c$ reversed. Full rotational symmetry is only recovered for $t_c \rightarrow 0$ in our model when $U,J_F=0$.
In that case, $[H_0+H_{SMOC},L]=0$, so that the total angular momentum, $L$, is a good quantum number, as it should. 
In this situation, we find that isotropic SOC ($\lambda_{xy}=\lambda_z=\lambda$) splits the $(2L+1) (2S+1)=6$ manifold ($L=1,S=1/2$) into a $j=1/2$ doublet 
with energy $E_0(1;j=1/2)=-\lambda$ and a $j=3/2$ quadruplet with energy $E_1(1;j=3/2)={\lambda \over 2}$.
This situation corresponds to removing the crystal field acting on the $d$-orbital manifold in transition metal oxides.

\subsection{Isolated triangular clusters with four electrons}

The basis states with $N=N_\uparrow+N_\downarrow = 4$ electrons includes states with
total spin $S_z=0$, ($N_\uparrow=2, N_\downarrow=2$), $S_z=1$ ($N_\uparrow=3, N_\downarrow=1$) and 
$S_z=-1$ ($N_\uparrow=1, N_\downarrow=3$).  Noting that basis states with total momentum $k'$ 
are equivalent to $k$ if they satisfy $k=k'\pm 3n$, we find that the basis states can be classified according 
to three possible values: $j=0, \pm 1$. Since the Hamiltonian does not mix sates with different $j$, the original $15\times 15$ matrix can be 
expressed in block diagonal form consisting of $5\times5$ matrices corresponding to $j=0,\pm 1$. 
We now explicitly show the classification of the $(k,\sigma)$ basis states according to $j=0,\pm 1$ 
and the analytical diagonalization of the matrices corresponding to each of the $j$-sectors.
We keep the  $|N;j,n \rangle$ classification of the basis states.  

\subsubsection{$j=0$ sector}

The three possible configurations with $k=\sigma=0$ are
\begin{eqnarray}
|4;0,1 \rangle &=& b^\dagger_{0 \uparrow} b^\dagger_{-1\uparrow} b^\dagger_{0\downarrow} b^\dagger_{1\downarrow} |0 \rangle
\nonumber 
\end{eqnarray}
\begin{eqnarray}
|4;0,2 \rangle &=& b^\dagger_{0 \uparrow} b^\dagger_{1\uparrow} b^\dagger_{0\downarrow} b^\dagger_{-1\downarrow} |0 \rangle
\nonumber \\
|4;0,3 \rangle &=& b^\dagger_{-1 \uparrow} b^\dagger_{1\uparrow} b^\dagger_{-1\downarrow} b^\dagger_{1\downarrow} |0 \rangle
\end{eqnarray}

There is only one configuration for either $k=-1, \sigma=1$,
\begin{equation}
|4;0,4 \rangle= b^\dagger_{0\uparrow } b^\dagger_{-1\uparrow} b^\dagger_{1\uparrow} b^\dagger_{-1\downarrow} |0\rangle,
\end{equation} 
or $k=1,\sigma=-1$ 
\begin{equation}
|4;0,5 \rangle= b^\dagger_{1\uparrow } b^\dagger_{0\downarrow} b^\dagger_{-1\downarrow} b^\dagger_{1\downarrow} |0\rangle.
\end{equation} 
Hence, the $j=0$ Hamiltonian reduces to a $5 \times 5$ matrix:
\\
\begin{widetext}
\(
H(4;j=0)=
  \begin{bmatrix}
    -2t_c +{4 U \over 3} -{7 J_F \over 6} -\lambda_z & {U-J_F/2 \over 3} &  -{U +J_F\over 3} & 0 & 0 \\
    {U -J_F/2\over 3}  & -2t_c+{4 U \over 3}-{7 J_F \over 6} +\lambda_z & -{U + J_F \over 3} & -{\lambda_{xy} \over \sqrt{2} }  & {\lambda_{xy} \over \sqrt{2}} \\
    -{U+J_F \over 3} & -{U +J_F \over 3} &  4 t_c+{4 U \over 3}-{5 J_F \over 3} & {\lambda_{xy} \over \sqrt{2}} & -{\lambda_{xy} \over \sqrt{2}} \\
     0  & -{\lambda_{xy} \over \sqrt{2}} & {\lambda_{xy} \over \sqrt{2} } & U+t_c-J_F+{\lambda_z \over 2}  & 0 \\
     0 & {\lambda_{xy} \over \sqrt{2}} &  -{\lambda_{xy} \over \sqrt{2}} & 0 & U+t_c-J_F+{\lambda_z \over 2} \\
  \end{bmatrix}
\)

\subsubsection{$j=-1$ sector}

We work in the basis
\begin{eqnarray}
|4;-1,1 \rangle &=&  b^\dagger_{0 \uparrow} b^\dagger_{-1\uparrow} b^\dagger_{-1\downarrow} b^\dagger_{1\downarrow} |0 \rangle
\nonumber \\
|4;-1, 2 \rangle &=&  b^\dagger_{-1\uparrow} b^\dagger_{1 \uparrow} b_{0 \downarrow} b_{-1\downarrow} |0 \rangle
\nonumber \\
|4;-1, 3 \rangle &=& b^\dagger_{0\uparrow}  b^\dagger_{1\uparrow} b^\dagger_{0\downarrow} b^\dagger_{1\downarrow} |0 \rangle,
\nonumber \\
|4;-1, 4 \rangle&=& b^\dagger_{0\uparrow} b^\dagger_{0\downarrow} b^\dagger_{-1\downarrow} b^\dagger_{1\downarrow} |0 \rangle,
\nonumber \\
|4;-1, 5  \rangle&=& b^\dagger_{0\uparrow} b^\dagger_{-1\uparrow} b^\dagger_{1\uparrow} b^\dagger_{1\downarrow} |0 \rangle.
\end{eqnarray}
The first three states have $k=-1=2$, $\sigma=0$, the fourth has $k=0$, $\sigma=-1$ and the fifth has $k=1=-2$, $\sigma=1$.
The  $j=-1$ Hamiltonian is
	\begin{equation}
	H(4;j=-1)=
	\begin{bmatrix}
	t _c+{4 U \over 3}-{7 J_F \over 6} -{\lambda_z \over 2} & {U -J_F/2\over 3} &  {U+J_F \over 3} & -{\lambda_{xy} \over \sqrt{2}} & 0 \\
	{U-J_F/2 \over 3}  & t_c+{4 U \over 3}-{7J_F \over 6}+{\lambda_z \over 2} & {U +J_F \over 3} & 0  & 0 \\
	{U+J_F \over 3} & {U +J_F \over 3} &  -2 t_c+ {4 U \over 3}-{5 J_F \over 3} & 0 &- {\lambda_{xy} \over \sqrt{2}} \\
	-{\lambda_{xy} \over \sqrt{2}} & 0 & 0 & -2 t_c+U-J_F& 0 \\
	0 & 0 &  -{\lambda_{xy} \over \sqrt{2}} & 0 & U+t_c-J_F-{\lambda_z \over 2} \\
	\end{bmatrix}
	\end{equation}
\end{widetext}

\subsubsection{$j=+1$ sector}

It is convenient to take the basis states as the time-reversed analogues of the  $j=-1$ sector:
\begin{eqnarray}
|4;+1,1 \rangle &=&  b^\dagger_{0 \uparrow} b^\dagger_{1\uparrow} b^\dagger_{-1\downarrow} b^\dagger_{1\downarrow} |0 \rangle
\nonumber 
\end{eqnarray}
\begin{eqnarray}
|4;+1, 2 \rangle &=&  b^\dagger_{-1\uparrow} b^\dagger_{1 \uparrow} b^\dagger_{0 \downarrow} b^\dagger_{1\downarrow} |0 \rangle,
\nonumber \\ 
|4;+1, 3 \rangle &=& b^\dagger_{0\uparrow}  b^\dagger_{-1\uparrow} b^\dagger_{0\downarrow} b^\dagger_{-1\downarrow} |0 \rangle,
\nonumber \\ 
|4;+1, 4 \rangle&=& b^\dagger_{0\uparrow} b^\dagger_{-1\uparrow} b^\dagger_{1\uparrow} b^\dagger_{0\downarrow} |0 \rangle,
\nonumber \\ 
|4;+1, 5  \rangle&=& b^\dagger_{-1\uparrow} b^\dagger_{0\downarrow} b^\dagger_{-1\downarrow} b^\dagger_{1\downarrow} |0 \rangle.
\end{eqnarray}
Thus one immediately sees that $H(4;j=+1)=H(4;j=-1)$. Hence, there is a double degeneracy of the 
eigenvalues $E_i(4;j=+1)=E_i(4;j=-1)$. 

For $\lambda=0$, the ground state is three-fold degenerate corresponding to the
$S=1$ triplet combination of the two unpaired spins in the cluster. These lowest three degenerate states correspond to $j=0, \pm 1$. 
From the above analysis we conclude that isolated clusters with four electrons 
can be described through the effective Hamiltonian given in Eq. (\ref{eq:trig})
where 
$D$ 
is an increasing function of SMOC as discussed in the main text.

\subsection{Isolated triangular clusters with three electrons}

The basis for  $N=3$ electrons consists of 20 configurations: 18 configurations with $S_z=1/2$ ($N_\uparrow=2,N_\downarrow=1$)
or $S_z=-1/2$ ($N_\uparrow =1, N_\downarrow=2$) and 2 configurations with $S_z=3/2$ ($N_\uparrow=3,N_\downarrow=0$)
or $S_z=-3/2$ ($N_\uparrow=0,N_\downarrow=3$).
The only allowed $j$ values for the cluster with $N=3$ electrons are 
$j=\pm {1 \over 2}, +{3 \over 2}$ with the largest ($8 \times 8$) matrix corresponding to $j=+{3 \over 2}$.  
The $j=-{3 \over 2}$ sector is not given here since the configurations are just the same
as the ones in the $j=+{3 \over 2}$ sector.

\subsubsection{$j =+ 3/2$}  

The configurations with $j=3/2$ are
\begin{eqnarray}
|3;+3/2,1 \rangle &=& b^\dagger_{0 \uparrow} b^\dagger_{1\uparrow} b^\dagger_{0\downarrow}  |0 \rangle
\nonumber \\
|3;+3/2,2 \rangle &=& b^\dagger_{0 \uparrow} b^\dagger_{-1\uparrow} b^\dagger_{-1\downarrow}  |0 \rangle
\nonumber \\
|3;+3/2,3 \rangle &=& b^\dagger_{-1 \uparrow} b^\dagger_{1\uparrow} b^\dagger_{1\downarrow} |0 \rangle
\nonumber \\
|3;+3/2,4 \rangle &=& b^\dagger_{1 \uparrow} b^\dagger_{0\downarrow} b^\dagger_{1\downarrow}  |0 \rangle
\nonumber \\
|3;+3/2,5 \rangle &=& b^\dagger_{0 \uparrow} b^\dagger_{-1\uparrow} b^\dagger_{1\uparrow} |0 \rangle
\nonumber \\
|3;+3/2,6 \rangle &=& b^\dagger_{0 \uparrow} b^\dagger_{0\downarrow} b^\dagger_{-1\downarrow}  |0 \rangle
\nonumber \\
|3;+3/2,7 \rangle &=& b^\dagger_{-1 \uparrow} b^\dagger_{-1\downarrow} b^\dagger_{1\downarrow}  |0 \rangle
\nonumber \\
|3;+3/2,8 \rangle &=& b^\dagger_{0 \downarrow} b^\dagger_{-1\downarrow} b^\dagger_{1\downarrow} |0 \rangle.
\end{eqnarray}
Yielding the  $8 \times 8$ Hamiltonian matrix
\\
\begin{widetext}
	
	\begin{eqnarray}
	&&H(3; 3/2)=\notag\\
	&&\begin{mbmatrix}
	-3t_c+{2U -5 J_F/2 \over 3}+{\lambda_z \over 2} & {U+J_F \over 3} &  {U+J_F \over 3} & -{\lambda_{xy} \over \sqrt{2}} & -{\lambda_{xy} \over \sqrt{2} } &  0   &   0  &   0 \\
	{U+J_F \over 3}  & {2 U -5 J_F/2 \over  3} & -{U +J_F \over 3} & 0  & 0 & {\lambda_{xy} \over \sqrt{2} } & -{\lambda_{xy} \over \sqrt{2} } &0  \\
	{U+J_F \over 3} & -{U+J_F \over 3} &  3 t_c+ {2 U -5 J_F/2 \over 3}-{\lambda_z \over 2} & -{\lambda_{xy} \over \sqrt{2} }& -{\lambda_{xy} \over \sqrt{2} } &   0    &  0  & 0  \\
	- {\lambda_{xy}  \over \sqrt{2}} & 0 & -{\lambda_{xy} \over \sqrt{2}} & {2 U -5 J_F/2 \over 3} & 0 &  {U +J_F \over 3} &  {U +J_F \over 3} & 0 \\
	-{\lambda_{xy} \over \sqrt{2}} & 0 &  -{\lambda_{xy} \over \sqrt{2} } & 0 & 0 &  0  &   0   & 0 \\
	0 & {\lambda_{xy} \over \sqrt{2}} & 0 & {U +J_F \over 3} & 0 & -3 t_c +{2 U  - 5 J_F/2 \over 3}+ {\lambda_z \over 2} & -{U+J_F \over 3} & -{\lambda_{xy} \over \sqrt{2}} \\
	0 & -{\lambda_{xy} \over \sqrt{2}} & 0 & {U+J_F \over 3} & 0 & -{U+J_F \over 3}  & 3t_c +{2 U -5 J_F/2 \over 3 }- {\lambda_z \over 2} & {\lambda_{xy} \over \sqrt{2}} \\
	0 & 0 & 0 & 0 & 0 & -{\lambda_{xy} \over \sqrt{2}}   & {\lambda_{xy} \over \sqrt{2}} &0 \\
	\end{mbmatrix}\notag\\
	\end{eqnarray}

\subsubsection{$j=\pm 1/2$} 

We take the basis
\begin{eqnarray}
|3;+1/2,1\rangle &=& b^\dagger_{0 \uparrow} b^\dagger_{0\downarrow} b^\dagger_{1\downarrow}  |0 \rangle
\nonumber \\
|3;+1/2,2 \rangle &=& b^\dagger_{1 \uparrow} b^\dagger_{-1\downarrow} b^\dagger_{1\downarrow}  |0 \rangle
\nonumber \\
|3;+1/2,3 \rangle &=& b^\dagger_{-1 \uparrow} b^\dagger_{0\downarrow} b^\dagger_{-1\downarrow} |0 \rangle
\nonumber \\
|3;+1/2,4 \rangle &=& b^\dagger_{-1 \uparrow} b^\dagger_{1\uparrow} b^\dagger_{0\downarrow}  |0 \rangle
\nonumber \\
|3;+1/2,5 \rangle &=& b^\dagger_{0 \uparrow} b^\dagger_{1\uparrow} b^\dagger_{-1\downarrow} |0 \rangle
\nonumber \\
|3;+1/2,6 \rangle &=& b^\dagger_{0 \uparrow} b^\dagger_{-1\uparrow} b^\dagger_{1\downarrow} |0 \rangle
\end{eqnarray}
and analogously for $j=-1/2$. 
The $6 \times 6$ Hamiltonian matrix reads
\\
	\(
	H(3;j=+1/2)=
	\begin{bmatrix}
	-3t_c+{2U \over 3}-{5 J_F \over 6}-{\lambda_z \over 2} & {U +J_F\over 3} &  {U+J_F \over 3} & 0& 0 & {\lambda_{xy} \over \sqrt{2} } \\
	{U+J_F \over 3}  & 3 t _c+ {2 U \over 3} -{5 J_F \over 6}+{\lambda_z \over 2} & -{U +J_F \over 3} & 0  & -{\lambda_{xy} \over \sqrt{2} } &  0 \\
	{U+J_F \over 3} & -{U+J_F  \over 3} & {2U \over 3}- {5 J_F \over 6}& 0 & 0  & 0  \\
	0  & 0 &0 & {2 U \over 3 }- {J_F \over 3} &  {U -J_F/2\over 3}  &-{U -J_F/2 \over 3} \\
	0  & -{\lambda_{xy} \over \sqrt{2}} &  0 & {U-J_F/2 \over 3} &  {2U \over 3}-J_F/3+\lambda_z & {U-J_F/2 \over 3} \\
	{\lambda_{xy} \over \sqrt{2}} & 0 &  0 & -{U-J_F/2 \over 3} & {U-J_F/2 \over 3} & {2 U \over 3}- {J_F \over 3} -\lambda_z \\
	\end{bmatrix}
	\)
Due to Kramers theorem the eigenstates, $E_n(3;j=1/2)=E_n(3;j=-1/2)$ and the energy levels for $E_n(3;j=3/2)$ are at least doubly 
degenerate. With no SMOC present, $E_n(3;j=\pm 1/2)=E_n(3;j=3/2)$ and the eigenstates are four-fold degenerate. 
However, when SMOC is present $E_n(3;j=\pm 1/2) \neq E_n(3;j=3/2)$ and the four-fold degeneracy is broken leading to 
two-fold degenerate levels.

\section{Expression for effective spin models from the canonical transformation of the $t$-$J$ model}\label{app:B}

In this appendix we model the $\ell$th trinuclear complex by the three site $t$-$J$ model, i.e.,
	\begin{eqnarray}
	H_{t-J}^{(\ell)}\equiv P_0\left[\sum_{\sigma,j=1}^{3}t_c\left(h^\dagger_{\ell j\sigma}h_{\ell (j+1)\sigma}+h^\dagger_{\ell j\sigma}h_{\ell (j-1)\sigma}\right)-\frac{J_c}{4}\sum\limits_{i\ne j\ne k=1}^{3}\sum_{\sigma,\sigma'} h_{\ell i\sigma} h^\dagger_{\ell j\sigma} (1- n_{\ell j\uparrow})(1- n_{\ell j\downarrow}) a_{\ell j\sigma'} a^\dagger_{\ell k\sigma'}
	\right]P_0, \notag\\
	\end{eqnarray}
where $h^\dagger_{\ell i\sigma}=a_{\ell i\sigma}$ creates an hole with spin $\sigma$ in the $i$th Wannier orbital and $P_0$ projects out states that contain empty sites. Note that it is important to retain the `three site' terms here, as we will need to consider states far from half-filling. For a single molecule the effective low-energy model, retaining only the three lowest energy states is given by Eq. (\ref{eq:trig}) with
\begin{eqnarray}
	D&=&  \frac{\lambda_z^2-\lambda_{xy}^2}{6 \left(2 t_c-J_c\right)}. \label{eq:Dapp}
\end{eqnarray}

	The $t$-$J$ model  of the interlayer coupling between neighbouring molecules $\ell$ and $m$ is
	\begin{eqnarray}
	H_{t-J}^c =  P_0\left[-t_z\sum_{\sigma}\sum_{j=1}^{3}\left(h^\dagger_{\ell j\sigma}h_{mj\sigma}+h^\dagger_{mj\sigma}h_{\ell j\sigma}\right)+J_z\sum_{j=1}^{3} \left( \hat{\bm S}_{\ell j}\cdot\hat{\bm S}_{mj} -\frac{\hat n_{\ell j}\hat n_{mj}}{4} \right)
	\right]P_0,
	\end{eqnarray}
where now three are no three site terms because of the topology of underlying tight-binding model [cf. Eq. (\ref{eq:tubeHa}) and Fig. \ref{fig:tubes}b]. Performing the canonical transformation described in section \ref{sec:canon} and  retaining quadratic terms in $t_z$, linear terms in $J_z$ (as $J_z$ is already quadratic in $t_z$) and quadratic terms in the SMOC (i.e., up to order $\lambda_z^2$, $\lambda_{xy}^2$, or $\lambda_{xy}\lambda_z$) yields an effective Hamiltonian described by Eq. (\ref{eq:c}) with
\begin{subequations}
\begin{eqnarray}
	\Delta D^{c} &=& -\frac{t_z^2}{81}\left[\frac{28t_c+J_c}{(2t_c-J_c)^3t_c}\lambda_z^2
	-\frac{24J_ct_c^3 +29J_c^2t_c^2 -17J_c^3t_c+2J_c^4}{2(4t_c-J_c)(2t_c-J_c)^3t_c^3}\lambda_{xy}^2
	\right], \label{eq:Dappc}
\\
J^{c}&=& \frac{J_z}{3}  \left[1
-\frac{1}{12(2 t_c-J_c)^2}\lambda_{z}^2
-\frac{J_c^2}{48(2 t_c-J_c)^2 t_c^2}\lambda_{xy}^2
\right]
\notag\\&&
+\frac{t_z^2}{81}\frac{36}{ 2t_c-J_c}\left[ 1  
+ \frac{2}{9(2t_c-J_c)^2}\lambda_z^2 
-\frac{160t_c^4 -48J_ct_c^3 -52J_c^2t_c^2 +26J_c^3t_c -3J_c^4}
{72(4t_c-J_c)(2t_c-J_c)^2t_c^3}\lambda_{xy}^2 
\right],
\\	
\Delta^c&=& 1 
+ \frac{ J_z}{48(2 t_c-J_c) t_{{z}}^2}
\left(
7\lambda_{z}^2
+
\frac{ 
	48J_ct_c^3 
	-12J_c^2t_c^2 
	-9J_c^3t_c 
	+2J_c^4
} 
{4 (4 t_c-J_c) t_c^3}\lambda_{xy}^2\right)
-\frac{1}{9 ({2 t_c-J_c})^2}
\left[
\lambda_{z}^2
+ \frac{24 J_c t_c^3 - 6 J_c^3 t_c + J_c^4}{8 (4 t_c-J_c) t_c^3}\lambda_{xy}^2
\right],\notag\\
\\
	P_{zz}&=&   
	\frac{4t_z^2}{9 (2t_c-J_c)^3} \lambda_z^2,
	\\
	P_{xx}&=&   
	\frac{t_z^2}{81}\frac{J_c^2(5t_c-J_c)}{ (2t_c-J_c)^3t_c^3}  \lambda_{xy}^2, 
	\\
	P_{zx}&=&   \frac{P_{xx}+P_{zz}}{2}.
	\end{eqnarray}
\end{subequations}

	The $t$-$J$ model  of the in-plane coupling between molecules $\ell$ and $m$ along a `1-bond' (cf. Fig. \ref{fig:decorated}) is
	\begin{eqnarray}
	H_{t-J}^{ab}=-t_g\sum_{\sigma} P_0\left( 
	\hat a_{\ell 1\sigma}^\dag \hat a_{m1\sigma} +
	\hat a_{m1\sigma}^\dag \hat a_{\ell 1\sigma} \right)P_0
	+ J_cP_0\left( {\bm S}_{\ell 1}\cdot{\bm S}_{m1} -\frac{\hat n_{\ell 1}\hat n_{m1}}{4} \right)
	P_0,
	\end{eqnarray}
	again the three site terms vanish because of the underlying tight-binding model [Eq. (\ref{eq:dumb})]. Performing the canonical transformation, adding in the 2- and 3-bonds, as described in section \ref{sec:canon}, and retaining quadratic terms in $t_g$, linear terms in $J_g$  and quadratic terms in the SMOC yields an effective Hamiltonian described by Eq. (\ref{eq:ab-compassXXZ}) with
\begin{subequations}
	\begin{eqnarray}
\Delta D^{ab} &=&
-\frac{t_g^2}{81}\left[\frac{30 t_c^2 -16 J_c t_c +2 J_c^2}{9 (4 t_c-J_c) (2 t_c-J_c)^2 t_c^2}\lambda_z^2
-\frac{96 J_c t_c^3 -212 J_c^2 t_c^2 +90 J_c^3 t_c -11 J_c^4}{36 (4 t_c-J_c) (2 t_c-J_c)^3 t_c^3}\lambda_{xy}^2 
\right], \label{Delta_g_pp}
	\\
	J^{ab}&=& \frac{J_g}{9} \left[1 - \frac{\lambda_z^2}{12(2t_c-J_c)^2} - \frac{J_c^2\lambda_{xy}^2}{48(2t_c-J_c)^2t_c^2}
\right] \notag\\&&
+   \frac{t_g^2}{81}\frac{4J_c}{t_c(2t_c-J_c)}\left[ 1  + \frac{(5t_c^2-5t_cJ_c+J_c^2)}{36(2t_c-J_c)^2}\frac{\lambda_z^2}{t_c^2} 
- {\frac{240 t_c^3 - 190 J_c t_c^2 + 53 J_c^2 t_c  -5 J_c^3}{36 (4 t_c-J_c) (2 t_c-J_c)^2}}
\frac{\lambda_{xy}^2}{t_c^2}
\right],
	\\
Q&=&\left[\frac{J_g}{9}\frac{J_c(8t_c+5J_c)}{144(2t_c-J_c)^2}
- \frac{t_g^2}{81}\frac{48t_c^2J_c-26t_cJ_c^2+3J_c^3}{18t_c(2t_c-J_c)^2(4t_c-J_c)}\right]
\frac{\lambda_{xy}^2}{t_c^2},
	\\
	\Delta^{ab}&=& 1 
	-
	\frac{J_g}{16 J_c (2 t_c-J_c) t_g^2}
	\left[
	(4t_c-J_c)\lambda_{z}^2
	-
	\frac{384 t_c^3 - 152 J_c t_c^2 + 3 J_c^3}{4 (4 t_c-J_c) t_c} \lambda_{xy}^2
	\right]
	\notag\\&&
	+
	\frac1{36 (J_c - 2 t_c)^2 t_c}
	\left[
	(7 t_c-J_c) \lambda_z^2
	-
	\frac{96 t_c^2 - 38 J_c t_c + 3 J_c^2}{4 t_c-J_c}\lambda_{xy}^2
	\right],
	\\
	J_{xz}^{ab}&=&\frac{1}{\sqrt{2}}\left[
	-\frac{J_g}{9}\left(\frac{J_c }{36 (2 t_c-J_c)^2 t_c}\right)
	+\frac{t_g^2}{81}\left(\frac{J_c(12t_c^2 -6 J_c t_c^2 + J_c^2)}{9 (4 t_c-J_c) (2 t_c-J_c)^3 t_c^2}\right)
	\right]\lambda_{xy}\lambda_z.
	\end{eqnarray}
\end{subequations}

\end{widetext}

\end{appendix}

\end{document}